\documentclass[a4paper,onecolumn,showpacs,notitlepage,floatfix,nofootinbib,superscriptaddress,prd]{revtex4-2}
\pdfoutput=1
\usepackage[utf8]{inputenc}

\usepackage[bookmarks, linktocpage, colorlinks = true, linkcolor = blue, urlcolor  = blue, citecolor = blue, anchorcolor = green, hyperindex = true, hyperfigures]{hyperref}

\usepackage[T1]{fontenc} 
\usepackage{bm} 

\usepackage{amsthm}

\usepackage{mathrsfs} 
\usepackage{subcaption}
\usepackage{mdframed}
\usepackage{enumitem}
\usepackage{xcolor}
\usepackage{subcaption}
\usepackage{mathtools}
\usepackage{tikz-cd}
\usepackage[normalem]{ulem}
\usepackage{amsmath}
\usepackage{amssymb}
\usepackage{amsfonts}
\usepackage{latexsym}

\usepackage{subcaption}
\usepackage{mdframed}
\usepackage{enumitem}
\usepackage{subcaption}
\usepackage{mathtools}
\usepackage{tikz-cd}
\usepackage{pict2e}
\newmdenv[skipabove=6mm]{kotak}   

\DeclareFontEncoding{LS1}{}{}
\DeclareFontSubstitution{LS1}{stix}{m}{n}
\DeclareSymbolFont{symbols2}{LS1}{stixfrak}{m}{n}
\DeclareMathSymbol{\lcangle}{\mathopen}{symbols2}{"E9}
\DeclareMathSymbol{\rcangle}{\mathclose}{symbols2}{"EA}

\makeatletter
\newsavebox{\@brx}
\newcommand{\llcangle}[1][]{\savebox{\@brx}{\(\m@th{#1\lcangle}\)}%
  \mathopen{\copy\@brx\kern-0.5\wd\@brx\usebox{\@brx}}}
\newcommand{\rrcangle}[1][]{\savebox{\@brx}{\(\m@th{#1\rcangle}\)}%
  \mathclose{\copy\@brx\kern-0.5\wd\@brx\usebox{\@brx}}}
\makeatother

\newcommand{\be}{\begin{eqnarray}}
\newcommand{\ee}{\end{eqnarray}}

\newcommand{\nl}{\nonumber \\}
\newcommand{\pd}{\partial}

\newcommand{\Tr}{{\rm Tr}}

\newcommand{\bul}{\overset{\underset{\bullet}{}}}

\newcommand{\lcurvyangle}{\llcangle}
\newcommand{\rcurvyangle}{\rrcangle}

\newtheorem*{fact*}{Fact}

\allowdisplaybreaks

\begin{document}

\title{{
    Exact quantization conditions and full transseries structures \\
    for ${\cal PT}$ symmetric anharmonic oscillators \\
  }
}


\author{Syo Kamata}
\email{skamata11phys@gmail.com}
\affiliation{Department of Physics, The University of Tokyo, 7-3-1 Hongo, Bunkyo-ku, Tokyo 113-0033, Japan}

\begin{abstract}
We study exact Wentzel–Kramers–Brillouin analysis (EWKB) for a ${\cal PT}$ symmetric quantum mechanics (QM) defined by the potential $V_{\cal PT}(x) = \omega^2 x^2 + g x^{2 K} (i x)^{\varepsilon}$ with $\omega \in {\mathbb R}_{\ge 0}$, $g \in {\mathbb R}_{>0}$ and $K, \varepsilon \in {\mathbb N}$ to clarify its perturbative/non-perturbative structure.
In our analysis, we mainly consider the massless cases, i.e., $\omega = 0$, and derive the exact quantization conditions (QCs) for arbitrary $(K,\varepsilon)$ including all perturbative/non-perturbative corrections.
From the exact QCs, we clarify full transseries structure of the energy spectra with respect to the inverse energy level expansion, and then formulate the Gutzwiller trace formula, the spectral summation form, and the Euclidean path-integral.
For the massive cases, i.e., $\omega > 0$, we show the fact that, by requiring existence of solution of the exact QCs, the path of analytic continuation in EWKB is uniquely determined for a given $N = 2K + \varepsilon$, and in consequence the exact QCs, the energy spectra, and the three formulas are all perturbative.
Similarities to Hermitian QMs and resurgence are also discussed as additional remarks.
\end{abstract}

\maketitle
\tableofcontents
\flushbottom

\section{Introduction}

Non-Hermitian quantum theories have important roles in a wide area of physics and provides rich physically interesting phenomena.
Those theories are also interesting topics from the view points of mathematical and computational physics, and those have been actively studied in recent years.
A ${\cal PT}$ symmetric theory is a particular class of non-Hermitian theories. 
A ${\cal PT}$ symmetric quantum mechanics (QM) was proposed in Refs.~\cite{Bender:1998ke,Bender:1998gh}, and its field theoretical generalization was also considered in Ref.~\cite{Bender:2018pbv}.
In high energy physics, study of the ${\cal PT}$ symmetric theories is currently one of the interesting subjects from various aspects, such as field theoretical understanding~\cite{Felski:2021evi,Bender:2021fxa}, beyond the standard model~\cite{Mavromatos:2021hpe,Romatschke:2022jqg,Romatschke:2022llf,Grable:2023paf,Weller:2023jhc,Romatschke:2023sce,Mavromatos:2024ozk}, and mathematical understanding~\cite{Dorey:2007zx,Emery:2019znd}.
See also Refs.~\cite{Bender:2019,Bender:2023cem}.

Due to broken Hermiticity, ${\cal PT}$ symmetric potentials can contain negative coupling, such as $V(x)= - g x^4$ with $g \in {\mathbb R}_{>0}$, and unstable at $x = \pm \infty$.
For this reason, the variable, $x$, is usually considered to be a complex value and defined as a (real) one-dimensional orbit on the complex $x$-plane to be consistent with ${\cal PT}$ invariance and to gain convergence of the wavefunctions.
The remarkable property of ${\cal PT}$ symmetric QMs is that, despite lack of Hermiticity in a Hamiltonian, energy spectra are real and bounded~\cite{Bender:1998ke,Dorey:2001uw,Jones:2006qs,Bender:2006wt}.
Then, a naive question arises; \textit{in the theoretical view point, how much and in what sense are PT symmetric QMs similar to/different from Hermitian QMs?}
There are several approaches to partially answer to this question, such as pseudo-Hermiticity~\cite{Mostafazadeh:2001jk,Mostafazadeh:2002pd,Mostafazadeh:2003gz}, ${\cal PT}/{\cal CPT}$ duality~\cite{Jones:2006qs,Bender:2006wt}, and the Ai-Bender-Sarkar conjecture~\cite{Ai:2022csx,Lawrence:2023woz,Kamata:2023opn}.
Each of them addressed quite important subjects for their theoretical and mathematical structures as well-defined quantum theories, i.e., Hilbert space and inner-product, energy spectra, and correspondence to analytic continuation of Hermitian QMs and their non-perturbative contributions.
They are also crucial for generalizations to ${\cal PT}$ symmetric field theories.

\begin{figure}[t]
\vspace{-1.5cm}
    \centering
    \includegraphics[scale=0.4]{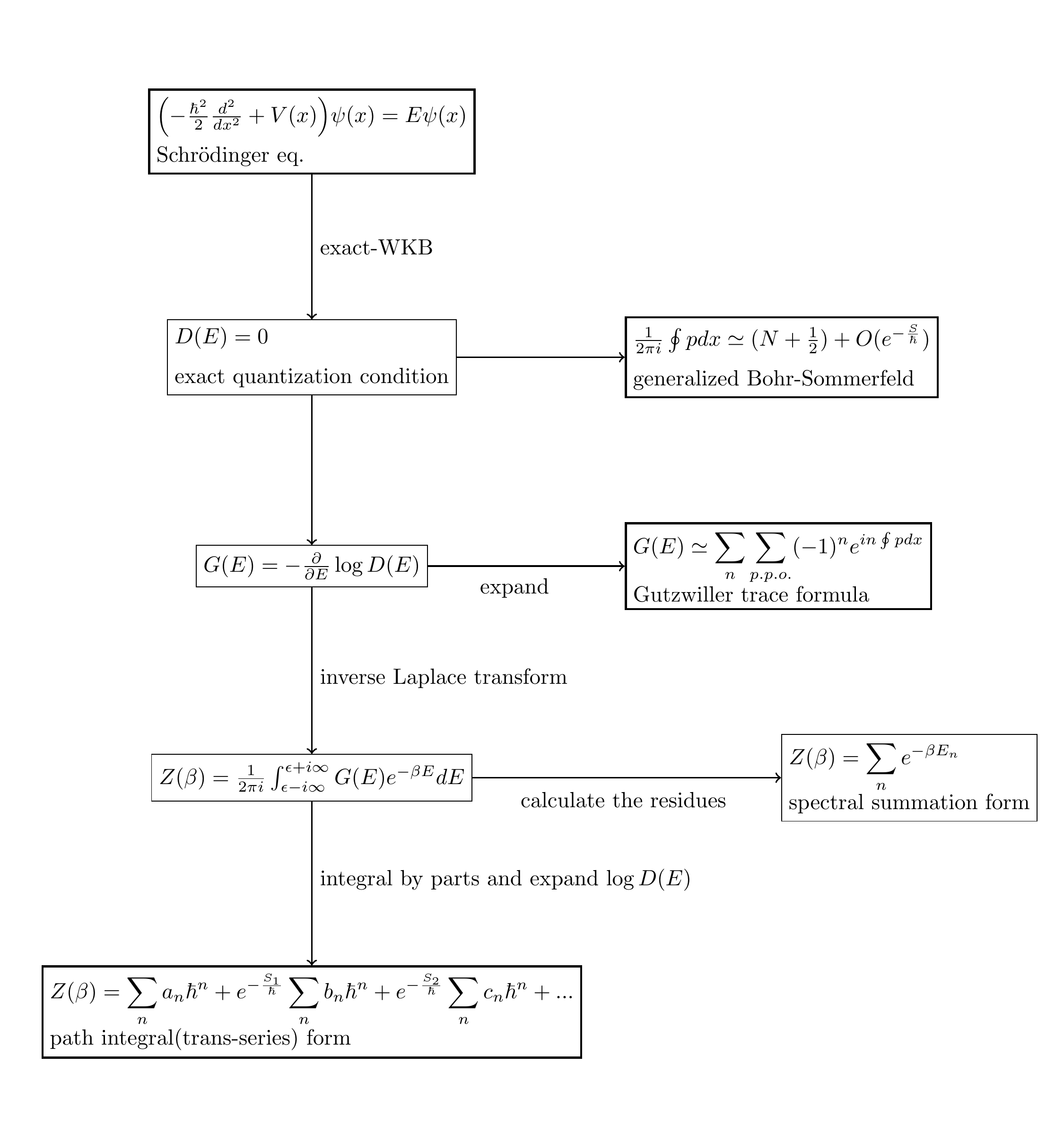}
    \vspace{-10mm}
    \caption{Flowchart of various formulas from an exact quantization condition in EWKB.
      This figure was originally shown in Ref.~\cite{Sueishi:2021xti}.
    }
    \label{fig:flowchart}
\end{figure}

In this paper, we study EWKB for the ${\cal PT}$ symmetric potential defined by the following Schr\"{o}dinger equation, ${\cal L} \psi = 0$:
\be
&& {\cal L} = - \hbar^2 \pd_x^2 + V(x) - E, \qquad \quad \hbar, E \in {\mathbb R}_{>0}, \, x \in {\mathbb C}, \nl 
&& V_{\cal PT}(x) = \omega^2 x^2 +  g x^{2 K} (i x)^{\varepsilon}, \qquad \omega \in {\mathbb R}_{\ge 0}, \  g \in {\mathbb R}_{>0}, \ K, \varepsilon \in {\mathbb N}, \label{eq:V_PT}
\ee
to clarify its perturbative/non-perturbative structure.
EWKB is formulated based on Borel resummation theory, and it is quite powerful to analyze non-perturbative physics in QMs.
We mainly consider the massless cases, i.e. $\omega =0$, and address the following issues:
\begin{enumerate}
\item[(i)] derivation of the exact quantization conditions for arbitrary $(K,\varepsilon)$ including all perturbative/non-perturbative corrections, 
\item[(ii)] clarification of full transseries structure of the energy spectra and their $(K,\varepsilon)$-dependence,
\item[(iii)] formulating the Gutzwiller trace formula, the spectral summation form, and the Euclidean path-integral from the exact quantization condition.
\end{enumerate}
We firstly try to derive the exact quantization conditions (QCs) which correspond to a generalization of the Bohr-Sommerfeld condition by using EWKB.
For application of EWKB to the massless monomial potentials, it is convenient to rescale $x$ as $x \rightarrow \left( \frac{E}{g} \right)^{1/N} x$ with $N := 2K + \varepsilon$ in Eq.(\ref{eq:V_PT}) and redefine the Schr\"{o}dinger operator, ${\cal L}$, as\footnote{
  Rescaling not only $x$ but also the time and the momentum as $(t,p) \rightarrow ((g E^{(N-2)/2})^{-1/N} t, E^{1/2} p)$ gives the same time-dependent Schr\"{o}dinger equation and uncertainty relation, i.e.
\be
i \hbar \frac{\pd}{\pd t} = \widehat{H} \ \rightarrow \ i \eta \frac{\pd}{\pd t} = \widetilde{H}, \qquad [x,p] = i \hbar \ \rightarrow \ [x,p] = i \eta,
\ee
where $\widetilde{H} := \widehat{H}/E$, and $\widetilde{H} \psi = \psi$.
}
\be
&& \frac{\cal L}{E} \rightarrow {\cal L} = - \eta^2 \pd_x^2 + Q, \qquad Q := x^{2K}(i x)^\varepsilon  -1,  \label{eq:L_res} \\
&& \eta := \frac{g^{1/N} \hbar}{E^{(N + 2)/(2N)}}. \label{eq:def_eta_E}
\ee
Thus, the wavefunction can be expanded by $\eta$, and all the parameters $(\hbar,g,E)$ appear in $\eta$ only.
$\varepsilon$ in the potential is a crucial parameter through this paper and is introduced as a deformation parameter of a domain of $x$ on the complex plane from the real axis.
In this sense, the potential has to be complexified, and non-trivial non-perturbative structures depending on $(N,K)$ are expected.
Since energy spectra including all perturbative/non-perturbative corrections can be obtained by solving the exact QCs, we then clarify full transseries structure of the energy spectra from the exact QCs.
Finally, we try to obtain the picture of Fig.~\ref{fig:flowchart} for arbitrary $(N,K)$.
Once constructing the exact QCs, one can construct various formulas, such as the Gutzwiller trace formula (GTF)~\cite{Gutzwiller1971}, the spectral summation form (SSF), and the Euclidean path-integral (EPI) from \textit{only} the exact QCs~\cite{Sueishi:2020rug}.
It would be helpful to see non-perturbative effects from the viewpoint of each the formula.

We would also briefly discuss the massive case, i.e. $\omega > 0$.
In the massive cases, the standard $\hbar$-expansion works for the energy spectra, and their transseries structures become much simpler than those of the massless cases.
It is because, in contrast to the massless cases, a suitable complex domain of $x$ is uniquely determined for a given $N$ by requiring existence of solution of the exact QCs.
As a result, the exact QCs, the energy spectra, and the formulas in Fig.~\ref{fig:flowchart} are all purely perturbative.
We show these facts.

This paper is organized as follows:
In Sec.~\ref{sec:setup_preparation}, we review EWKB.
In Sec.~\ref{sec:quant_cond}, we construct exact QCs for the massless case with arbitrary $(N,K)$ using EWKB based on the $\eta$-expansion.
In Sec.~\ref{sec:energy_trans}, by using the exact QCs, we consider the transseries structure of the energy spectra with respect to the inverse energy level expansion.
In Sec.~\ref{sec:varous_formula}, we construct the three formulas, such as the Gutzwiller trace formula, the spectral summation form, and the Euclidean path-integral, using the exact QCs.
In Sec.~\ref{sec:mass_cases}, we discuss the massive cases.
In Sec.~\ref{sec:comments}, we make some comments on similarities to Hermitian QMs and resurgence.
Sec.~\ref{sec:summary} is devoted to summary and conclusion.
Technical computations, such as construction of the ${\cal CPT}$ inner-product, derivation of ${\bf Fact}$ in Sec.~\ref{sec:mass_cases}, and the alien calculus in Sec.~\ref{sec:resurgence_analysis} are summarized in Apps.~\ref{sec:pseudo_Herm}, \ref{sec:mass_borel}, and \ref{sec:Alien_energy}, respectively.

This study is a generalization of earlier works analyzed by the (standard) WKB analysis, e.g., Refs.~\cite{Bender:1998ke,Bender:1998gh,bender1999complex,bender2013advanced,Bender:2019,Ai:2022csx}, in $\varepsilon \in {\mathbb N}$.
Many parts of our analyses in this paper are based on Refs.~\cite{Sueishi:2020rug,Sueishi:2021xti,Kamata:2021jr,Kamata:2023opn}.


\section{Exact WKB analysis} \label{sec:setup_preparation}

In this section, we review EWKB in our setup in Sec.~\ref{sec:EWKB_connect}, and then explain Voros symbols and the Delabaere-Dillinger-Pham formula (DDP) formula which have a key role for Borel resummation in EWKB in Sec.~\ref{sec:voros_DDP}.
There are many nice reviews for Borel resummation theory and EWKB.
See, for example, Refs.~\cite{Ec1,delabaere1994introduction,Sauzin:1405,Marino:2012zq,Dorigoni2019,Aniceto:2018bis,delabaere2016divergent,Costin2008} and Refs.~\cite{Berry,BPV,Silverstone,Takei1,Takei2,Schafke1,Alvarez1,Zinn-Justin:2004vcw,Dunne:2013ada,Dunne:2014bca}, respectively.

\subsection{EWKB ansatz and connection formula} \label{sec:EWKB_connect}
Through this paper, we perform EWKB to the ${\cal PT}$ symmetric QM in Eq.(\ref{eq:V_PT}).
In the ${\cal PT}$ symmetric QM, the variable, $x$, is extend to a complex value and can be taken to be a (real) one-dimensional orbit to converge the wavefunction at $|x| = \infty$ on the complex $x$-plane.
Parity and time-reversal transformations are defined as ${\cal P}: x \rightarrow - x$ and ${\cal T}: (i, x ) \rightarrow  ( -i , \bar{x})$, where $\bar{x}$ is the complex conjugation of $x$, respectively, so that ${\cal PT}$ symmetry, i.e. ${\cal PT}: (i,x) \rightarrow (-i,-\bar{x})$, gives a constraint to the domain of $x$.
Despite the constraint from ${\cal PT}$ symmetry, in general the domain is not uniquely determined because there are a number of asymptotic domains to converge the wavefunction.
The asymptotic domains can be classified by a pair, $(N, K)$ (or $(K, \varepsilon)$) with $N := 2K + \varepsilon$, by being continuously deformed by the change of $\varepsilon$ with a fixed $K$.
By this manner, we take the following subspace on the complex $x$-plane as the domain of $x$:
\be   
&& \gamma_{{\cal PT}_{(N,K)}} = \Theta(-s) s e^{+i \theta{(N,K)}} s +  \Theta(s) s e^{-i \theta{(N,K)}}, \qquad (s \in {\mathbb R}) \nl
&& \theta{(N,K)} := \frac{\pi (N-2K)}{2 (N+2)} = \frac{\pi\varepsilon}{2 (N+2)}, \label{eq:gam_PT}
\ee
where  $\Theta(s)$ is the step function.
Notice that the domain can be continuously deformed not to change the result when performing the analytic continuation in EWKB.

In order to find the picture in Fig.~\ref{fig:flowchart}, we firstly have to obtain a generalized QC denoted by ${\frak D}$ by taking the following procedure in EWKB:
\begin{enumerate}
\item  drawing a Stokes graph by preparing an ansatz to the wavefunction,
\item and then, performing analytic continuation along the path, $\gamma_{\cal PT}$, in Eq.(\ref{eq:gam_PT}) to obtain a monodromy matrix.  
  One of the components corresponds to the QC by imposing normalizability to the wavefunction.
\end{enumerate}
The above process is the same for any values of $(N,K)$.
In the below, we would explain the procedure for the massless cases, i.e., $\omega = 0$ in Eq.(\ref{eq:V_PT}), but for the massive cases the similar analysis works by replacing $\eta$ with $\hbar$ as an expansion parameter.
\\  \par
In our analysis, we use $\eta$ for the expansion parameter by beginning with Eq.(\ref{eq:L_res}) and assume the following EWKB ansatz:
\be
\psi_a(x,\eta) = \sigma (\eta)\exp \left[ \int_a^x  dy \, S(y,\eta) \right], \qquad S(x,\eta) \sim \sum_{\ell \in  {\mathbb N}_0 -1} S_{\ell}(x) \eta^{\ell} \quad \mbox{as} \quad \eta \rightarrow 0_+, \label{eq:ansatz_wave}
\ee
where $\sigma(\eta)$ is the integration constant generally depending on $\eta$,  and $a \in {\mathbb C}$ is a normalization point for the wavefunction on the complex $x$-plane.
The coefficients, $S_{\ell}(x)$, are determined order by order from the Riccati equation given by Eq.(\ref{eq:L_res}) as
\be
S(x,\eta)^2 + \pd_x S(x,\eta) = \eta^{-2} Q(x), \qquad 
\ee
where $Q(x)$ is defined in Eq.(\ref{eq:L_res}).
Explicitly, it can be written down as
\be
&& S_{-1}(x) = \pm \sqrt{Q(x)}, \qquad S_{0}(x) = - \frac{\pd_x \log Q(x)}{4}, \nl
&& S_{+1}(x) = \pm \frac{1}{8 \sqrt{Q(x)}} \left[\pd_x^2 \log Q(x) - \frac{(\pd_x \log Q(x))^2}{4} \right], \qquad \cdots.
\ee
The formal expansion, $S(x,\eta)$, can be decomposed into two parts as the odd- and even-power expansions as
\be
&& S_{\rm od}(x, \eta) = \sum_{\ell \in {\mathbb N}_0} S_{2\ell-1}(x) \eta^{2\ell -1}, \qquad S_{-1}  = \sqrt{Q(x)}, \label{eq:Sod} \\
&& S_{\rm ev}(x, \eta) = \sum_{\ell \in {\mathbb N}_0} S_{2\ell}(x) \eta^{2\ell } = - \frac{1}{2} \pd_{x} \log S_{\rm od}(x,\eta), \label{eq:Sev}
\ee
and $S_{\rm ev}(x, \eta)$ can be expressed by $S_{\rm od}(x,\eta)$.
As a result, the wavefunction (\ref{eq:ansatz_wave}) can be expressed as
\be
\psi_{a \pm}(x,\eta) &=& \frac{\sigma_{\pm} (\eta)}{\sqrt{S_{\rm od}(x,\eta)}} \exp \left[ \pm \int_a^x  dy \, S_{\rm od}(y,\eta) \right] \nl
 &=& \sigma_{\pm}(\eta) \exp \left[ \pm \frac{1}{\eta} \int_{a}^{x} dx^\prime \, S_{{\rm od},-1}(x^\prime) \right] \sum_{n \in {\mathbb N}_0} \psi_{a \pm,n}(x) \eta^{n+\frac{1}{2}}.  \label{eq:psi_apm}
\ee
where $\pm$ corresponds the two independent transseries solutions of the Schr\"{o}dinger equation.
Borel resummation is a composite operation of Borel transform ${\cal B}$ and Laplace integral ${\cal L}_\theta$, i.e., ${\cal S}_\theta : = {\cal L}_\theta \circ {\cal B}$.
These operations to the wavefunction are defined as
\be
&& {\cal B}[\psi_{a\pm}](x, \xi) := \frac{\psi_{a \pm,n}(x)}{\Gamma(n+\frac{1}{2})}(\xi \pm \xi_0(x))^{n - \frac{1}{2}} = \psi_{B,a\pm}(x, \xi), \qquad \xi_0(x) := \sqrt{Q_0(x)}, \\
&& {\cal L}_{\theta}[\psi_{B,a\pm}](x, \eta) := \int^{\infty e^{i \theta} }_{\mp \xi_0} d \xi \, e^{-\frac{\xi}{\eta}} \psi_{B,a\pm}(x,\xi),
\ee
In the below, we take a vectorial from for the wavefunction, as $\psi_a := (\psi_{a +},\psi_{a -})^{\top}$.

A Stokes graph holds all information of Borel summability of the wavefunction and can be drawn by a specific form of $\eta^{-1} \int dx \, S_{{\rm od},-1}(x)$, i.e., the leading order of $\int dx \, S_{\rm od}(x,\eta)$.
Turning points are defined from a potential as ${\rm TP} := \{ x \in {\mathbb C}  \, | \, Q(x)  = 0 \}$, and in our case it is given by
\be
   {\rm TP} = \{ e^{ \pi i \frac{4 n - N + 2 K}{2 N}} \in {\mathbb C} \, | \, n \in \{ 1, \cdots, N \} \}. \label{eq:TP_def}
\ee
We attach labels to each turning point as $a_1,\cdots,a_N$ in such a way that
\be
   {\rm Re}[a_1] \le \cdots \le {\rm Re}[a_N], \quad \mbox{and} \quad
   {\rm Im}[a_{n}] < {\rm Im}[a_{n+1}] \ \ \mbox{if} \ \ {\rm Re}[a_{n}] = {\rm Re}[a_{n+1}].
\ee
By this manner, all the labels are uniquely determined.
These turning points are also used for the normalization point in the wavefunction.
After obtaining turning points, we find Stokes lines which emerge from each the turning point.
Those are defined as
\be
{\rm Im} \left[ \eta^{-1} \int_a^{x} dy \, S_{-1}(y) \right] = 0 \quad \mbox{with} \quad \eta \in {\mathbb C}, \ a \in {\rm TP}.
\ee
If ${\rm Re} \left[ \eta^{-1} \int_a^{x} dy \, S_{-1}(y) \right]$ is monotonically increasing (resp. decreasing) as going far away from the turning point along the Stokes line, we attach a label, ``$+$'' (resp. ``$-$''), to the line.

When performing analytic continuation along a certain path on the complex $x$-plane, one has to glue wavefunctions on each domain separated by Stokes lines.
In EWKB, it is carried out by introducing connection matrices.
One has to be careful that an infinitesimally small phase has to be introduced to $\eta$ before performing the analytic continuation if Stokes a phenomenon occurs at $\arg(\eta) = 0$.
Otherwise, the wavefunction is Borel non-summable on the entire complex domain.
A specific form of the connection matrices generally depend on a type of the turning point, e.g., how many Stokes lines emerge from it.
When it is a simple turning point called as Airy-type Stokes graph, the connection matrix for crossing the Stokes line anti-clockwise is given by
\be
M_{+} =
\begin{pmatrix}
  1 &&  i \\
  0 && 1
\end{pmatrix}, \qquad M_{-} =
\begin{pmatrix}
  1 && 0 \\
  i && 1
\end{pmatrix},
\ee
where $M_+$ or $M_-$ is chosen by the label of the Stokes line determined by the behavior of ${\rm Re} \left[ \eta^{-1} \int_a^{x} dy \, S_{-1}(y) \right]$, and their inverse correspond to crossing them clockwise.
Notice that the connection matrix is determined in such a way that \textit{the Borel resummed wavefunction} is continuous at a point on the Stokes line, i.e.
\be
 {\cal S}_{\theta}[\psi_a^{\rm I}(x_* + 0_-)] = {\cal S}_{\theta}[\psi_a^{\rm I}(x_* + 0_+)], \qquad  \psi_a^{\rm I}(x_* + 0_+) := M_\pm \psi_a^{\rm II}(x_* + 0_+),
\ee
where $\psi^{\rm I,II}$ are wavefunctions defined on certain domains, ${\rm I}$ and ${\rm II}$, separated by the Stokes line, and $x_*$ is a point on the Stokes line that we are crossing.
In addition, we assumed that $x_* + 0_-$ and $x_* + 0_+$ belong to the ${\rm I}$ and ${\rm II}$ domains, respectively.
When there exists a number of turning points, one has to change the normalization point to an appropriate turning point for next crossing the Stokes line.
This is performed by acting the normalization matrix as
\be
&& \psi_{a_{n_1}} = N_{a_{n_1},a_{n_2}}\psi_{a_{n_2}}, \qquad a_{n_1}, a_{n_@} \in {\rm TP}, \nl
&& N_{a_{n_1},a_{n_2}} :=
\begin{pmatrix}
  e^{\int_{a_{n_1}}^{a_{n_2}} dx \, S_{\rm od}(x,\eta)}  &&  0 \\
0  &&  e^{-\int_{a_{n_1}}^{a_{n_2}} dx \, S_{\rm od}(x,\eta)} 
\end{pmatrix}.
\ee
In addition, the Airy-type Stokes graph has a branch-cut.
When the wavefunction goes through the branch-cut, the effect to the wavefunction can be expressed by a branch-cut matrix $T$ defined as
\be
T : =
\begin{pmatrix}
0  & & -i \\
-i  & & 0
\end{pmatrix}, \qquad M_\pm T = T M_\mp, \qquad N_{a_{n_1},a_{n_2}} T = T N^{-1}_{a_{n_1},a_{n_2}}.
\ee
which swaps the components of the wavefunction.
One can compute a monodromy matrix, denoted by ${\cal M}$, by taking the path in Eq.(\ref{eq:gam_PT}) from $s = -\infty$ to $+\infty$ and appropriately taking the connection matrices, normalization matrices, and branch-cut matrices:
\be
   {\cal S}_{\theta} [\psi^{(s = -\infty)}_a](e^{- i \theta(N,K)} s, \hbar) = {\cal S}_{\theta} [{\cal M}] \cdot {\cal S}_{\theta} [\psi_a^{(s = +\infty )}](e^{- i \theta(N,K)} s,\hbar), \quad (s \gg 1)
\ee
where ${\cal S}_\theta[\psi^{(s = +\infty)}]$ is the (Borel resummed) solution of the Schr\"{o}dinger equation in the domain corresponding to $s \gg 1$, and ${\cal M}$ consists of $M_\pm$, $N_{a_{n_1},a_{n_2}}$, and $T$.
As a result, the Borel resummed wavefunction, ${\cal S}_{\theta} [\psi^{(s = -\infty)}_a]$, is analytic-continued from $s = -\infty$ to $+\infty$.
Imposing normalizability to the wavefunction requires vanishing one of the components in ${\cal M}$, which corresponds to the QC, ${\frak D} = 0$.
Since ${\frak D}$ is a function of $E$, solving ${\frak D} = 0$ in terms of $E$ gives an energy spectrum.
\\ 

\begin{figure}[t]
  \begin{center}
    \begin{tabular}{cc}
      \begin{minipage}{0.5\hsize}
        \begin{center}
          \includegraphics[clip, width=75mm]{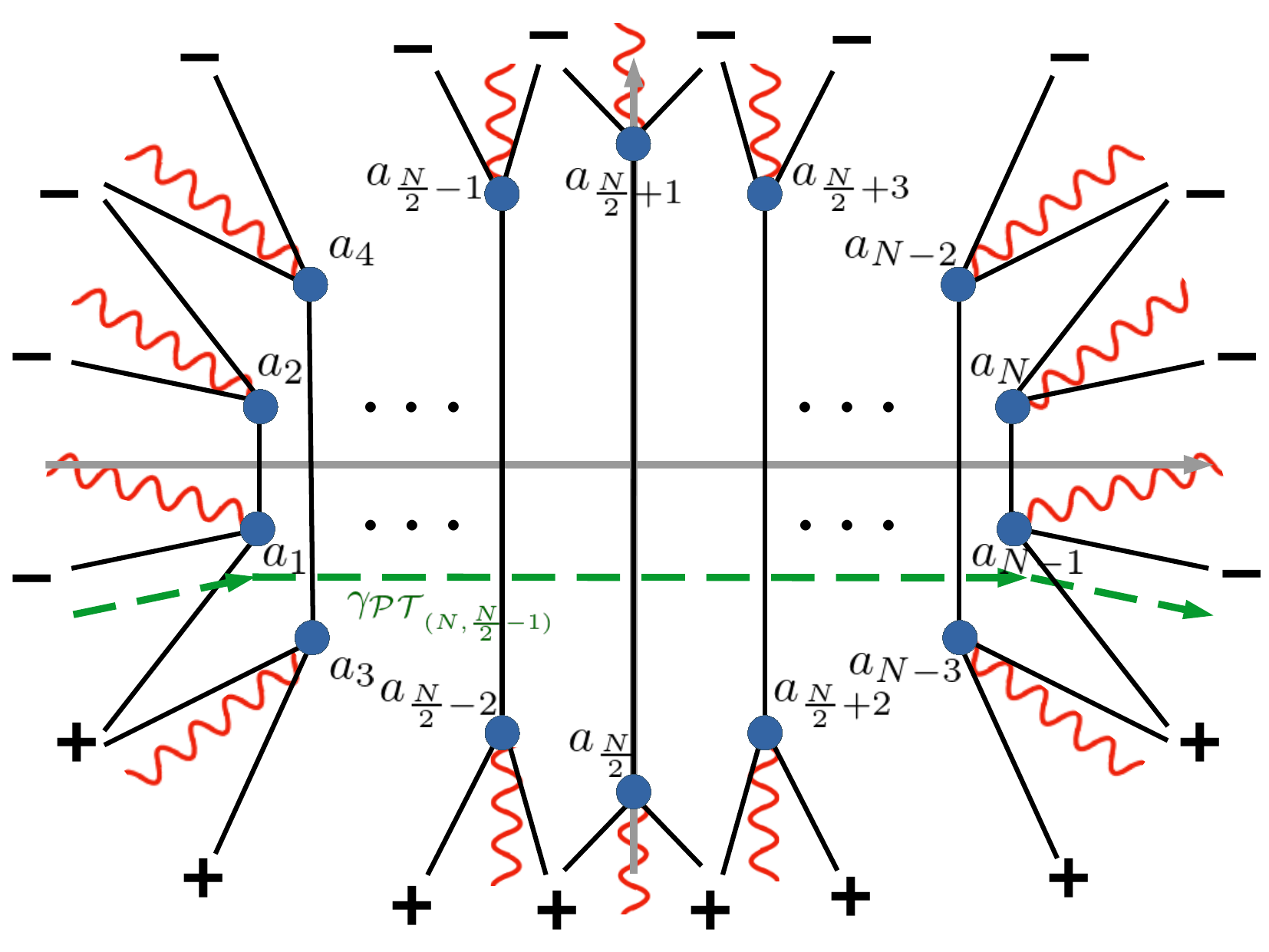}
          \hspace{1.6cm} (E-1) $K \in 2{\mathbb N}$ and $\varepsilon \in 4{\mathbb N}_0 + 2$
        \end{center}
      \end{minipage}
      \begin{minipage}{0.5\hsize}
        \begin{center}
          \includegraphics[clip, width=75mm]{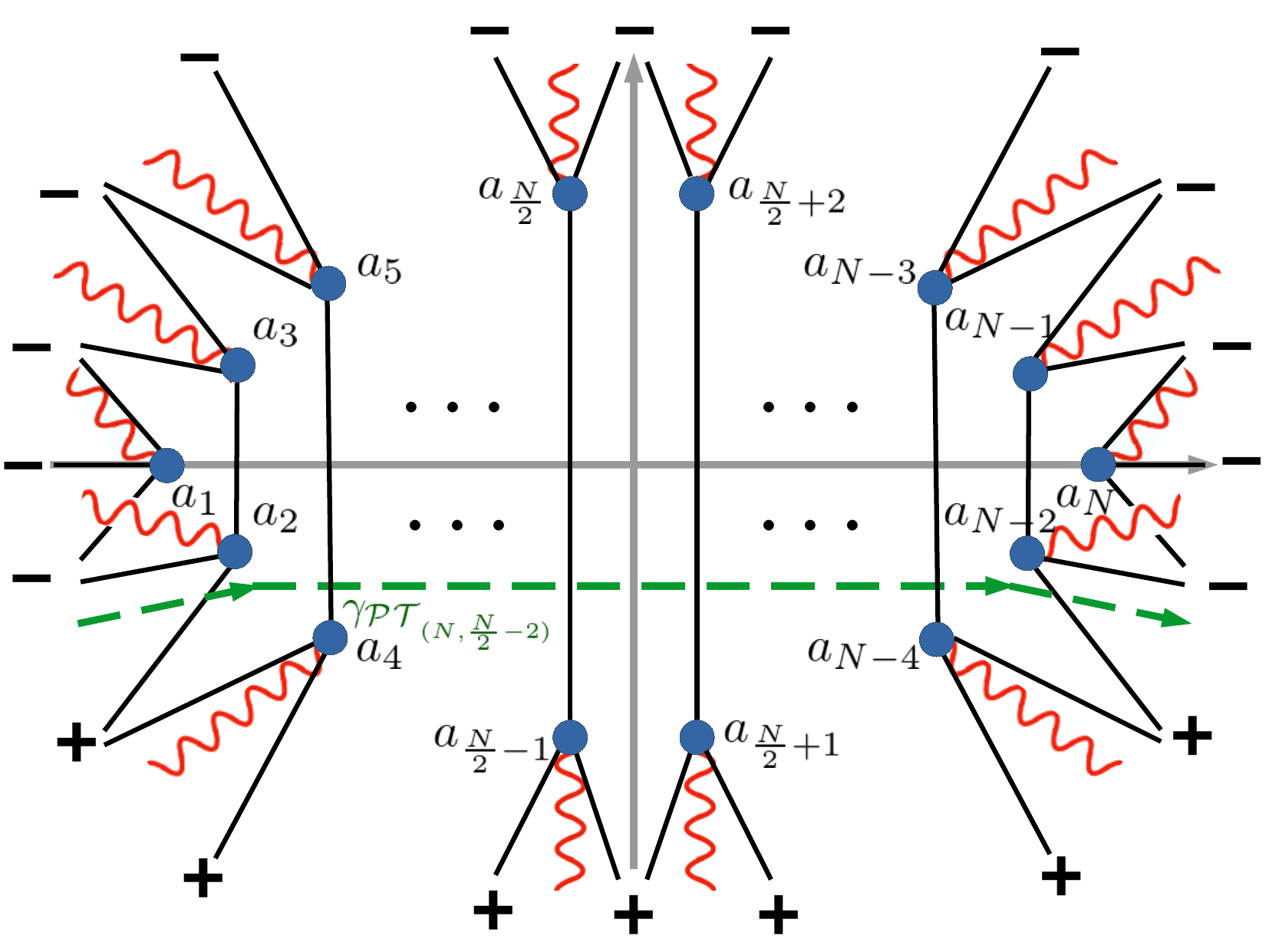} 
          \hspace{1.6cm} (E-2) $K \in 2{\mathbb N}_0 + 1$ and $\varepsilon \in 4{\mathbb N}$
        \end{center}
      \end{minipage} \\ \\
      \begin{minipage}{0.5\hsize}
        \begin{center}
          \includegraphics[clip, width=75mm]{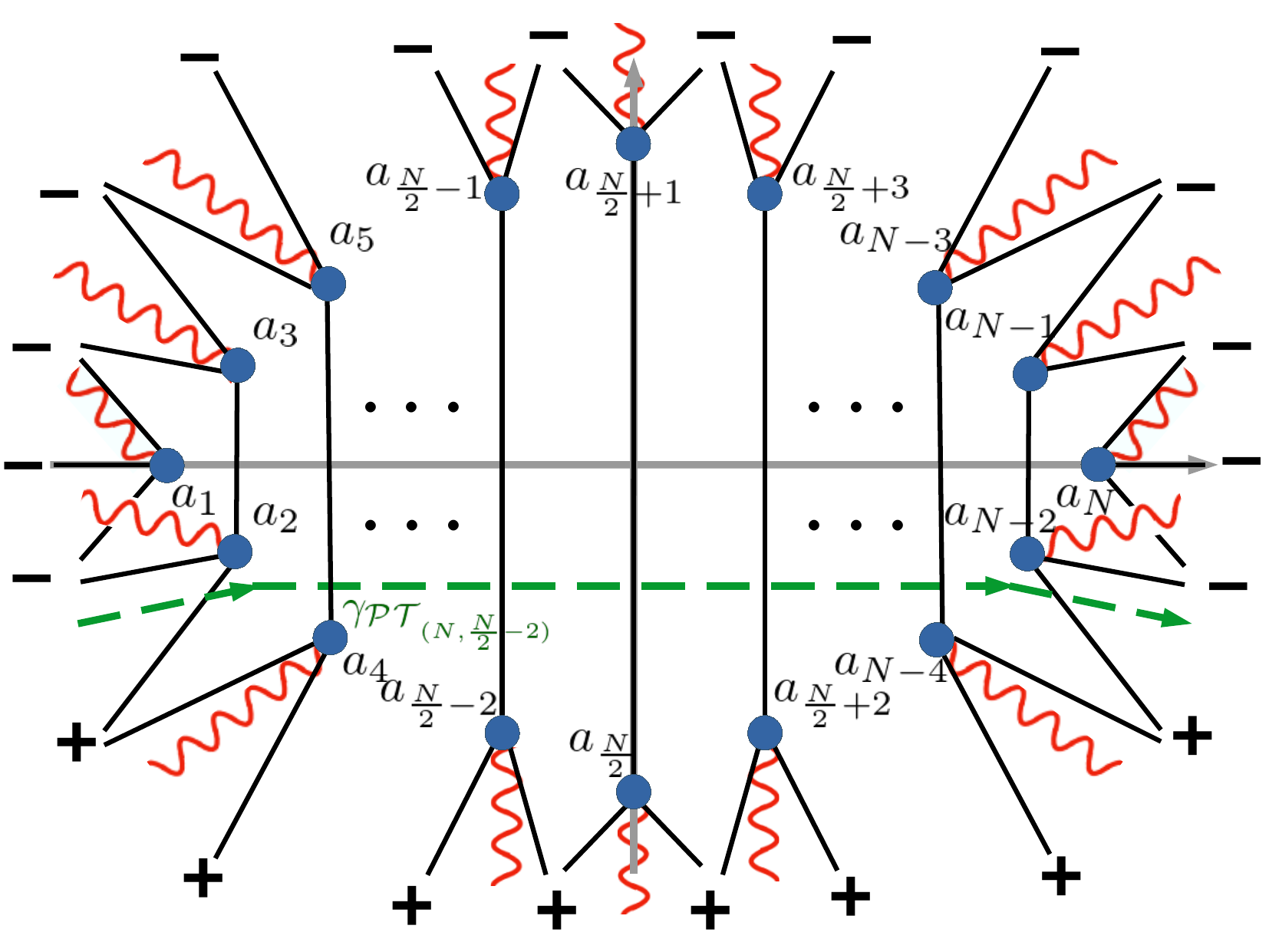}
          \hspace{1.6cm} (E-3) $K \in 2{\mathbb N}$ and $\varepsilon \in 4{\mathbb N}$
        \end{center}
      \end{minipage}
      \begin{minipage}{0.5\hsize}
        \begin{center}
          \includegraphics[clip, width=75mm]{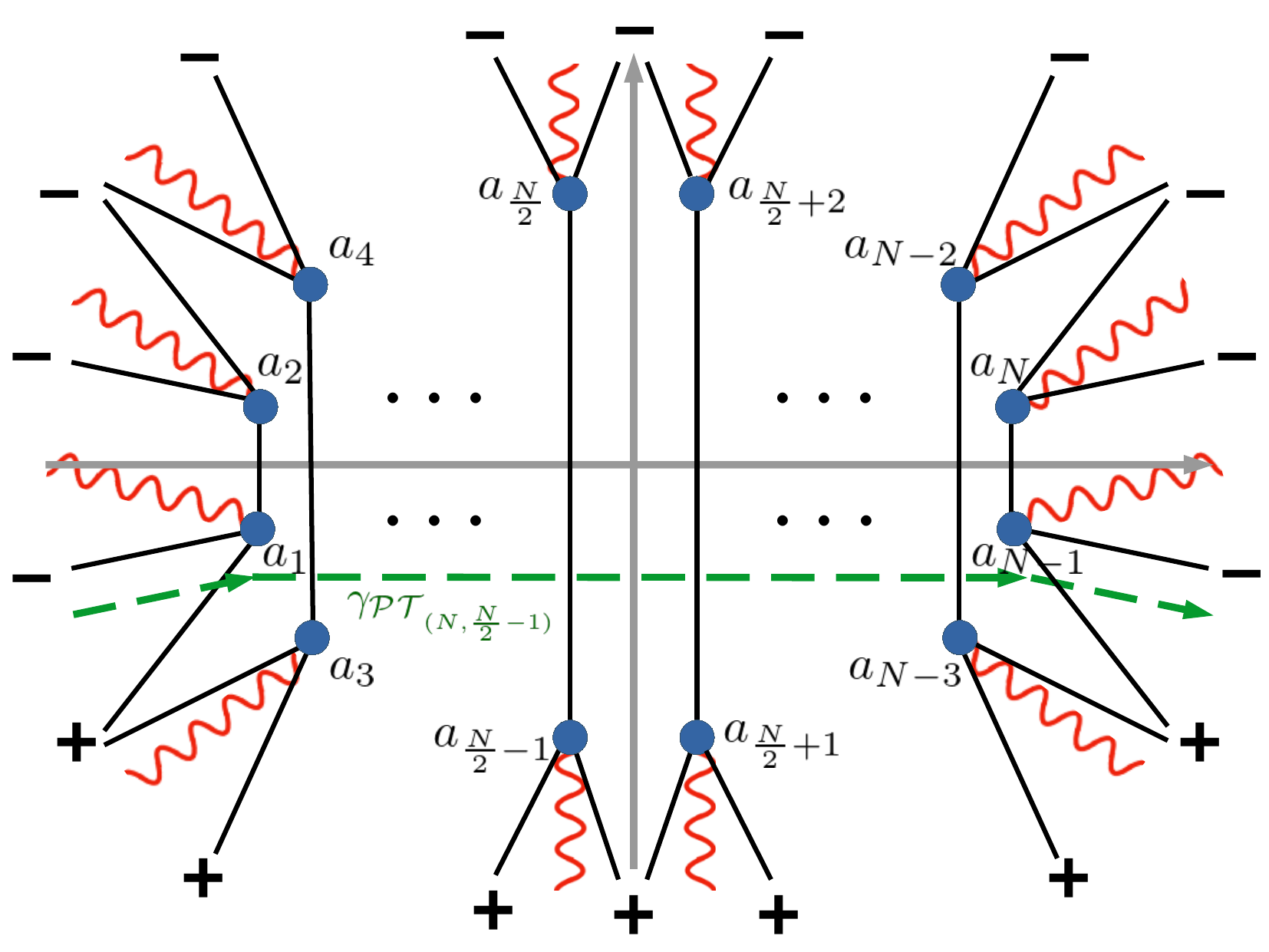} 
          \hspace{1.6cm} (E-4) $K \in 2{\mathbb N}_0 +1$ and $\varepsilon \in 4{\mathbb N}_0 + 2$
        \end{center}
      \end{minipage}       
    \end{tabular} 
    \caption{Stokes graphs given by even $N$ for $\arg(\eta) = 0$.
      The gray arrows are the real and imaginary axes.
      The blue dots, black lines, and red waves denote turning points, Stokes lines, and branch-cuts, respectively.
      The green lines denote the path of analytic continuation in Eq.(\ref{eq:gam_PT}) which is the nearest to the real axis under the condition of $(K,\varepsilon)$.
    }
    \label{fig:Stokes_shape_even}
  \end{center}
\end{figure}

\begin{figure}[t]
  \begin{center}
    \begin{tabular}{cc}
            \begin{minipage}{0.5\hsize}
        \begin{center}
          \includegraphics[clip, width=75mm]{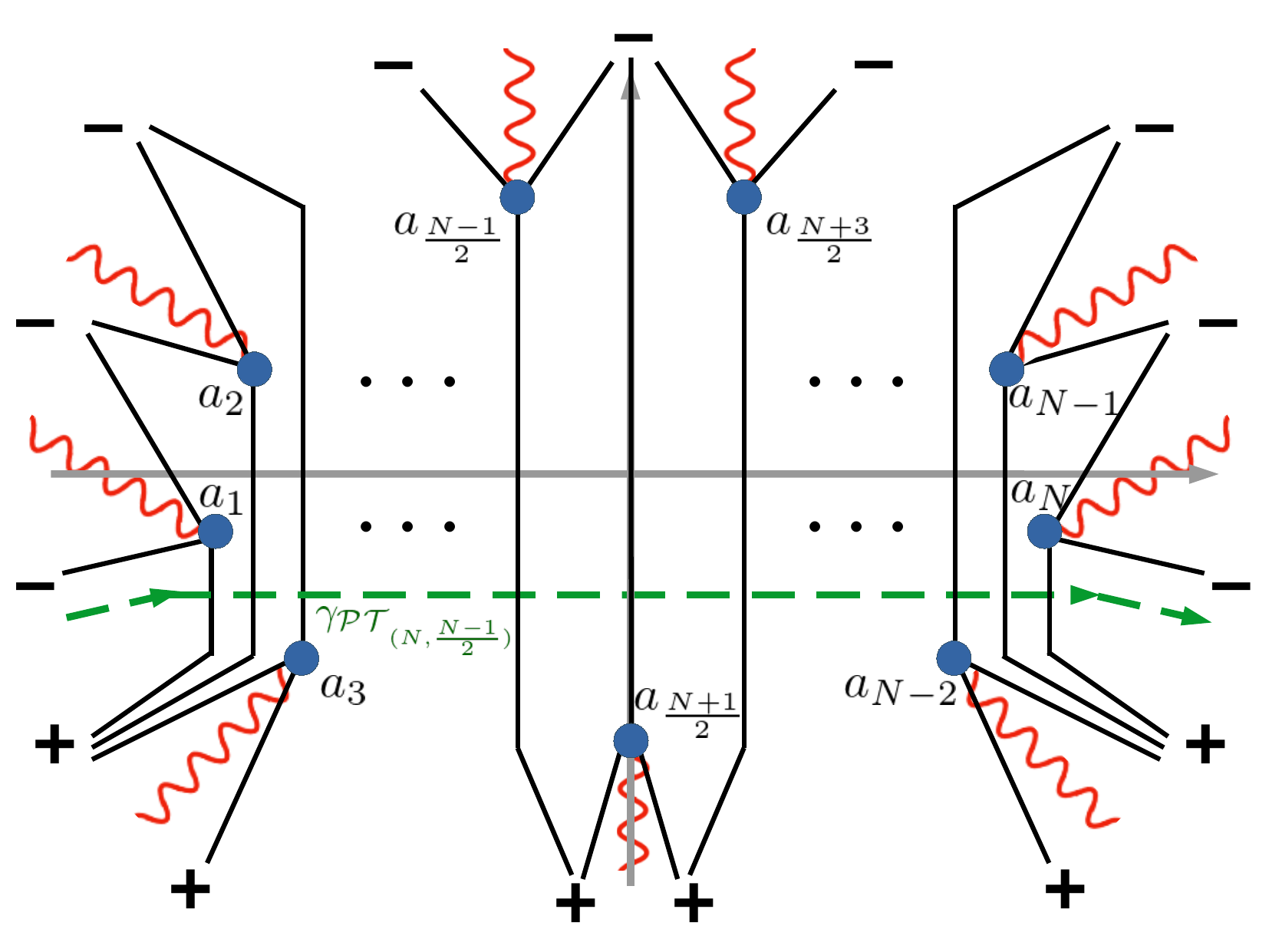}
          \hspace{1.6cm} (O-1) $K \in 2{\mathbb N}$
        \end{center}
      \end{minipage}
      \begin{minipage}{0.5\hsize}
        \begin{center}
          \includegraphics[clip, width=75mm]{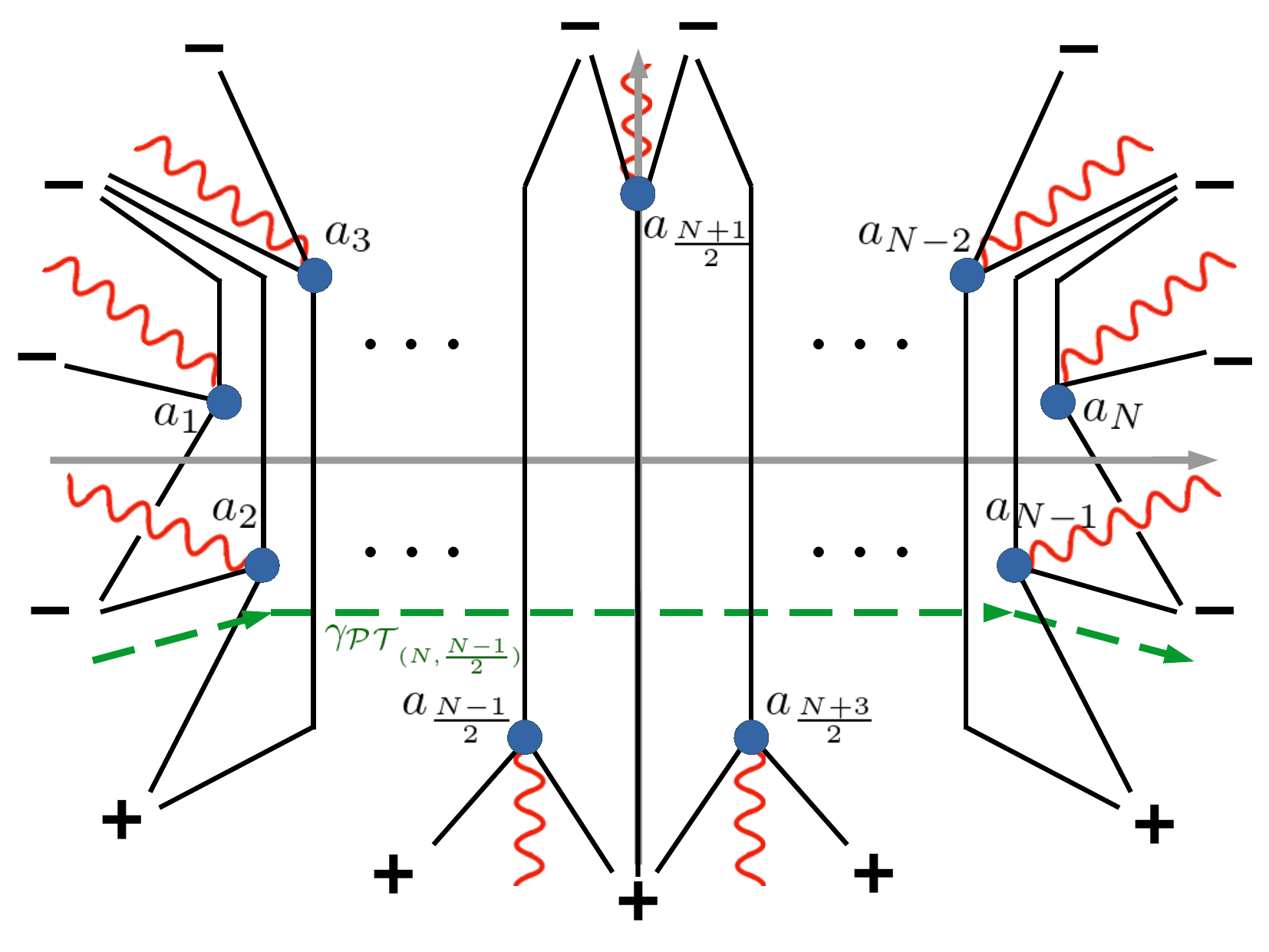} 
          \hspace{1.6cm} (O-2) $K \in 2{\mathbb N}_0 + 1$
        \end{center}
      \end{minipage} 
    \end{tabular} 
    \caption{Stokes graphs given by odd $N$ for $\arg(\eta) = 0$.
      The gray arrows are the real and imaginary axes.
      The blue dots, black lines, and red waves denote turning points, Stokes lines, and branch-cuts, respectively.
      The green lines denote the path of analytic continuation in Eq.(\ref{eq:gam_PT}) which is the nearest to the real axis under the condition of $K$.
    }
    \label{fig:Stokes_shape_odd}
  \end{center}
\end{figure}

It is worth to see Stokes graphs that we will consider in this paper.
Shapes of the Stokes graphs can be also classified by $(N,K)$ (or $(K,\varepsilon)$) as follows:
\begin{itemize}
\item Even $N$
\begin{itemize}
\item[(E-1)]
  \underline{$K \in 2{\mathbb N}$ and $\varepsilon \in 4{\mathbb N}_0+2$ $(N \in 4 {\mathbb N} + 2)$:} \\
  There exists a pair of turning points $(a_{\frac{N}{2}},a_{\frac{N}{2}+1})$ such that ${\rm Re}[a_{\frac{N}{2}}] = {\rm Re}[a_{\frac{N}{2}+1}]= 0$ and ${\rm Im}[a_{\frac{N}{2}}] = -{\rm Im}[a_{\frac{N}{2}+1}]$ < 0.
\item[(E-2)]
  \underline{$K \in 2{\mathbb N}_0+1$ and $\varepsilon \in 4{\mathbb N}$ $(N \in 4 {\mathbb N} + 2)$:} \\
  There exists a pair of turning points $(a_{1},a_{N})$ such that ${\rm Re}[a_{1}] = - {\rm Re}[a_{N}] < 0$ and ${\rm Im}[a_{1}] = {\rm Im}[a_{N}]$ = 0.  
\item[(E-3)]
  \underline{$K \in 2{\mathbb N}$ and $\varepsilon \in 4{\mathbb N}$ $(N \in 4 {\mathbb N})$:} \\
  There exist pairs of turning points $(a_{1},a_{N})$ and $(a_{\frac{N}{2}},a_{\frac{N}{2}+1})$ such that ${\rm Re}[a_{1}] = - {\rm Re}[a_{N}] < 0$ and ${\rm Im}[a_{1}] = {\rm Im}[a_{N}]$ = 0, and ${\rm Re}[a_{\frac{N}{2}}] = {\rm Re}[a_{\frac{N}{2}+1}]= 0$ and ${\rm Im}[a_{\frac{N}{2}}] = -{\rm Im}[a_{\frac{N}{2}+1}]$<0, respectively.
\item[(E-4)]
  \underline{$K \in 2{\mathbb N}_0 + 1$ and $\varepsilon \in 4{\mathbb N}_0 + 2$ $(N \in 4 {\mathbb N})$:} \\
  Such a pair on the real and the imaginary axes does not exists.
\end{itemize} 
\item Odd $N$
\begin{itemize}
\item[(O-1)]
  \underline{$K \in 2{\mathbb N}$:} \\
    There exists a turning point $a_{\frac{N+1}{2}}$ such that ${\rm Re}[a_{\frac{N+1}{2}}] = 0$ and ${\rm Im}[a_{\frac{N+1}{2}}] < 0$.
\item[(O-2)]    
  \underline{$K \in 2{\mathbb N}_0+1$:} \\
  There exists a turning point $a_{\frac{N+1}{2}}$ such that ${\rm Re}[a_{\frac{N+1}{2}}] = 0$ and ${\rm Im}[a_{\frac{N+1}{2}}] > 0$.
\end{itemize}
\end{itemize}
The schematic figures for the even and odd $N$ cases are shown in Figs.~\ref{fig:Stokes_shape_even} and \ref{fig:Stokes_shape_odd}, respectively.
Due to the ${\mathbb Z}_{N}$ symmetry given by
\be
{\mathbb Z}_N \ : \  x \rightarrow e^{2 \pi i \frac{n}{N} } x \quad \mbox{with} \quad n \in \{ 0,\cdots, N-1 \} \label{eq:ZN_symm}
\ee
in the potential (\ref{eq:L_res}), the turning points distribute as a regular polygon.
In our convention, we choose branch-cuts in such a way that the labels of Stokes lines are all ``$-$'' (resp. ``$+$'') if the associated asymptotic domains are above (resp. below) the path of analytic continuation in Eq.(\ref{eq:gam_PT}).
By this manner, the lower component of the wavefunction, $\psi_{a-}$,  needs to be zero in the limit that $s \rightarrow \pm \infty$, and thus, one can find the QC from the resulting monodromy matrix ${\cal M}$ obtained by analytic continuation along $\gamma_{\cal PT}$ as ${\frak D} :\propto {\cal M}_{12} = 0$.
For even $N$, the ${\mathbb Z}_2$ symmetry defined by
\be
{\mathbb Z}_2 :  x \rightarrow - x \quad \subset {\mathbb Z}_N \quad \mbox{if} \ \ \mbox{$N$ is even}, \label{eq:Z2_symm}
\ee
remains in the potential as a subgroup of the ${\mathbb Z}_N$ symmetry in Eq.(\ref{eq:ZN_symm})\footnote{
  To avoid confusion, we distinguish this ${\mathbb Z}_2$ symmetry from the ${\cal P}$ symmetry.
  It is because the domain of $x$ on the complex plane has been determined by the constraint from the ${\cal PT}$ symmetry and is not invariant under the ${\cal P}$ symmetry.
}.
As we can see later, this ${\mathbb Z}_2$ symmetry has a crucial role for non-perturbative structure in the QCs.



\subsection{Voros symbol and Delabaere-Dillinger-Pham formula} \label{sec:voros_DDP}
\begin{figure}[t]
    \centering
    \includegraphics[scale=0.30]{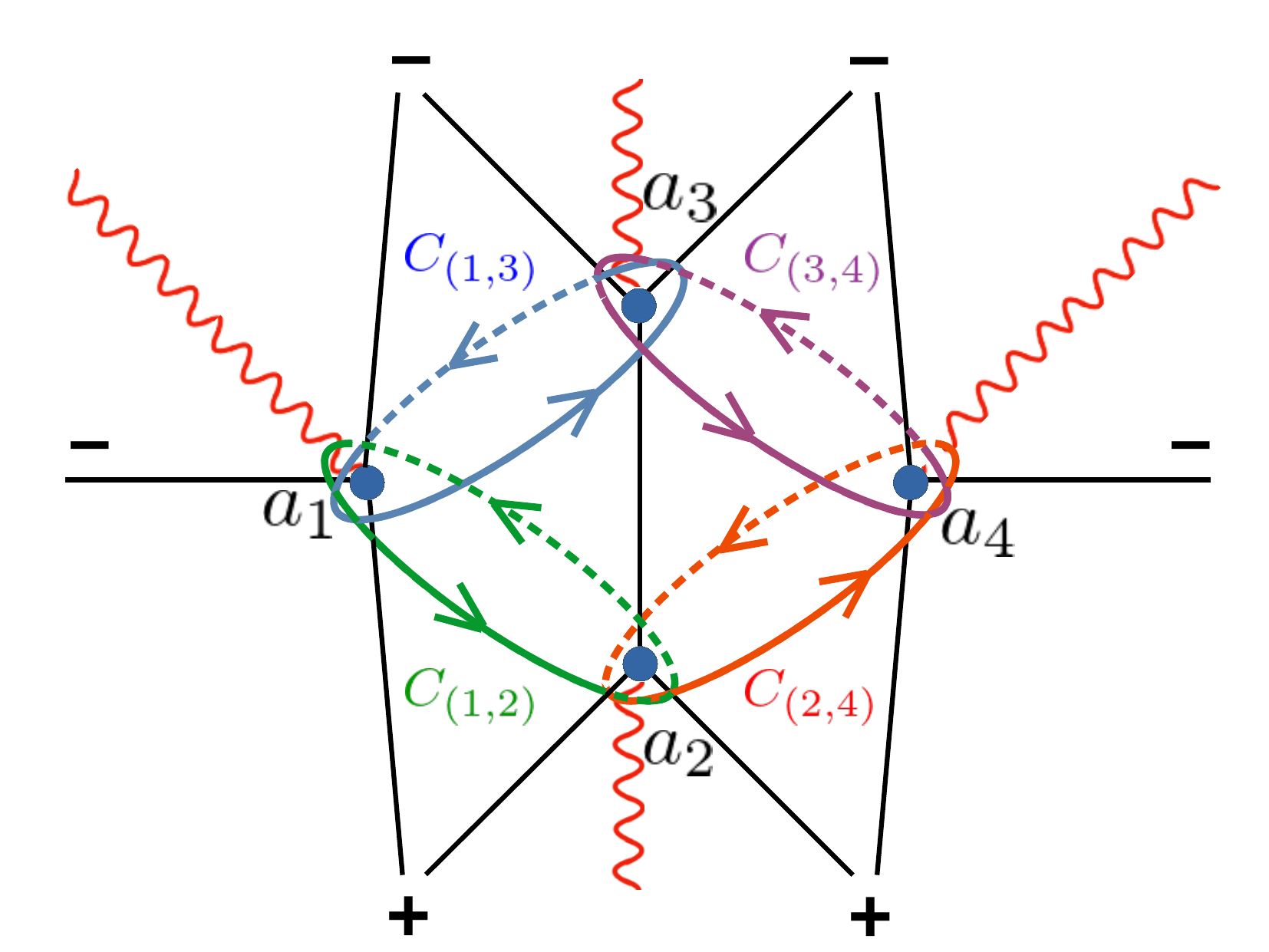}
    \caption{Example of cycles defined in Eq.(\ref{eq:C_cycle}).
      In this graph, the set of non-perturbative cycles defined by Eq.(\ref{eq:def_CsetNP}) is ${\bf C}_{{\rm NP},\theta} = \{ C_{(2,3)}\}$, where $ C_{(2,3)} = C_{(1,2)}^{-1} \cdot C_{(1,3)} = C_{(2,4)} \cdot C_{(3,4)}^{-1}$.
      The intersection numbers defined in Eq.(\ref{eq:int_sec}) are $\langle C_{(1,2)}, B \rangle = \langle C_{(1,3)}, B \rangle = \langle C_{(2,4)}, B \rangle = \langle C_{(3,4)}, B \rangle = +1$, where $B:=C_{(2,3)}$.
    }
    \label{fig:def_cycles}
\end{figure}

In EWKB, the QCs are generally expressed by Voros symbols (periodic cycles)~\cite{Voros1983}, and one has to take care of their perturbative/non-perturbative relations when a Stokes phenomenon happens at $\arg(\eta) = 0$.
The situation always arises for any even $N$.

A cycle, $C_{(n_1, n_2)}$, is defined as a contour integration of  $S_{\rm od}(x,\eta)$ going around two turning points as
\be
C_{(n_1, n_2)}(\eta) &:=& \exp \left[ \oint^{a_{n_2}}_{a_{n_1}} dx \, S_{\rm od}(x,\eta) \right]. \qquad (a_{n_1}, a_{n_2} \in {\rm TP}) \label{eq:C_cycle}
\ee
These cycles generally satisfy 
\be
C_{(n_1,n_2)} = C_{(n_2,n_1)}^{-1} , \qquad C_{(n_1,n_2)} = C_{(n_1,n_3)} \cdot C_{(n_3,n_2)}.
\ee
We show an example of the cycles in Fig.~\ref{fig:def_cycles}.
In our case, thanks to the ${\mathbb Z}_N$ symmetry (\ref{eq:ZN_symm}), all cycles with fixed $(N,K)$ can be written by the same formal expansion $\phi(e^{i \vartheta} \eta)$ with a complex phase $\vartheta$ depending on a turning point which $C_{(n_1,n_2)}$ goes around.
Explicitly, it can be written as
\be
C_{(n_1, n_2)}(\eta) &=& \exp \left[ \phi(e^{i (\arg(a_{n_2}) + \frac{\pi}{2}) }\eta) - \phi(e^{i (\arg(a_{n_1}) + \frac{\pi}{2}) }\eta)   \right],  \label{eq:Cn1n2}
\ee
where $\phi(e^{ i \vartheta} \eta)$ is given by
\be
\phi(e^{i \vartheta} \eta) &=& \sum_{n \in 2 {\mathbb N}_0 -1} v_{n} (e^{i \vartheta} \eta)^{n} \nl  
&=& \sum_{n \in 2 {\mathbb N}_0 -1} v_{n}\cos (n \vartheta) \eta^{n} + i \sum_{n \in 2 {\mathbb N}_0 -1} v_{n} \sin (n \vartheta) \eta^{n},
\ee
with the real coefficients $v_{n \in {2 {\mathbb N}_0}-1}$ for all $n$ given by
\be
&& v_{-1} = \frac{\pi^{1/2} \Gamma \left(1 + \frac{1}{N}\right)}{\Gamma \left(\frac{3}{2} + \frac{1}{N}\right)}, \qquad  v_1 = \frac{\pi^{1/2} N \Gamma \left( 2 - \frac{1}{N} \right)}{12 \Gamma \left(\frac{1}{2} - \frac{1}{N} \right)}, \nl
&& v_3 = \frac{\pi^{1/2} N \left(2 N^2+N-3\right) \Gamma \left(2 -\frac{3}{N} \right)}{1440 \Gamma \left(- \frac{1}{2} -\frac{3}{N} \right)}, \qquad \cdots.
\label{eq:cof_v}
\ee

When a Stokes phenomenon occurs at a certain $\theta := \arg(\eta)$, in particular $\theta = 0$, on the Stokes graph, the effect has to be taken into account to write down the energy spectrum from the QCs obtained by taking $\arg(\eta) = \theta+0_\pm$.
The Delabaere-Dillinger-Pham (DDP) formula gives perturbative/non-perturbative relations among the cycles, which enable us to achieve the purpose~\cite{DDP2,DP1}.
Here, we define a set of cycles, denoted by ${\bf C}_{{\rm NP},\theta}$, having a degeneracy of Stokes lines from two turning points at $\theta$.
Schematically, 
\be
   {\bf C}_{{\rm NP},\theta} := \left\{ C_{(n_1,n_2 \ne n_1)} \, | \, a_{n_1} \leftrightarrow
 a_{n_2}  \, \, \mbox{connected by degenerated Stokes lines}, \  a_{n_1}, a_{n_2} \in {\rm TP} \right\}. \label{eq:def_CsetNP}
\ee
Notice that ${\bf C}_{{\rm NP},\theta} = \emptyset$ if no Stokes phenomenon happens at $\theta$.
For arbitrary cycles given by $C_{(n_1,n_2)} \notin {\bf C}_{{\rm NP},\theta}$ and $\widetilde{C}_{(n_1,n_2)} \in {\bf C}_{{\rm NP},\theta}$, the DDP formula can be expressed as a one-parameter group Stokes automorphism ${\frak S}_{\theta}^{\nu \in {\mathbb R}}$ as
\be
&&  {\frak S}_{\theta}^\nu [C_{(n_1,n_2)}]= C_{(n_1,n_2)} \prod_{\widetilde{C} \in {\bf C}_{{\rm NP},\theta}} (1 + \widetilde{C})^{ \nu \langle C_{(n_1,n_2)}, \widetilde{C} \rangle},   \\
&& {\frak S}_{\theta}^\nu [\widetilde{C}_{(n_1,n_2)}] = \widetilde{C}_{(n_1,n_2)},
\ee
where $\langle A, B \rangle$ is the intersection number between two cycles, $A$ and $B$, which is defined as
\be
\langle \rightarrow \,,\,  \uparrow \rangle = \langle \leftarrow \, , \, \downarrow \rangle = +1, \qquad  \langle \leftarrow \, , \, \uparrow  \rangle =  \langle \rightarrow \, , \, \downarrow  \rangle = -1. \label{eq:int_sec}
\ee

When one computes a QC (or monodromy matrix) at $\theta$ with a Stokes phenomenon, a transseries of the QC depends on $\theta + 0_\pm$, i.e., ${\frak D}^{\theta + 0_+} \ne {\frak D}^{\theta + 0_-}$.
In order to eliminate the discontinuity at $\theta$, we formulate \textit{median resummation} using the Stokes automorphism.
The Stokes automorphism taking $\nu = 1$ compensates a discontinuity caused by the Stokes phenomenon at $\theta$ as
\be
{\cal S}_{\theta+0_+} = {\cal S}_{\theta+0_-} \circ {\frak S}_{\theta}^{\nu = 1},
\ee
and the median resummation ${\cal S}_{{\rm med},\theta}$ gives a Borel resummed form without the discontinuity as
\be
   {\cal S}_{{\rm med},\theta} := {\cal S}_{\theta+0_+} \circ {\frak S}_{\theta}^{\nu = -1/2} = {\cal S}_{\theta+0_-} \circ {\frak S}_{\theta}^{\nu = +1/2}.
\ee
The transseries eliminated the discontinuity, ${\frak D}^{\theta}$, can be uniquely determined by the Stokes automorphism as
\be
   {\frak D}^{\theta}   = {\frak S}^{\nu = +1/2}_{\theta}[{\frak D}^{\theta + 0_+}] = {\frak S}^{\nu = -1/2}_{\theta}[{\frak D}^{\theta + 0_-}], \label{eq:D0_nodisc}
\ee
which is derived by
\be
&& {\cal S}_{\theta+0_+}[{\frak D}^{\theta + 0_+}] = {\cal S}_{\theta+0_-} \circ {\frak S}_{\theta}^{\nu = 1}[{\frak D}^{\theta + 0_+}] =  {\cal S}_{\theta+0_-}[{\frak D}^{\theta + 0_-}] \nl 
&\Rightarrow \quad&  {\frak S}_{\theta}^{\nu = +1/2}[{\frak D}^{\theta + 0_+}] = {\frak S}_{\theta}^{\nu = -1/2}[{\frak D}^{\theta + 0_-}].
\ee
Thus, one finds that
\be
   {\cal S}_{{\rm med},\theta}[{\frak D}^{\theta}] =  {\cal S}_{\theta+0_+}[{\frak D}^{\theta + 0_+}] = {\cal S}_{\theta+0_-}[{\frak D}^{\theta + 0_-}] \mathop\sim^{|\eta| \rightarrow 0_+} {\frak D}^{\theta}.
\ee
Notice that the Stokes automorphism acts to a function of the cycles, $F(C_{(n_1,n_2)},C_{(n_3,n_4)},\cdots)$, as a homomorphism:
\be
   {\frak S}^{\nu}_{\theta} [F(C_{(n_1,n_2)},C_{(n_3,n_4)},\cdots)] =  F({\frak S}^{\nu}_{\theta}[C_{(n_1,n_2)}],{\frak S}^{\nu}_{\theta}[C_{(n_3,n_4)}],\cdots).
\ee

Cycles in ${\bf C}_{{\rm NP},\theta}$ defined in Eq.(\ref{eq:def_CsetNP}) do not all affect a QC through the DDP formula (or Stokes automorphism), and only elements having non-zero intersections with cycles consisting of the QC are important.
In the below discussions, we suppose that only cycles relevant to the QC would be given when we provide ${\bf C}_{{\rm NP},\theta}$.

In this paper, we call an exact QC the object ${\frak D}^{\theta}$ to satisfy Eq.(\ref{eq:D0_nodisc}).

\section{Exact quantization conditions} \label{sec:quant_cond} 
In this section, we construct exact QCs for a given $(N,K)$ using EWKB.
We firstly demonstrate the $K=1$ cases in Sec.~\ref{sec:QCK1}, and then show the generalization to $K>1$ in Sec.~\ref{sec:QCKgen}.

\begin{figure}[tbp]
  \begin{center}
    \begin{tabular}{cc}
      \begin{minipage}{0.5\hsize}
        \begin{center}
          \includegraphics[clip, width=75mm]{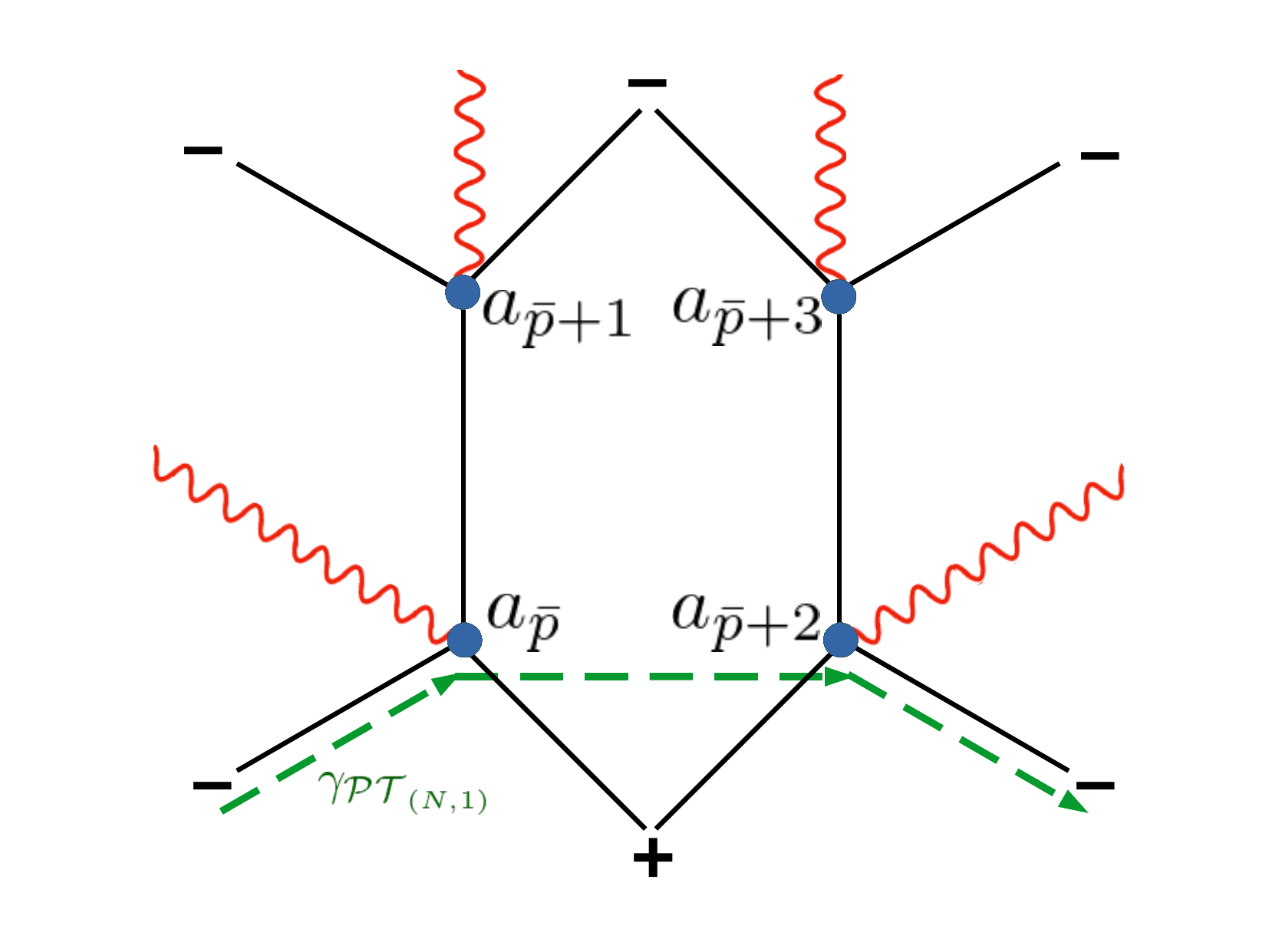}
          \hspace{1.6cm} (a) Even $N$
        \end{center}
      \end{minipage}
      \begin{minipage}{0.5\hsize}
        \begin{center}
          \includegraphics[clip, width=75mm]{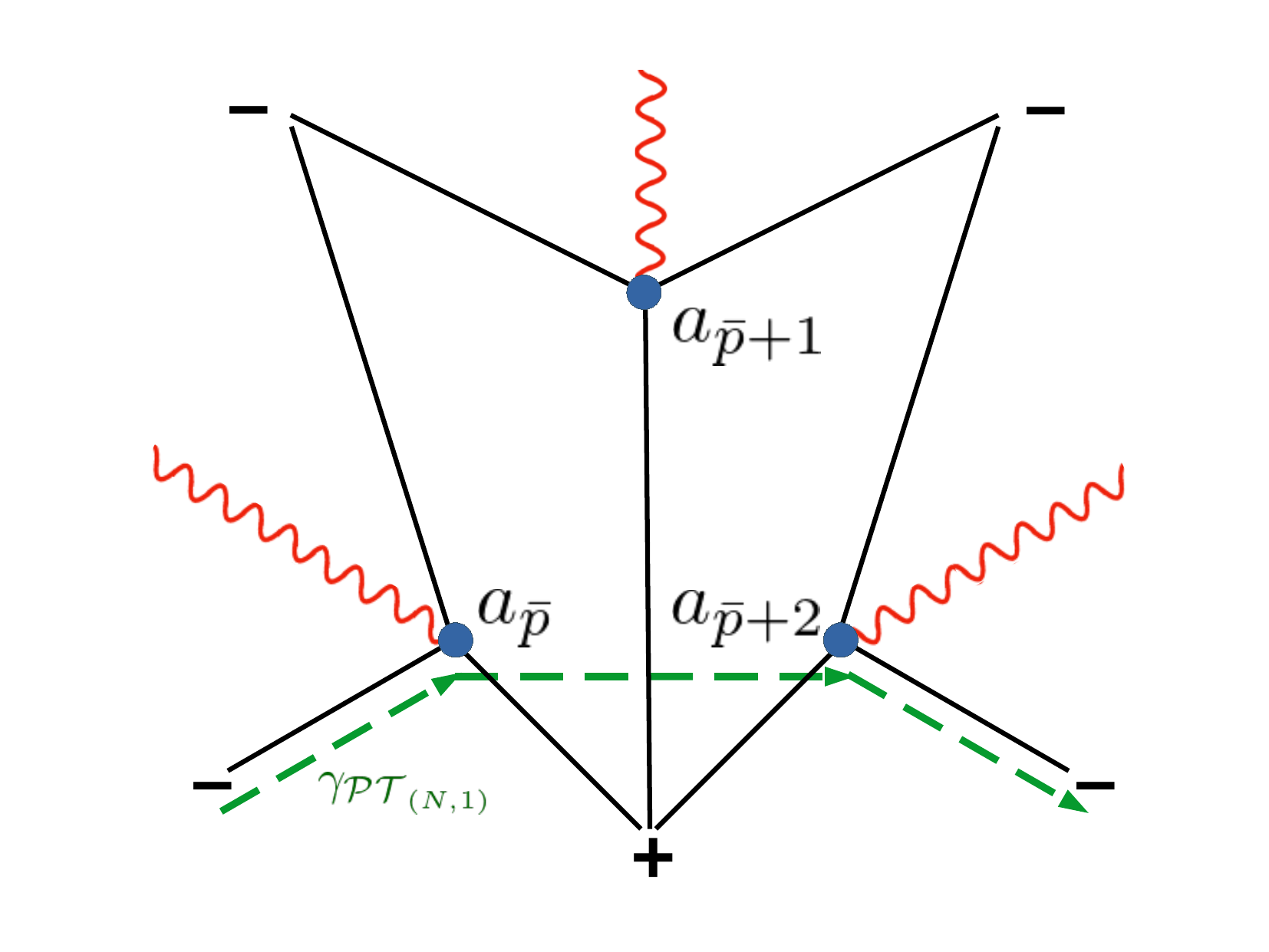} 
          \hspace{1.6cm} (b) Odd $N$
        \end{center}
      \end{minipage} 
    \end{tabular} 
    \caption{Stokes graphs given by $K=1$ for $\arg(\eta)=0$.
      The green lines denote the path of analytic continuation.
    }
    \label{fig:Stokes_K1}
  \end{center}
\end{figure}

\subsection{The $K=1$ cases} \label{sec:QCK1}
We consider the $K=1$ cases.
The Stokes graphs and the paths of analytic continuation in Fig.~\ref{fig:Stokes_K1}.
Since the structure of the Stokes graphs depends on even or odd $N$ (or $\varepsilon$), we individually consider the two cases.

\subsubsection{Even $N$} \label{sec:evenN_K1}
We consider the even $N$ cases.
In this case, a Stokes phenomenon occur at $\arg(\eta)=0$, so that one has to take care of a discontinuity in QCs by using the DDP formula.
As we can see later, the resulting exact QC contains only a perturbative cycle.

\begin{figure}[tbp]
  \begin{center}
    \begin{tabular}{cc}
      \begin{minipage}{0.5\hsize}
        \begin{center}
          \includegraphics[clip, width=75mm]{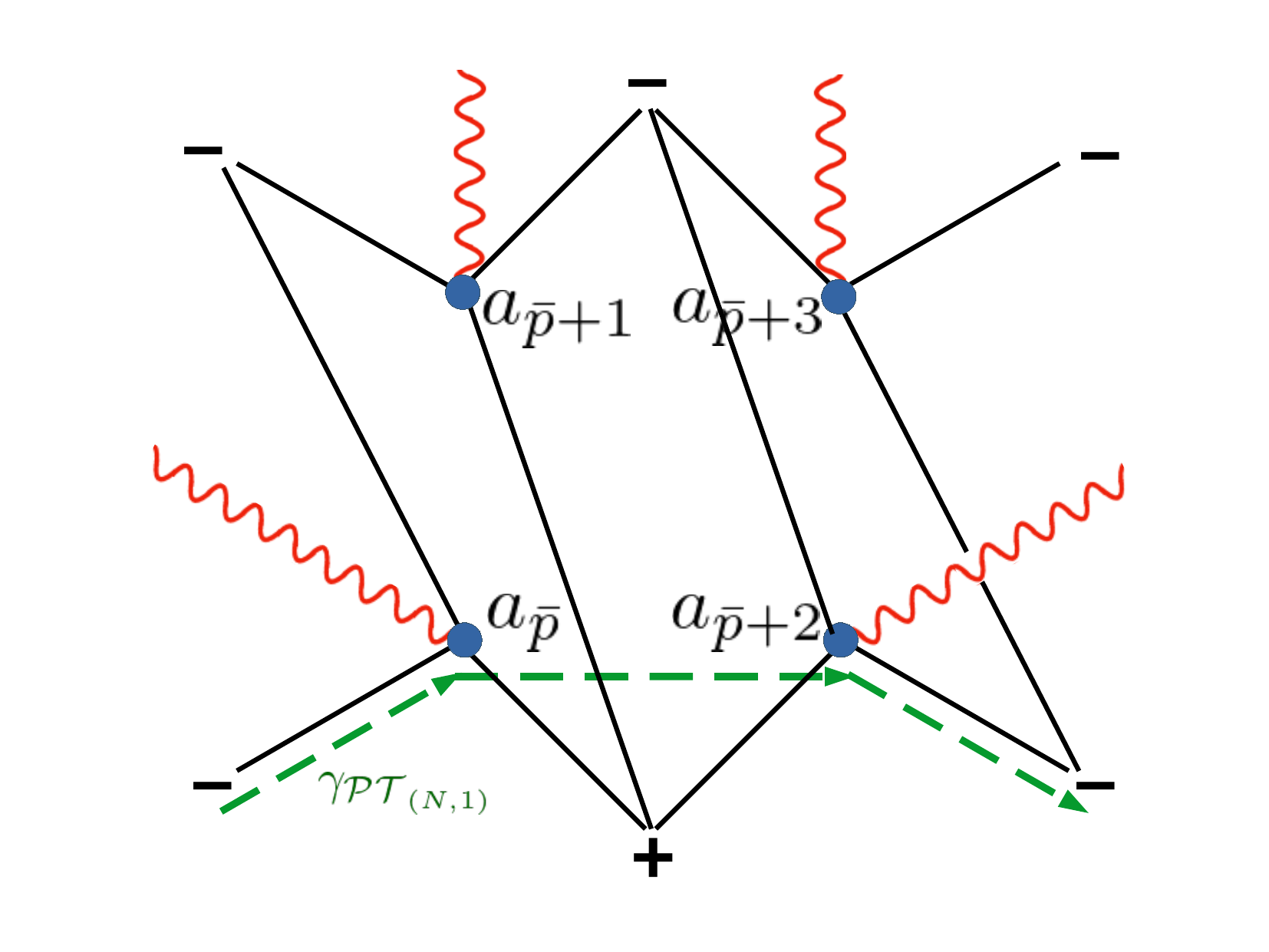}
          \hspace{1.6cm} (a) $\arg(\hbar) = 0_+$
        \end{center}
      \end{minipage}
      \begin{minipage}{0.5\hsize}
        \begin{center}
          \includegraphics[clip, width=75mm]{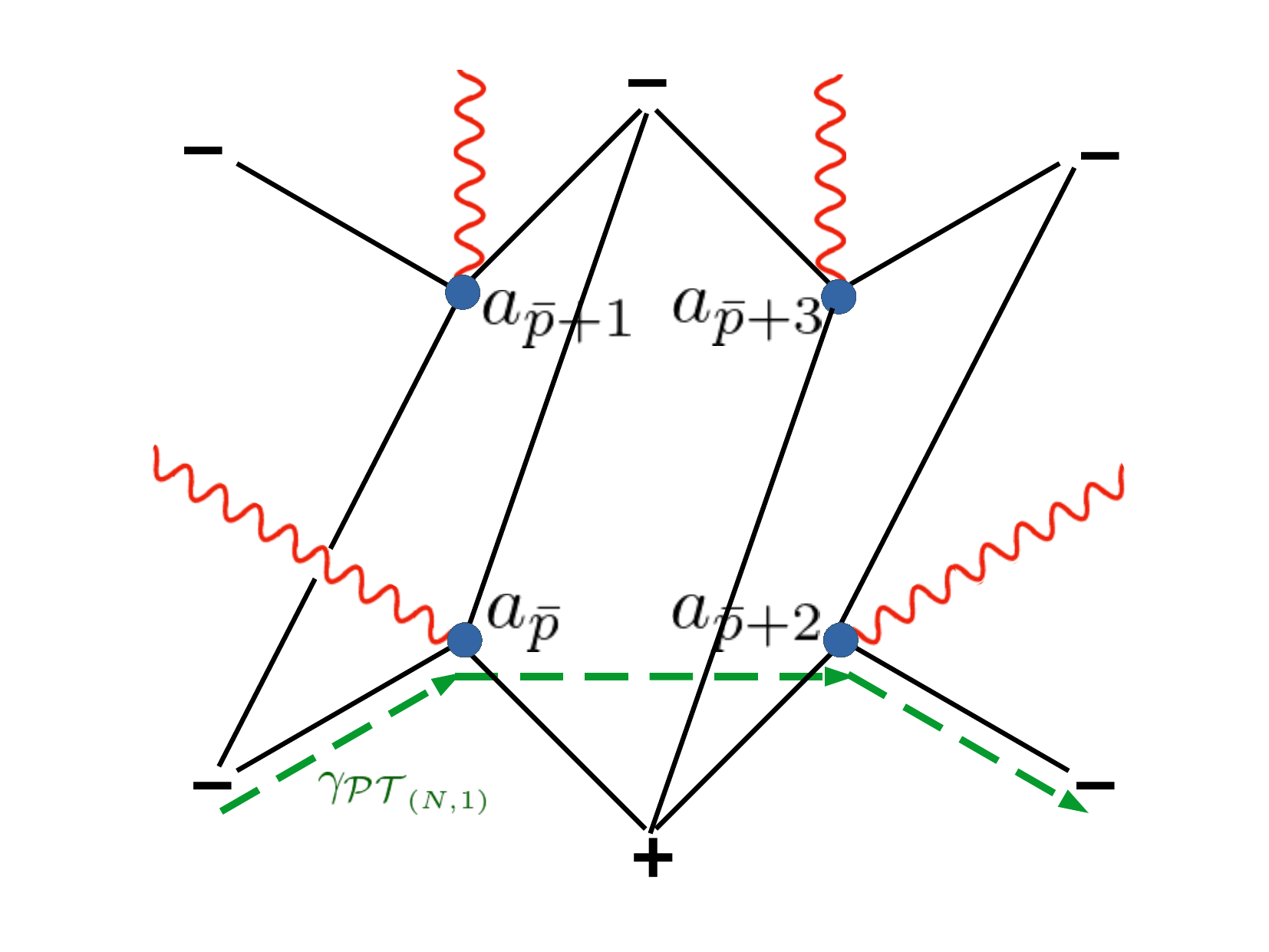} 
          \hspace{1.6cm} (b) $\arg(\hbar) = 0_-$
        \end{center}
      \end{minipage} 
    \end{tabular} 
    \caption{Stokes graphs given by even $N$ with $K=1$ for $\arg(\eta)=0_\pm$.
      The green lines denote the path of analytic continuation.
    }
    \label{fig:Stokes_NevenK1_arg0pm}
  \end{center}
\end{figure}

In order to see this fact, we firstly identify a perturbative cycle.
For finding the QC, we perform analytic continuation along a certain path given by Eq.(\ref{eq:gam_PT}) with $K=1$.
From the path, $\gamma_{{\cal PT}_{(N,1)}}$, and the location of turning points (\ref{eq:TP_def}) in (E-2)(E-4) of Fig.~\ref{fig:Stokes_shape_even}, one can see that the perturbative cycle is given as $C_{(\bar{p},\bar{p}+2)}$ consisting of the two turning points located at 
\be
a_{\bar{p}} = -i  e^{-\frac{\pi}{N} i}, \qquad a_{\bar{p}+2} = -i e^{+\frac{\pi}{N} i}. \label{eq:a_barp} 
\ee
The specific form of $C_{(\bar{p},\bar{p}+2)}$ is given by Eq.(\ref{eq:Cn1n2}) as
\be
C_{(\bar{p},\bar{p}+2)} = \exp \left[ \phi(e^{+\frac{\pi}{N} i}\eta) - \phi(e^{-\frac{\pi}{N} i} \eta)   \right] 
= \exp \left[ 2  i \sum_{n \in 2 {\mathbb N}_0 -1} v_{n} \sin \frac{\pi n}{N} \cdot \eta^{n} \right], \label{eq:C_pp_even_K1}
\ee
where the coefficients $v_{n \in 2 {\mathbb N}_0-1}$ are given in Eq.(\ref{eq:cof_v}).
Indeed, this cycle is a pure oscillation without an exponential damping factor.

Since the Stokes graph has a Stokes phenomenon at $\arg(\eta) = 0$, we introduce an infinitesimally small phase to $\eta$ before finding monodromy matrices.
The Stokes graphs for $\arg(\eta)=0_\pm$ are drawn in Fig.~\ref{fig:Stokes_NevenK1_arg0pm}.
Then, by taking the path of analytic continuation, $\gamma_{{\cal PT}_{(N,1)}}$, one finds the following monodromy matrices depending on $\arg(\eta) = 0_\pm$:
\be
{\cal M}^{0_+} &=& M_+ N_{\bar{p},\bar{p}+1} M_+ N_{\bar{p}+1,\bar{p}+2} M_+ N_{\bar{p}+2,\bar{p}}, \\
{\cal M}^{0_-} &=& M_+ N_{\bar{p},\bar{p}+3} M_+ N_{\bar{p}+3,\bar{p}+2} M_+ N_{\bar{p}+2,\bar{p}}.
\ee
Here, we used the shorten notation for the normalization matrices as $N_{a_{n_1},a_{n_2}} \rightarrow N_{n_1,n_2}$.
The QCs are extracted from ${\frak D}^{0_\pm} \propto {\cal M}^{0_\pm}_{12} = 0$ by normalizability of the wavefunction, and those can be expressed by the cycles as
\be
&& {\frak D}^{0_+} \propto 1 + \frac{C_{(\bar{p},\bar{p}+2)}}{1+C_{(\bar{p},\bar{p}+1)}}, \qquad {\frak D}^{0_-} \propto 1 + C_{(\bar{p},\bar{p}+2)}(1 + C_{(\bar{p}+2,\bar{p}+3)}
).
\ee
A set of non-perturbative cycles can be found from the Stokes graph, and its subset relevant to the above QCs is given by
\be
   {\bf C}_{{\rm NP},\theta = 0} = \{C_{(\bar{p},\bar{p}+1)}, C_{(\bar{p}+2,\bar{p}+3)} \}.
\ee
Notice that $C_{({\bar{p}},\bar{p}+1)}= C_{({\bar{p}+2},\bar{p}+3)}$ because of the ${\mathbb Z}_2$ symmetry in Eq.(\ref{eq:Z2_symm}).
Due to the Stokes phenomenon, the DDP formula is non-trivial and obtained by counting intersection numbers among the cycles as
\be
&& {\frak S}_0^{\nu} [C_{(\bar{p},\bar{p}+2)}] = C_{(\bar{p},\bar{p}+2)} \prod_{n=0}^1 (1 + C_{({\bar{p}+2n},\bar{p}+2n+1)} )^{\nu}, \nl
&& {\frak S}_0^{\nu} [C_{({\bar{p}+2n},\bar{p}+2n+1)}] = C_{({\bar{p}+2n},\bar{p}+2n+1)} \quad \mbox{for} \quad n = 0, 1.
\ee
From Eq.(\ref{eq:D0_nodisc}), the exact QC removed the discontinuity is given by
\be
   {\frak D}^0 \propto 1 + C_{(\bar{p},\bar{p}+2)} \sqrt{\frac{1 + C_{(\bar{p}+2,\bar{p}+3)}}{1 + C_{(\bar{p},\bar{p}+1)}}} = 1 + C_{(\bar{p},\bar{p}+2)}. \label{eq:D0_even_K1}
\ee
Here, we used $C_{({\bar{p}},\bar{p}+1)} = C_{({\bar{p}+2},\bar{p}+3)}$.
The contributions from the two cycles, $C_{({\bar{p}},\bar{p}+1)}$ and $C_{({\bar{p}+2},\bar{p}+3)}$, are canceled to each other by the ${\mathbb Z}_2$ symmetry, and the remaining cycle, $C_{({\bar{p}},\bar{p}+2)}$, is a pure oscillation, as is shown in Eq.(\ref{eq:C_pp_even_K1}).
Therefore, the exact QC includes only the perturbative contribution.

\subsubsection{Odd $N$} 
Then, we consider the odd $N$ cases.
The procedure to derive the exact QCs is the same to Sec.~\ref{sec:evenN_K1}, but the main difference from the even $N$ cases is that Stokes phenomena do not happen at $\theta = 0$, and thus ${\bf C}_{{\rm NP},\theta = 0} = \emptyset$.
Hence, one does not have to care of discontinuities on the Borel plane.

The monodromy matrix is obtained by taking the path of analytic continuation in Eq.(\ref{eq:gam_PT}) with $K=1$, which is given by
\be
{\cal M}   &=& M_+ N_{{\bar{p}},{\bar{p}+1}} M_+ N_{{\bar{p}+1},{\bar{p}+2}} M_+ N_{{\bar{p}+2},{\bar{p}}},
\ee
where the turning points, $a_{\bar{p}}$ and $a_{\bar{p}+2}$ are given by Eq.(\ref{eq:a_barp}).
Imposing normalizability to the wavefunction yields the exact QC, ${\frak D} \propto {\cal M}_{12} = 0$, which takes the form that
\be
   {\frak D} \propto 1  + C_{(\bar{p},\bar{p}+2)} + C_{(\bar{p}, {\bar{p}+1})}. \label{eq:D0_odd_K1} 
\ee
The exact QC contains two cycles, $C_{(\bar{p},\bar{p}+2)}$ and $C_{(\bar{p},\bar{p}+1)}$, and the former and the latter correspond to perturbative and non-perturbative contributions, respectively.
$C_{(\bar{p},\bar{p}+2)}$ has the same form to Eq.(\ref{eq:C_pp_even_K1}), and $C_{(\bar{p}, {\bar{p}+1})}$ is expressed as
\be
C_{(\bar{p}, \bar{p}+1)} &=& \exp \left[ \phi(-\eta) - \phi(e^{-\frac{\pi}{N}i }  \eta)   \right] \nl
&=& \exp \left[ - \sum_{n \in 2 {\mathbb N}_0 -1} v_{n} \left( \cos  \frac{\pi n}{N} + 1 \right) \cdot \eta^{n} + i \sum_{n \in 2 {\mathbb N}_0 -1} v_{n} \sin  \frac{\pi n}{N} \cdot \eta^{n}  \right],
\ee
where the coefficients $v_{n \in 2{\mathbb N}_0-1}$ are given by Eq.(\ref{eq:cof_v}).
In order to specify the perturbative/non-perturbative structure of the exact QC, it is helpful to replace the cycles with $P$ and $B$ which are purely oscillating and exponentially damping, respectively.
By these symbols, Eq.(\ref{eq:D0_odd_K1}) can be expressed by
\be
&& {\frak D} \propto \frac{1}{C_{(\bar{p},\bar{p}+2)}^{1/2}} \left[ 1 + C_{(\bar{p},\bar{p}+2)} + C_{(\bar{p}, {\bar{p}+1})} \right]
= P^{-1/2} + P^{+1/2} + B, \label{eq:D0_odd_K1_2} \\
&& P := \exp \left[ 2 i \sum_{n \in 2 {\mathbb N}_0 -1} v_{n} \sin  \frac{\pi n}{N} \cdot \eta^{n}  \right], \\
&& B:= \exp \left[ - \sum_{n \in 2 {\mathbb N}_0 -1} v_{n} \left( \cos  \frac{\pi n}{N} + 1 \right) \cdot \eta^{n}  \right], \label{eq:Bcycle_odd_K1} 
\ee
where $C_{(\bar{p},\bar{p}+2)} = P$, and $C_{(\bar{p},\bar{p}+1)} = B P^{1/2}$.
Since ${\cal K}[P] = P^{-1}$ and ${\cal K}[B] = B$, where ${\cal K}$ is the complex conjugate operator, the exact QC can take the form invariant under complex conjugation, and thus the energy spectrum is also expected to be real.

\subsection{Generalization to $K>1$} \label{sec:QCKgen}

We consider the generalization to $K>1$.
Once fixing the values of $(N,K)$, one can count the number of cycles in the QC because the number of cycles roughly corresponds to that of the normalization matrices, $N_{{n_1},{n_2}}$ in the QC.
In other words, one has to deal only with cycles intersecting with the path of analytic continuation, $\gamma_{\cal PT}$.
Notice that Eq.(\ref{eq:gam_PT}) implies that taking a larger $K$ with a fixed $N$ makes a path closer to the real axis.
This fact implies that the number of relevant cycles in the  QC becomes larger as taking larger $K$, and as a result  a specific form of the QC and its transseries structure become more complicated.

By looking to the number of the normalization matrices, $N_{{n_1},{n_2}}$, intersecting with the path of analytic continuation (\ref{eq:gam_PT}) in the Stokes graphs in Figs.~\ref{fig:Stokes_shape_even} and \ref{fig:Stokes_shape_odd}, one can identify the number of  relevant turning points to the QCs for a given $(N,K)$ as 
\be
\mbox{\# of relevant turning points} = 
\begin{cases}
  2 K + 2 & \mbox{for even $N$}  \\
  2 K + 1 & \mbox{for odd $N$}
\end{cases},
\ee
and cycles in the QCs consist of these turning points.
In addition, turning points corresponding to a perturbative cycle, $(a_{\bar{p}},a_{p})$, can be found as
\be
&& a_{\bar{p}} = -i e^{- \pi i \frac{K}{N}}, \qquad a_{p} = -i e^{+ \pi i \frac{K}{N}}, \qquad p = {\bar p} + 2 K. \label{eq:apbarp_K}
\ee
Notice that ${\rm Re}[a_p] = - {\rm Re}[a_{\bar{p}}]>0$ and ${\rm Im}[a_{\bar{p}}] = {\rm Im}[a_p] <0$, and it is a consequence of ${\cal PT}$ symmetry, ${\cal PT}: x \rightarrow - \bar{x} $. 
From Eq.(\ref{eq:apbarp_K}), the perturbative cycle is given by
\be
P &:=& C_{(\bar{p}, p = \bar{p}+2K)} = \exp \left[ \phi(e^{+\frac{\pi K}{N} i}\eta) - \phi(e^{-\frac{\pi K}{N} i} \eta)   \right] \nl
&=& \exp \left[ 2  i \sum_{n \in 2 {\mathbb N}_0 -1} v_{n} \sin \frac{\pi K n}{N} \cdot \eta^{n} \right], \label{eq:C_pp_Kgen}
\ee
where the coefficients $v_{n \in 2{\mathbb N}_0 - 1}$ are the same to Eq.(\ref{eq:cof_v}).
In contrast, non-perturbative parts in the QCs are quite non-trivial.
Those non-perturbative cycles indeed depend on the values of $(N,K)$, and their structure on the Borel plane has difference between even and odd $N$, as we considered for $K=1$.
In the below, we find cycle-representations of the exact QCs for arbitrary $(N,K)$.
Details of their perturbative/non-perturbative structure would be discussed in Sec.~\ref{sec:energy_trans}.

For even $N$, a Stokes phenomenon occurs at $\arg(\eta)=0$, so that one has to take care of discontinuities in the QCs, ${\frak D}^{0_\pm}$, obtained from the monodromy matrices, ${\cal M}^{0_\pm}$.
These can be found by taking the path in Eq.(\ref{eq:gam_PT}) as
\be
{\cal M}^{0_+} &=&  M_+ N_{a_{\bar{p}},a_{\bar{p}+1}} \left[ \prod_{\ell=1}^{K-1} M_+ N_{a_{\bar{p}+2\ell-1},a_{\bar{p}+2 \ell}} M_-^{-1} N_{a_{\bar{p}+2\ell},a_{\bar{p}+2 \ell+1}} \right] \nl
&& \times M_+ N_{a_{\bar{p}+2K-1}, a_{\bar{p}+2K}} M_+ N_{a_{\bar{p}+2K},a_{\bar{p}}}, \\
{\cal M}^{0_-} &=&  M_+ N_{a_{\bar{p}},a_{\bar{p}+3}} \left[ \prod_{\ell=1}^{K-1} M_+ N_{a_{\bar{p}+2\ell+1},a_{\bar{p}+2 \ell}} M_-^{-1} N_{a_{\bar{p}+2\ell},a_{\bar{p}+2 \ell+3}} \right] \nl
   && \times M_+ N_{a_{\bar{p}+2K + 1}, a_{\bar{p}+2K}} M_+ N_{a_{\bar{p}+2K},a_{\bar{p}}}.
\ee
Imposing normalizability to the wavefunction requires ${\cal M}^{0_\pm}_{12} = 0$, and one can find the QCs as
\be
&&   {\frak D}^{0_\pm} \propto  \sum_{(n_1,\cdots, n_K) \in \{0,1\}^{K}}  \left[ \prod_{\ell=1}^{K-1}  {\frak D}^{(\ell)0_\pm}_{n_\ell, n_{\ell + 1}}  \right] {\frak D}^{(K)0_\pm}_{n_K}, \\
&& {\frak D}^{(\ell \in \{1,2,\cdots, K-1\})0_\pm}_{n_\ell,n_{\ell + 1}} := \left( \widetilde{C}^{0_\pm}_{(\bar{p}+2\ell-2,\bar{p}+2\ell+1)} \right)^{n_\ell} \left( \delta_{n_{\ell+1},0} + \widetilde{B}^{-1}_{(\bar{p}+2\ell+1,\bar{p}+2\ell)} \cdot \delta_{n_{\ell+1},1} \right)^{n_\ell}, \nl
&&    {\frak D}^{(K) 0_\pm}_{n_K}:= \left( \widetilde{C}^{0_\pm}_{(\bar{p}+2K-2,\bar{p}+2K)} \right)^{n_K}, 
\ee
where $\bar{p}$ is the label associated with $a_{\bar p}$ in Eq.(\ref{eq:apbarp_K}), and 
\be
&& \widetilde{C}_{(\bar{p}+2 \ell-2,\bar{p}+2 \ell+1)}^{0_\pm} := \frac{C_{(\bar{p}+2 \ell-2,\bar{p}+2 \ell+1)}}{\prod_{n=0}^1 (1+B_{(\bar{p}+2 \ell+2n-2,\bar{p}+2 \ell+2n-1)})^{(1 \pm 1)/2}}, \\
&& \widetilde{C}_{(\bar{p}+2K-2,\bar{p}+2K)}^{0_\pm} := C_{(\bar{p}+2K-2,\bar{p}+2K)} \frac{(1+ B_{(\bar{p}+2K,\bar{p}+2K+1)})^{(1 \mp 1)/2}}{(1+B_{(\bar{p}+2K-2,\bar{p}+2K-1)})^{(1 \pm 1)/2}}, \\
&& \widetilde{B}^{-1}_{(\bar{p}+2 \ell + 1,\bar{p}+2 \ell) } := \frac{1+B_{(\bar{p}+2 \ell,\bar{p}+2 \ell+1)}}{B_{(\bar{p}+2 \ell,\bar{p}+2 \ell+1)}} \quad \in {\mathbb R}_{>0}. 
\ee
with $\ell \in \{ 1,2,\cdots,K-1 \}$.
Here, we defined symbols, $B_{(\bar{p}+2 \ell,\bar{p}+2 \ell+1)} := C_{(\bar{p}+2 \ell,\bar{p}+2 \ell+1)}$, to emphasize non-perturbative cycles with degeneracies of the Stokes lines at $\arg(\eta)=0$.
From them, the set of non-perturbative cycles relevant to th DDP formula is given by
\be
   {\bf C}_{{\rm NP},\theta=0} = \{ B_{(\bar{p},\bar{p}+1)}, B_{(\bar{p}+2,\bar{p}+3)}, \cdots, B_{(p-2,p-1)}, B_{(p,p+1)}\}, \qquad p = \bar{p} + 2K.
\ee
One can easily construct the DDP formula by counting the intersection number of the other cycles with the $B$-cycles, and it is given by
\be
&& {\frak S}_0^{\nu} [C_{(\bar{p}+2\ell-2,\bar{p}+2 \ell+1)}] = C_{(\bar{p}+2\ell-2,\bar{p}+2 \ell+1)} \prod_{n=0}^1 (1 + B_{(\bar{p}+2\ell+2n-2,\bar{p}+2 \ell + 2n-1)})^{\nu}, \nl
&& {\frak S}_0^{\nu} [B_{(\bar{p}+2 \ell+2n-2,\bar{p}+2 \ell + 2n-1)}] = B_{(\bar{p}+2 \ell+2n-2,\bar{p}+2 \ell + 2n-1)} \quad \mbox{for} \quad n = 0, 1, \label{eq:DDP_evenN_Kgen}
\ee
with $\ell \in \{1,2,\cdots, K\}$.
Eliminating the discontinuity by the DDP formula as ${\frak D}^0 \propto {\frak S}_0^{\pm 1/2}[{\frak D}^{0_\pm}]$ leads to the exact QC as
\be
&& {\frak D}^{0} \propto  \frac{1}{P^{1/2}} \sum_{(n_1,\cdots, n_K) \in \{0,1\}^{K}}  \left[ \prod_{\ell=1}^{K-1}  {\frak D}^{(\ell)}_{n_\ell, n_{\ell + 1}}  \right] {\frak D}^{(K)}_{n_K}, \nl
&& {\frak D}^{(\ell \in \{1,2,\cdots, K-1\})}_{n_\ell,n_{\ell + 1}} := \left( \widetilde{C}^0_{(\bar{p}+2\ell-2,\bar{p}+2\ell+1)} \right)^{n_\ell} \left( \delta_{n_{\ell+1},0} + \widetilde{B}^{-1}_{(\bar{p}+2\ell+1,\bar{p}+2\ell)} \cdot \delta_{n_{\ell+1},1} \right)^{n_\ell}, \nl
&& {\frak D}^{(K)}_{n_K}:= \left( \widetilde{C}^0_{(\bar{p}+2K-2,\bar{p}+2K)} \right)^{n_K}, \label{eq:D0_evenN_Kgen} 
\ee
where $\delta_{n_1,n_2}$ is the Kronecker delta, and $\widetilde{C}^0_{(\bar{p}+2 \ell-2,\bar{p}+2 \ell+1)}$ and $\widetilde{C}^0_{(\bar{p}+2K-2,\bar{p}+2K)}$ are defined as
\be
 \widetilde{C}^0_{(\bar{p}+2 \ell-2,\bar{p}+2 \ell+1)} &:=& {\frak S}_{0}^{\pm 1/2}[\widetilde{C}_{(\bar{p}+2 \ell-2,\bar{p}+2 \ell+1)}^{0_\pm}] \nl
 &=& \frac{C_{(\bar{p}+2 \ell-2,\bar{p}+2 \ell+1)}}{\prod_{n=0}^1 \sqrt{1 + B_{(\bar{p}+2 \ell+2n -2,\bar{p}+2 \ell+2n-1)}}}, \\
\widetilde{C}^0_{(\bar{p}+2K-2,\bar{p}+2K)} &:=& {\frak S}_{0}^{\pm 1/2}[\widetilde{C}^{0_\pm}_{(\bar{p}+2K-2,\bar{p}+2K)}] \nl
&=& C_{(\bar{p}+2K-2,\bar{p}+2K)} \sqrt{\frac{1 + B_{(\bar{p}+2K,\bar{p}+2K+1)}}{1 + B_{(\bar{p}+2K-2,\bar{p}+2K-1)}}}.
\ee
In Eq.(\ref{eq:D0_evenN_Kgen}), we multiplied $P^{-1/2}$ to make ${\frak D}^{0}$ invariant under complex conjugation, i.e., ${\cal K}[{\frak D}^0] = {\frak D}^0$.
It is notable that, due to the ${\mathbb Z}_N$ symmetry in Eq.(\ref{eq:ZN_symm}),
the cycles in the QC are not all independent on each other and have relations that
\be
 {\rm Re}[\log C_{(\bar{p}+2 n-2,\bar{p}+2 n + 1)}] &=& \frac{1}{2} \left( \log B_{(\bar{p}+2 n-2,\bar{p}+2 n-1)}  + \log B_{(\bar{p}+2 n,\bar{p}+2 n + 1)} \right) \nl
 &=& {\rm Re}[\log C_{(\bar{p}+2K-2 n,\bar{p}+ 2 K-2 n + 3)}],  \label{eq:C_evenN_rel1} \\
 {\rm Im}[\log C_{(\bar{p}+2 n-2,\bar{p}+2 n + 1)}] &=& {\rm Im}[\log C_{(\bar{p}+2K-2 n,\bar{p}+ 2 K-2 n + 3)}], \\
   \log B_{(\bar{p}+2 n-2,\bar{p}+2 n - 1)} &=& \log B_{(\bar{p}+2K-2 n+2,\bar{p}+ 2 K-2 n + 3)} \in {\mathbb R}_{>0}, \label{eq:C_evenN_rel2} 
\ee
for $n \in \{ 1,2 \cdots,\lfloor K/2 \rfloor +1 \}$.

The generalization for odd $N$ is simpler than even $N$ because of lack of the ${\mathbb Z}_2$ symmetry in Eq.(\ref{eq:Z2_symm}), i.e., no Stokes phenomenon at $\arg(\eta) = 0$.
Hence, one does not need to care of the DDP formula.
Performing analytic continuation along the path, $\gamma_{\cal PT}$, in Eq.(\ref{eq:gam_PT}) yields the exact QC as
\be
&& {\frak D} \propto  \frac{1}{P^{1/2}} \sum_{(n_1,\cdots, n_K) \in \{0,1\}^{K}}  \left[ \prod_{\ell=1}^{K-1} {\frak D}^{(\ell)}_{n_\ell, n_{\ell + 1}} \right] {\frak D}^{(K)}_{n_K}, \nl
&&  {\frak D}^{(\ell \in \{ 1,2, \cdots, K-1 \})}_{n_\ell, n_{\ell + 1}}:=  C^{n_{\ell}}_{(\bar{p}+2 \ell-2,\bar{p}+2\ell-1)} ( 1 + C_{(\bar{p}+2\ell-1,\bar{p}+2\ell)} \cdot \delta_{n_{\ell+1}, 1} )^{n_\ell}, \nl
&& {\frak D}^{(K)}_{n_K} :=  C_{(\bar{p}+2K-2,\bar{p}+2K-1)}^{n_{K}} \left( 1 +  C_{(\bar{p}+2K-1,\bar{p}+2K)} \right)^{n_{K}}. \label{eq:D0_oddN_Kgen}
\ee
Due to the ${\mathbb Z}_N$ symmetry in Eq.(\ref{eq:ZN_symm}), those cycles have dependencies to each other such that
\be
&& {\rm Re}[\log C_{(\bar{p}+n-1,\bar{p}+n)}] = - {\rm Re}[\log C_{(\bar{p}+2K-n,\bar{p}+2K-n+1)}], \label{eq:C_oddN_rel} \\
&& {\rm Im}[\log C_{(\bar{p}+n-1,\bar{p}+n)}] = {\rm Im}[\log C_{(\bar{p}+2K-n,\bar{p}+2K-n+1)}],
\ee
for $n \in \{1,2,\cdots,K \}$.

\section{Energy spectra and their transseries structure} \label{sec:energy_trans}
In this section, we find transseries structure of the energy spectra from the exact QCs constructed in Sec.~\ref{sec:quant_cond}.
A slightly non-trivial issue is that, according to Eq.(\ref{eq:L_res}), the energy solution should take the form that
\be
E_{k} = e(k) \, (g^{1/N} \hbar)^{2N/(N+2)}, \quad (k \in {\mathbb N}_0) \label{eq:Ek_c}
\ee
where $k$ is an energy level, and $e(k)$ is a dimensionless function depending only on $k$.
This means that $\eta$ has no dependence of $\hbar$ in total, and thus, our main task is to find the functional form of $e(k)$.
Although it is quite tough to obtain the exact analytic function, finding its transseries solution is in principle possible by using the (inverse) energy level, $k$, as an expansion parameter~\cite{BPV,Bucciotti:2023trp}.

Before looking to the structure, we briefly explain the relation between the $\eta$- and $k$-expansions in (E)WKB analyses.
Although the $\hbar$-expansion seems not to work, expanding $k$ around $k = +\infty$ is compatible with (E)WKB analyses.
As we will describe precisely later, we employ an ansatz for $\eta$ which is an expansion in terms of $\kappa^{-1}$ as $\kappa \rightarrow +\infty$, where $\kappa := \pi(k + 1/2)$ with $k \in {\mathbb N}_0$.
By taking the ansatz as $\eta^{-1} \sim \sum_{\ell \in {2\mathbb N}_{0}-1} c_{\ell} \kappa^{-\ell}$, we recursively determine the coefficients, $c_{\ell \in {2\mathbb N}_{0}-1}$, by solving the QCs.
Since the $\eta$-expansion is nothing but an expansion of $c(k)^{-1} = e(k)^{-(N+2)/(2N)}$ as $c(k) \rightarrow +\infty$, the result of $c_{\ell}$ can be converted to a transseries of $e(k)$ by Eq.(\ref{eq:def_eta_E}).
In addition, topology of the Stokes graphs generated by $\eta$ and $\kappa^{-1}$ are the same if $c_{-1}$ is a non-zero real value.
The relation between the two graphs can be seen from Eqs.(\ref{eq:L_res})(\ref{eq:def_eta_E}) by multiplying $\eta^{-2}$ to ${\cal L}$ and substituting $\eta^{-1} \sim c_{-1} \kappa^{-1} + c_{1} \kappa^{1} + \cdots$ into it, one finds
\be
&& \eta^{-2} {\cal L} \sim - \pd_x^2 + \kappa^{2} \widetilde{Q}, \nl
&&  \widetilde{Q} := \eta^{-2} \kappa^{-2} Q \sim \left[ c_{-1}^2  + 2 c_{-1} c_{1} \kappa^{-2} + O(\kappa^{-4}) \right] Q. \label{eq:tilde_Q}
\ee
By assuming $c_{-1} \ne 0$, the leading and the higher orders of $\kappa^{-1}$ in $\widetilde{Q}$ can be regarded as a some sort of classical part and quantum deformations, respectively, in the sense of  $\hbar$-expansion.
Therefore, from the viewpoint of  the $\kappa^{-1}$-expansion, solving the QC is identical to determining the specific potential form of $\widetilde{Q}$, which is a kind of inverse problems to the standard (E)WKB method to determine energy spectra using the $\hbar$-expansion.

In the below, we clarify transseries structure of the energy spectra for arbitrary $(N,K)$.
Since the dependence of $K$ in the energy solutions is in general  quite complicated, we explicitly write down the solutions only for $K=1$ in Sec.~\ref{sec:Energy_K1} and derive transseries ansatz of the energy spectra only in Sec.~\ref{sec:Energy_Kgen}.
By using the ansatz, the same procedure in Sec.~\ref{sec:Energy_K1} works to find the energy solutions for arbitrary $(N,K)$.
This study is a direct generalization of the results found by the standard WKB analysis, e.g., in Refs.~\cite{Bender:1998ke,Bender:1998kf,Alvarez1,Bender_2001,Kamata:2023opn}.

\subsection{The $K=1$ cases} \label{sec:Energy_K1}

\subsubsection{Even $N$}
We consider the energy solution for even $N$ and write down it as a transseries by solving the exact QC in Eq.(\ref{eq:D0_even_K1}).
Taking ${\frak D}^0 = 0$ gives
\be
1 + C_{(\bar{p},\bar{p}+2)} =  0 &\quad \Rightarrow \quad&  \sum_{n \in 2 {\mathbb N}_0 -1} v_{n} \sin \frac{\pi n}{N} \cdot \eta^{n} = \pi \left( k + \frac{1}{2} \right). \qquad (k \in {\mathbb Z})  \label{eq:D0_even_K1_2}
\ee
where the coefficients, $v_{n \in 2{\mathbb N}_0-1}$, are defined in Eq.(\ref{eq:cof_v}).
When choosing a suitable ansatz for $\eta$, one has to be careful that it should be consistent with asymptotics of the EWKB ansatz which is the $\eta$-expansion as $\eta \rightarrow 0_+$.
Since the leading term in Eq.(\ref{eq:D0_even_K1_2}) is $O(\eta^{-1})$, the ansatz of $\eta^{-1}$ has to be the large $k$-expansion.
For this reason, we prepare ansatz for $\eta$ as\footnote{
  The transseries of $\eta$ is available from that of $\eta^{-1}$, which begins with $O(\kappa^{-1})$ and can be written as
  \be
  \eta  \sim \sum_{\ell \in 2 {\mathbb N}_0+1}d^{(0)}_\ell \kappa^{-\ell} \quad \mbox{as} \quad \kappa \rightarrow +\infty.
  \ee
  The coefficients $d^{(0)}_{\ell}$ are uniquely determined from $c^{(0)}_{\ell}$ in Eq.(\ref{eq:def_eta_even_K1}).
}
\be
&& \eta^{-1} = \frac{E^{(N+2)/(2N)}}{g^{1/N} \hbar} \sim \sum_{\ell \in 2 {\mathbb N}_0-1}c^{(0)}_\ell \kappa^{-\ell} \quad \mbox{as} \quad \kappa \rightarrow +\infty, \label{eq:def_eta_even_K1} \\
&& \kappa = \kappa(k) := \pi \left(  k  + \frac{1}{2} \right), \qquad (k \in {\mathbb N}_0) \label{eq:def_kappa}
\ee
where $k$ is the energy level, and we assume that $e^{(0)}_{-1} > 0$ to make the energy spectrum positive when taking a non-negative $k$. 
Substituting Eq.(\ref{eq:def_eta_even_K1}) into Eq.(\ref{eq:D0_even_K1_2}) determines the coefficients, $e^{(0)}_{\ell \in 2{\mathbb N}_0 - 1}$, recursively, and the solution is given by
\begin{small}
\be
c^{(0)}_{-1} &=& \frac{\Gamma \left( \frac{3}{2} + \frac{1}{N} \right)}{\pi^{1/2} \Gamma \left( 1 + \frac{1}{N} \right) \sin \frac{\pi}{N}}, \nl
c^{(0)}_{1} &=& \frac{\pi^{1/2} N \Gamma \left( 2 -\frac{1}{N} \right) \sin \frac{\pi}{N}}{12 \Gamma \left(\frac{1}{2} -\frac{1}{N} \right)}, \nl
 c^{(0)}_{3} &=& \frac{\pi ^{3/2} N \Gamma \left( 1 + \frac{1}{N} \right) \sin^2 \frac{\pi }{N}}{1440 \Gamma \left(\frac{3}{2} + \frac{1}{N} \right)^2} \nl
 && \cdot \left[ \frac{\left(2 N^2+N-3\right) \Gamma \left(2-\frac{3}{N}\right) \Gamma \left(1 + \frac{1}{N} \right) \sin \frac{3 \pi}{N}}{\Gamma \left( -\frac{1}{2} -\frac{3}{N} \right)} - \frac{10 N \Gamma \left(2 - \frac{1}{N} \right)^2 \Gamma \left(\frac{3}{2} + \frac{1}{N} \right) \sin \frac{\pi}{N}}{\Gamma \left( \frac{1}{2} -\frac{1}{N} \right)^2} \right], \nl
 \vdots && \label{eq:cofe_pt_K1} 
\ee
\end{small}
From Eq.(\ref{eq:def_eta_even_K1}), one can find the transseries energy solution as
\be
\frac{E}{(g^{1/N} \hbar)^{\frac{2N}{N+2}}} \sim \kappa^{\frac{N-2}{N+2}}  \sum_{\ell \in 2 {\mathbb N}_0-1} e^{(0)}_{\ell}  \kappa^{- \ell}, \qquad e^{(0)}_{\ell \in 2{\mathbb N}_0-1} \in {\mathbb R},
\ee
where the coefficients $e^{(0)}_{\ell}$ can be found from $c^{(0)}_{\ell}$ as
\begin{small}
\be
\widetilde{e}^{(0)}_{-1} &=& 1, \nl 
\widetilde{e}^{(0)}_{1} &=&
\frac{\pi (N-1) N \cot \frac{\pi}{N} \sin^2 \frac{\pi}{N}}{3 (N+2)^2}, \nl
\widetilde{e}^{(0)}_{3} &=& 
\frac{(N-1) N \sin^3 \frac{\pi}{N}}{720 (N+2)^4} \nl
&& \cdot \left( \frac{(N+2)^3 (2 N+3)  \Gamma \left(2-\frac{3}{N}\right) \Gamma \left(1+\frac{1}{N}\right)^2 \Gamma \left(\frac{1}{N}\right) \Gamma \left(-\frac{1}{2} -\frac{1}{N} \right)^2 \cos^2 \frac{\pi}{N}  \sin \frac{3 \pi}{N}}{\Gamma \left( \frac{3}{2} + \frac{1}{N}\right) \Gamma \left(-\frac{1}{2} -\frac{3}{N}\right)} \right. \nl
&& \left. \qquad - 20 \pi^2 (N-1) (N+6) \cot^2 \frac{\pi}{N} \sin \frac{\pi}{N}\right),  \nl
 \vdots &&
 \label{eq:pt_energy_K1} 
\ee 
\end{small}
where $\widetilde{e}^{(0)}_\ell := e^{(0)}_\ell /e^{(0)}_{-1}$ normalized by $e^{(0)}_{-1} = \pi^{-\frac{N}{N+2}} \left( \frac{\Gamma \left( \frac{3}{2} + \frac{1}{N} \right)}{\Gamma \left( 1 + \frac{1}{N} \right) \sin \frac{\pi}{N}} \right)^{\frac{2N}{N+2}}$.
The energy solution does not contain non-perturbative sectors because the exact QC only contains the perturbative cycle.
Notice that the resulting energy spectrum is positive real.

\subsubsection{Odd $N$} \label{sec:energy_oddN_K1}
Next, we find the energy solutions for odd $N$.
As different from the even $N$ cases in Eq.(\ref{eq:D0_even_K1}), the exact QCs for odd $N$ contain the non-perturbative contributions, denoted by $B$ in Eq.(\ref{eq:D0_odd_K1_2}).
The same ansatz to Eq.(\ref{eq:def_eta_even_K1}) works for the perturbative sector, but we need to identify higher transmonomials corresponding the non-perturbative sectors and add them to the ansatz of $\eta$.

The higher transmonomial can be easily found using the leading order of $\kappa^{-1}$ in $\eta$. 
Since the perturbative sector is derived from $P$ as
\be
&& 
1 + P \sim 0 \quad \Rightarrow \quad \cos \left[ v_{-1} \sin  \frac{\pi}{N} \cdot \eta^{-1} \right] \sim 0, 
\ee
one can find the leading order of $\eta^{-1}$ as
\be
&& \eta^{-1} \sim \frac{\kappa}{v_{-1} \sin \frac{\pi}{N}} = \frac{ \Gamma \left(\frac{3}{2} + \frac{1}{N} \right)}{\pi^{1/2} \Gamma \left(1 + \frac{1}{N}\right) \sin \frac{\pi}{N}} \kappa, 
\ee
where $\kappa$ is defined in Eq.(\ref{eq:def_kappa}) with the energy level, $k$.
From Eq.(\ref{eq:Bcycle_odd_K1}), the exponential damping factor, i.e., higher transmonomial, is obtained from the leading order of $B$ as
\be
B &\sim& \exp \left[ - v_{-1} \left( \cos  \frac{\pi}{N} + 1 \right) \cdot \eta^{-1} \right] \, \sim \, \exp \left[ - S_{1} \kappa \right],  \\ \nl
S_1 &:=& \frac{\cos \frac{\pi}{N} + 1}{\sin \frac{\pi}{N} } \ \in {\mathbb R}_{>0}.  \label{eq:damp_S1}
\ee
Hence, in order to obtain a closed form for all algebraic operations in the exact QC, the transseries ansatz for $\eta$ has to take the following form:
\be
\eta^{-1} &=& \frac{E^{(N+2)/(2N)}}{g^{1/N} \hbar} \nl
&\sim& \sum_{\ell \in 2 {\mathbb N}_0-1} c^{(0)}_\ell 
\kappa^{-\ell} + \sum_{n \in {\mathbb N}} \sum_{\ell \in {\mathbb N}_0} c^{(n)}_{\ell} \sigma^n \kappa^{-\ell} \quad \mbox{as} \quad \kappa \rightarrow +\infty, \label{eq:ansatz_eta_oddN_K1} \\
\sigma &:=& \sin \kappa \cdot e^{- S_1 \kappa} = (-1)^k e^{- S_1 \kappa}. 
\ee
Substituting the ansatz into Eq.(\ref{eq:D0_odd_K1_2}) and solving ${\frak D}=0$ recursively determine the coefficients, $c^{(n)}_{\ell} \in {\mathbb R}$.
The perturbative coefficients, $c^{(0)}_{\ell \in 2{\mathbb N}_0-1}$, are the same to Eq.(\ref{eq:cofe_pt_K1}), and the first two non-perturbative sectors, $c^{(n=1,2)}_{\ell \in {\mathbb N}_0}$, can be written down as
\begin{small}
\be
c^{(1)}_0 &=& \frac{\Gamma \left(\frac{3}{2} + \frac{1}{N}\right)}{2 \pi^{1/2} \Gamma \left(1 + \frac{1}{N}\right) \sin \frac{\pi}{N}}, \nl
c^{(1)}_1 &=&  -\frac{\pi^{1/2} N \Gamma \left(2 - \frac{1}{N} \right)}{12 \Gamma \left(\frac{1}{2} -\frac{1}{N} \right)} \left(\cos \frac{\pi}{N} + 1\right), \nl
c^{(1)}_2 &=& \frac{\pi^{1/2} N \Gamma \left(2-\frac{1}{N}\right) \sin \frac{\pi}{N} }{72 \Gamma \left( \frac{1}{2} - \frac{1}{N}\right)^2 \Gamma \left(\frac{3}{2} + \frac{1}{N} \right)}  \nl
&& \cdot \left[ 2 \pi N \Gamma \left(2-\frac{1}{N}\right) \Gamma \left(1 + \frac{1}{N} \right) \cos^4 \frac{\pi}{2 N} -3 \Gamma \left(\frac{1}{2} - \frac{1}{N} \right) \Gamma \left(\frac{3}{2} + \frac{1}{N} \right) \right], \nl
\vdots&& \\
c^{(2)}_0 &=& - \frac{\Gamma \left(\frac{3}{2}+\frac{1}{N}\right)}{8 \pi^{1/2} \Gamma \left(1+\frac{1}{N}\right) \sin^2 \frac{\pi}{2 N}} ,\nl
c^{(2)}_1 &=& \frac{\pi^{1/2} N \Gamma \left(2-\frac{1}{N}\right) \cos^3 \frac{\pi }{2 N}}{6 \Gamma \left(\frac{1}{2}-\frac{1}{N}\right) \sin \frac{\pi}{2 N}},\nl
c^{(2)}_2 &=& \frac{\pi^{1/2} N \Gamma \left(2-\frac{1}{N}\right) \cos^2 \frac{\pi }{2 N}}{72 (N+2) \Gamma \left(\frac{1}{2}-\frac{1}{N}\right)} \left[ 9 (N+2) - \frac{8 \pi  (N-1) \cos^3 \frac{\pi}{2 N} \cos \frac{\pi }{N}}{\sin\frac{\pi }{2 N}} \right], \nl
\vdots && 
\ee
\end{small}
The higher non-perturbative sectors, $c_{\ell \in {\mathbb N}_0}^{(n>2)}$, can be also determined in the similar way.
The energy spectrum is obtained from the above results from Eq.(\ref{eq:ansatz_eta_oddN_K1}) and holds the similar non-perturbative structure to that of $\eta$, which is expressed by
\be
&& \frac{E}{(g^{1/N} \hbar)^{\frac{2N}{N+2}}} \sim \kappa^{\frac{N-2}{N+2}} \left[ \sum_{\ell \in 2 {\mathbb N}_0-1} e^{(0)}_{\ell}  \kappa^{- \ell} +  \sum_{n \in {\mathbb N}} \sum_{\ell \in {\mathbb N}_0} \sigma^{n} e^{(n)}_{\ell} \kappa^{- \ell} \right], 
\quad e^{(n)}_{\ell} \in {\mathbb R}. 
\ee
The perturbative coefficients, denoted by $e_{\ell}^{(0)}$, are the same to Eq.(\ref{eq:pt_energy_K1}), and the first two non-perturbative sectors, $e_{\ell}^{(n = 1,2)}$, can be written down as
\begin{small}
\be
\widetilde{e}^{(1)}_{0} &=&\frac{N}{N+2}, \nl
\widetilde{e}^{(1)}_{1} &=& \frac{\pi  (N-1) N \cos \frac{\pi}{N}}{3 (N+2)^2} \left( \cos \frac{\pi}{N} + 1 \right), \nl
\widetilde{e}^{(1)}_{2} &=& - \frac{2 \pi  (N-1) N \sin \frac{\pi}{N} \cos \frac{\pi}{N}}{9 (N+2)^3}  \left( 3 -  \pi  (N-1) \cos^4 \frac{\pi }{2 N} \cot \frac{\pi }{N} \right), \nl
\vdots && \\
\widetilde{e}^{(2)}_{0} &=& \frac{1}{4} \left(\frac{\pi^{-1/2} \Gamma \left(\frac{1}{2}+\frac{1}{N}\right)}{\Gamma \left(1+\frac{1}{N}\right) \sin \frac{\pi}{N}}-\frac{2 N \cot \frac{\pi }{2 N}}{N+2}\right), \nl
\widetilde{e}^{(2)}_{1} &=& - \frac{N}{12 (N+2)^3} \left( 8 \pi^{-1/2} N^2 \Gamma \left(2-\frac{1}{N}\right) \Gamma \left(\frac{3}{2}+\frac{1}{N}\right) \cos \frac{\pi}{N}  \cos^2  \frac{\pi}{2 N} \right. \nl
&& \left. - 3 (N-2) - 16 \pi  (N-1) \cos^4 \frac{\pi }{2 N} \cot \frac{\pi}{N} \right), \nl
\widetilde{e}^{(2)}_{2} &=& \frac{N \sin \frac{2 \pi}{N}}{144 (N+2)^3} \left[ \frac{2 \pi^{1/2} N^2 \Gamma \left(2-\frac{1}{N}\right) }{\Gamma \left(-\frac{1}{2}-\frac{1}{N} \right)} \left( \frac{3 (N+6)}{\cos \frac{\pi}{N}} - \frac{\pi  (N-1) \sin^3 \frac{\pi}{N}}{\sin^4 \frac{\pi}{2 N}} \right) \right. \nl
&& \left. - \pi (N-1) \left( 12 (N-2) \cot \frac{\pi}{2 N} + \frac{\pi  (N-1) \sin^3 \frac{\pi}{N} \sin \frac{2 \pi}{N}}{2\sin^6 \frac{\pi}{2 N}} \right) \right], \nl
\vdots && 
\ee
\end{small}
where $\widetilde{e}^{(n)}_\ell := e^{(n)}_\ell /e^{(0)}_{-1}$ are normalized coefficients divided by $e^{(0)}_{-1} = \pi^{-\frac{N}{N+2}} \left( \frac{\Gamma \left( \frac{3}{2} + \frac{1}{N} \right)}{\Gamma \left( 1 + \frac{1}{N} \right) \sin \frac{\pi}{N}} \right)^{\frac{2N}{N+2}}$.
Notice that, the same as the even $N$ cases, the resulting energy spectrum is positive real.

\subsection{Generalization to $K>1$} \label{sec:Energy_Kgen}
We consider the generalization to $K>1$ from Eqs.(\ref{eq:D0_evenN_Kgen})(\ref{eq:D0_oddN_Kgen}).
Since the analysis for the perturbative part is almost the same to the $K=1$ cases, one can straightforwardly find the perturbative coefficients of $\eta^{-1}$ and the energy spectrum, i.e., $c_{\ell}^{(0)}$ and $e_{\ell}^{(0)}$.
These can be expressed from the results of $K=1$, Eqs.(\ref{eq:cofe_pt_K1})(\ref{eq:pt_energy_K1}), by replacing $\sin \frac{\pi n}{N}$ as $\sin \frac{\pi n}{N}\rightarrow \sin \frac{\pi K n}{N}$ (but not for $\cos \frac{\pi n}{N}$ and $\cot \frac{\pi n}{N}$).
Therefore, the perturbative sector has been already solved.

In the below, we investigate the non-perturbative structure in their transseries.
Since transseries structure of the energy spectra are essentially the same as that of $\eta$, we mainly address derivation of a transseries ansatz of $\eta$.
We would not write down specific forms of the non-perturbative coefficients for $K>1$ because they highly depend on the value of $(N,K)$, but construction of the ansatz is indeed sufficient to see properties of the energy solutions, such as the non-perturbative structure and the spectral reality.

Firstly, we see the even $N$ cases.
The number of independent non-perturbative sectors in the energy spectra can be found by counting that of independent exponential damping factors of the cycles contained in the exact QCs (\ref{eq:D0_evenN_Kgen}) using Eqs.(\ref{eq:C_evenN_rel1})(\ref{eq:C_evenN_rel2}).
It is determined as
\be
\mbox{\# of independent NP sectors} =
\begin{cases}
  0 & \mbox{for $K=1$} \\
  \lfloor K/2\rfloor + 1 & \mbox{otherwise}
\end{cases}. \label{eq:num_of_S_even}
\ee
We should remind that the case of $K=1$ is special because, as we saw in Sec.~\ref{sec:evenN_K1}, their contributions are canceled to each other.
It is notable that the number is determined only by $K$ and irrelevant to $N$.
Such a non-perturbative sector can be classified by its damping ratio such as $S_1$ for $K=1$ in Eq.(\ref{eq:damp_S1}).
For $K>1$ in even $N$, the damping ratios are defined as
\be
&& S_n := - \frac{\sin \arg(a_{\bar{p}+2n}) - \sin \arg(a_{\bar{p}})}{\sin \frac{\pi K}{N}} = \frac{\cos \frac{\pi (K-2n)}{N} - \cos \frac{\pi K}{N}}{\sin \frac{\pi K}{N}}, \qquad (n \in \{ 1, 2,\cdots, \lfloor K/2 \rfloor \}) \nl
&& S_{\lfloor K/2 \rfloor + 1} :=   \frac{\sin \arg (a_{\bar{p}+1}) - \sin \arg (a_{\bar{p}})}{\sin \frac{\pi K}{N}} = \frac{2 \cos \frac{\pi K}{N}}{\sin \frac{\pi K}{N}}. \label{eq:Sn_evenN}
\ee
By these ratios, the leading orders of the cycles in the exact QCs (\ref{eq:D0_evenN_Kgen}) can be identified as
\be
&&  {\rm Re}[\log C_{(\bar{p}+2 n-2,\bar{p}+2 n + 1)}] \nl
&& \sim
   \begin{cases}
     - (S_1 + S_{\lfloor K/2 \rfloor +1}) \kappa + O(\kappa^{-1}) &  \mbox{for $n=1$}\\
     - (S_{n-1} + S_{n} + S_{\lfloor K/2 \rfloor +1}) \kappa + O(\kappa^{-1}) &  \mbox{for $n \in \{ 2,\cdots, \lfloor K/2 \rfloor \}$}\\
     - (2 S_{\lfloor K/2 \rfloor} + S_{\lfloor K/2 \rfloor +1}) \kappa + O(\kappa^{-1}) &  \mbox{for $n=\lfloor K/2 \rfloor + 1$ if $K \in 2{\mathbb N}_0+1$}\\
   \end{cases},  \\ \nl
   &&   {\rm Re}[\log C_{(\bar{p}+ 2K-2,\bar{p}+2 K)}] \sim - S_1 \kappa + O(\kappa^{-1}), \\ \nl
   && \log B_{(\bar{p}+2 n-2,\bar{p}+2 n - 1)} \nl
   && \sim
   \begin{cases}
     - S_{\lfloor K/2 \rfloor + 1} \kappa + O(\kappa^{-1})  & \mbox{for $n=1$}\\
     - (2 S_{n-1} +  S_{\lfloor K/2 \rfloor + 1}) \kappa + O(\kappa^{-1}) & \mbox{for $n \in \{ 2, \cdots, \lfloor K/2 \rfloor +1 \}$}
   \end{cases}.
\ee
Hence, one can construct the ansatz of $\eta^{-1}$ as
\be
\eta^{-1} &=& \frac{E^{(N+2)/(2N)}}{g^{1/N} \hbar} \nl
&\sim&   \sum_{\ell \in 2{\mathbb N}_0-1} c^{({\bf 0})}_\ell \kappa^{-\ell} + \sum_{\substack{{\bf n} \in {\mathbb N}_{0}^{\lfloor K/2  \rfloor +1}\\ |{\bf n}|>0}} \sum_{\ell \in {\mathbb N}_0} \bm{\sigma}^{\bf n} c^{({\bf n})}_\ell \kappa^{-\ell} \quad \mbox{as} \quad \kappa \rightarrow +\infty, \label{eq:eta_evenN_Kgen} \\
\bm{\sigma}^{\bf n} &:=& \prod_{j =1}^{\lfloor K/2 \rfloor +1} \sigma_{(j)}^{n_j}, \qquad \sigma_{(j)}:= \sin \kappa \cdot  e^{-S_j \kappa} = (-1)^{k} e^{-S_j \kappa}. \label{eq:sig_evenN_Kgen}
\ee
Notice that, from Eqs.(\ref{eq:D0_evenN_Kgen})(\ref{eq:C_evenN_rel1})(\ref{eq:C_evenN_rel2}), one can find that $c^{(0,\cdots,0,n)}_{\ell} =0$ for any $\ell \in {\mathbb N}_0$ and $n \in {\mathbb N}$.

Then, we consider the odd $N$ cases.
In the similar way to the even $N$ cases, counting the number of independent non-perturbative sectors in the exact QCs (\ref{eq:D0_oddN_Kgen}) using Eq.(\ref{eq:C_oddN_rel}) gives
\be
\mbox{\# of independent NP sectors} = K.
\ee
Similar to the situation for even $N$, the number is determined only by $K$.
By defining the damping ratios as
\be
S_n &:=&  (-1)^{n+1}  \frac{\sin \arg(a_{\bar{p}+n}) - \sin \arg (a_{\bar{p}})}{\sin \frac{\pi K}{N}} =  \frac{\cos \frac{\pi (K - n)}{N} - (-1)^{n} \cos \frac{\pi K}{N}}{\sin \frac{\pi K}{N}}, \qquad (n \in \{ 1, \cdots, K \})  \label{eq:Sn_oddN}
\ee
the leading orders of the cycles can be written as
\be
{\rm Re}[\log C_{(\bar{p}+n-1, \bar{p}+n)}] \sim
\begin{cases}
  - S_1 \kappa + O(\kappa^{-1}) & \mbox{for $n=1$} \\
  (-1)^{n} (S_{n-1}+S_n) \kappa + O(\kappa^{-1}) & \mbox{for $n \in \{2,3,\cdots,K\}$}
\end{cases}.
\ee
In consequence, one can find the following transseries ansatz for $\eta^{-1}$:
\be
\eta^{-1} &=& \frac{E^{(N+2)/(2N)}}{g^{1/N} \hbar} \nl
&\sim&   \sum_{\ell \in 2{\mathbb N}_0-1} c^{({\bf 0})}_\ell \kappa^{-\ell} + \sum_{\substack{{\bf n} \in {\mathbb N}_{0}^{K}\\ |{\bf n}|>0}} \sum_{\ell \in {\mathbb N}_0} \bm{\sigma}^{\bf n} c^{({\bf n})}_\ell \kappa^{-\ell} \quad \mbox{as} \quad \kappa \rightarrow +\infty, \label{eq:eta_oddN_Kgen} \\
\bm{\sigma}^{\bf n} &:=& \prod_{j =1}^{K} \sigma_{(j)}^{n_j}, \qquad \sigma_{(j)}:= \sin \kappa \cdot  e^{-S_j \kappa} = (-1)^{k} e^{-S_j \kappa}.
\ee
It is remarkable that the number of the non-perturbative sectors is roughly twice as the even cases with the same $K$ due to lack of the ${\mathbb Z}_2$ symmetry in Eq.(\ref{eq:Z2_symm}).
\\ \par
\begin{figure}[tbp]
  \begin{center}
    \begin{tabular}{cc}
      \begin{minipage}{0.5\hsize}
        \begin{center}
          \includegraphics[clip, width=75mm]{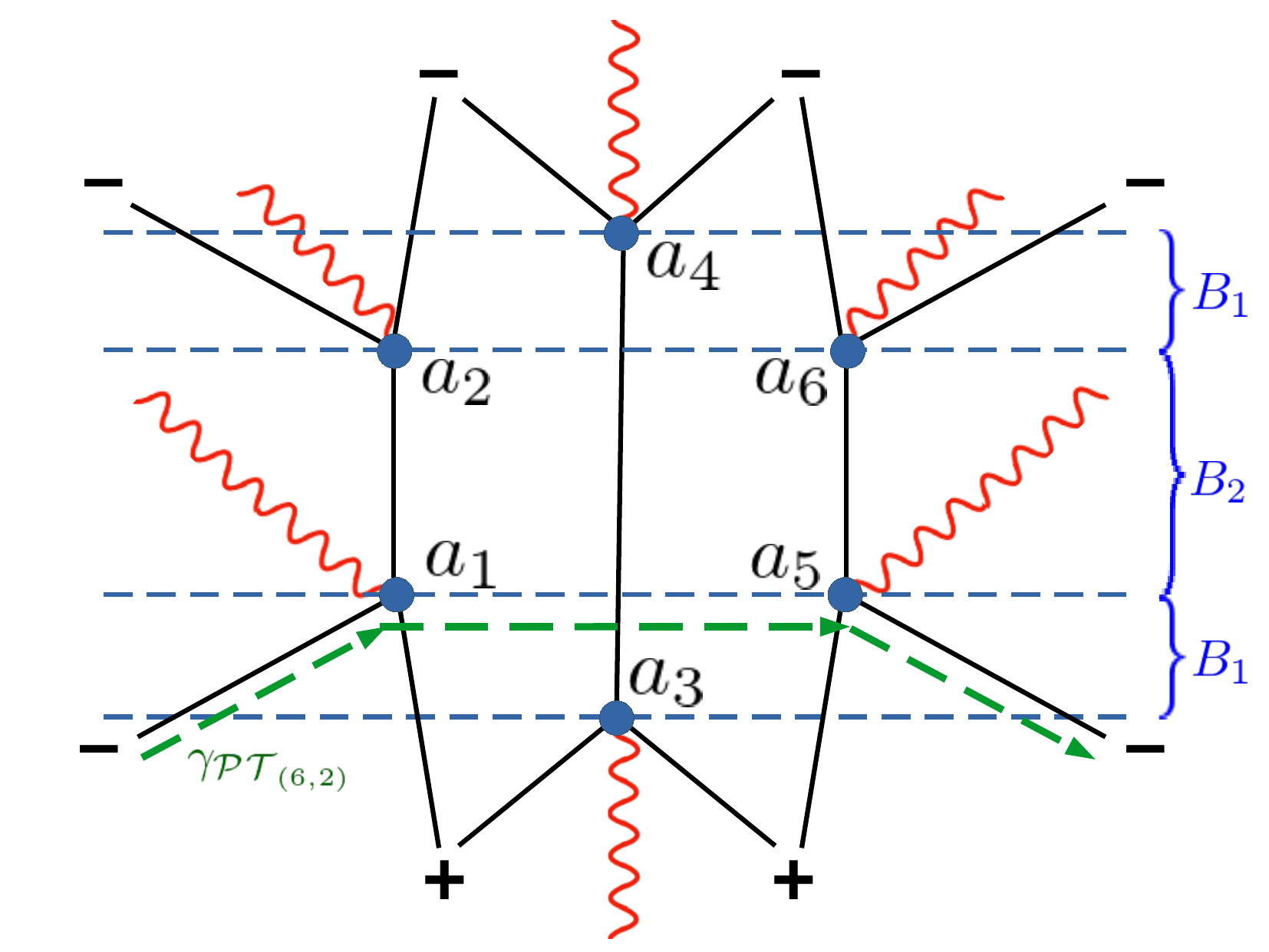}
          \hspace{1.6cm} (a) $(N,K) = (6,2)$
        \end{center}
      \end{minipage}
      \begin{minipage}{0.5\hsize}
        \begin{center}
          \includegraphics[clip, width=75mm]{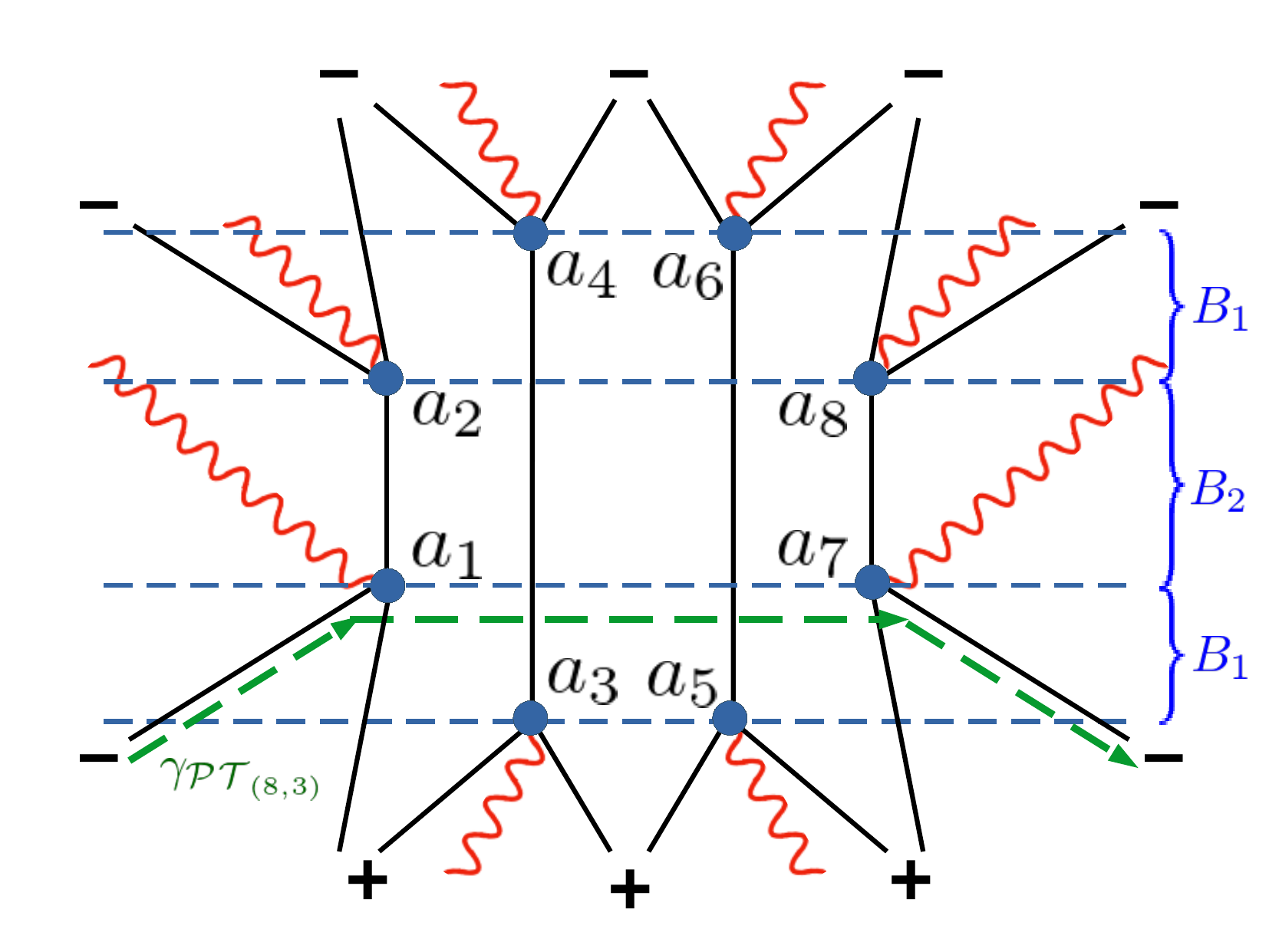} 
          \hspace{1.6cm} (b) $(N,K) = (8,3)$
        \end{center}
      \end{minipage} \\ \\
      \begin{minipage}{0.5\hsize}
        \begin{center}
          \includegraphics[clip, width=75mm]{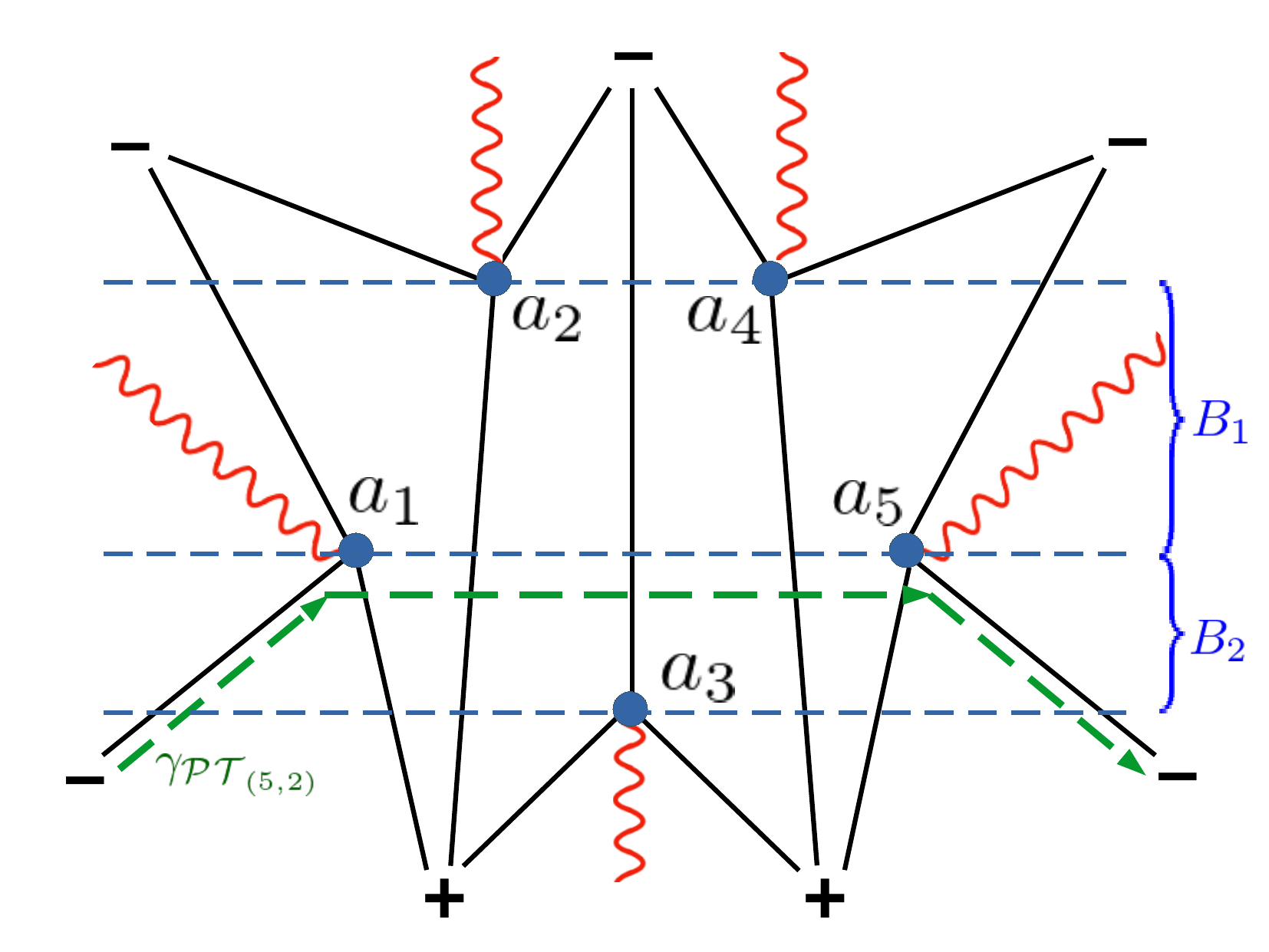}
          \hspace{1.6cm} (c) $(N,K) = (5,2)$
        \end{center}
      \end{minipage}
      \begin{minipage}{0.5\hsize}
        \begin{center}
          \includegraphics[clip, width=75mm]{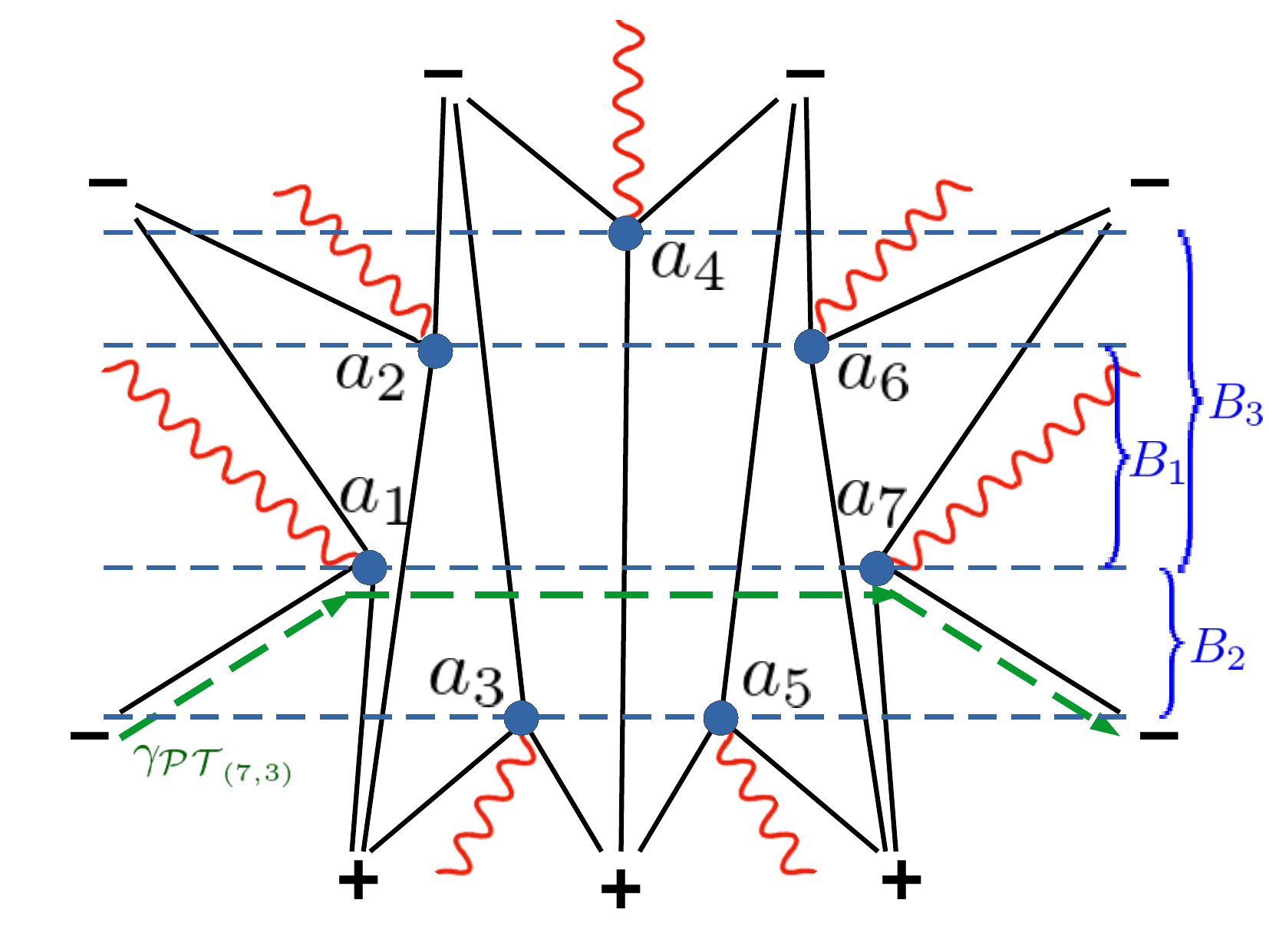}
          \hspace{1.6cm} (d) $(N,K) = (7,3)$
        \end{center}
      \end{minipage}      
    \end{tabular} 
    \caption{Stokes graphs with $\arg(\eta) = 0$ for some $(N,K)$ shown in
      Example 1-4.
      The green lines denote the path of analytic continuation.
      For even $N$, (a) and (b), degeneracies of the Stokes lines occur between the turning points in complex conjugated pairs.
      The length labeled by $B_{n}$ corresponds to half of the integration contour in the exponent of the associated $B$-cycles in Eqs.(\ref{eq:QC_N6K2_2})(\ref{eq:QC_N8K3_2})(\ref{eq:QC_N5K2_2})(\ref{eq:QC_N7K3_2}).
    }
    \label{fig:examples}
  \end{center}
\end{figure}

 
We below show some examples of the exact QCs obtained by Eqs.(\ref{eq:D0_evenN_Kgen})(\ref{eq:D0_oddN_Kgen}) and their reexpressions.
In these examples, we use the symbols, $B_{n \in {\mathbb N}}$, to denote non-perturbative cycles such that $B_{n} \propto e^{-S_n \kappa}$ with the damping ratios $S_n$ given by Eqs.(\ref{eq:Sn_evenN}) and (\ref{eq:Sn_oddN}) for even and odd $N$, respectively.
From the reexpressions, one can see consistency with the ansatz in Eqs.(\ref{eq:eta_evenN_Kgen})(\ref{eq:eta_oddN_Kgen}) and the spectral reality.
The corresponding Stokes graphs are shown in Fig.~\ref{fig:examples}.

\subsubsection*{{Example 1: $(N,K) = (6,2)$}} \label{sec:N6_K4} 
We show the case of $(N,K) = (6,2)$.
Eq.(\ref{eq:D0_evenN_Kgen}) leads to
\be
   {\frak D}^{0} &\propto& \frac{1}{P^{1/2}} \left[ 1 + C_{(1,5)} \sqrt{\frac{1+B_{(5,6)}}{1+B_{(1,2)}}} + \frac{C_{(1,4)}}{\sqrt{1 + B_{(1,2)}} \sqrt{1 + B_{(3,4)}}} + \frac{C_{(3,5)} + C_{(3,6)}}{\sqrt{1 + B_{(3,4)}} \sqrt{1 + B_{(5,6)}}} \right]. \nl \label{eq:QC_N6K2}
\ee
Redefining the cycles by splitting into the damping and oscillation parts, it can be reexpressed by
\be
   {\frak D}^{0} &\propto& P^{-1/2} + P^{+1/2} + \frac{B_1 + 2 B_1 B_{2}}{\sqrt{1 + B_{2}} \sqrt{1 + B_{1}^2 B_{2}}}. \label{eq:QC_N6K2_2}
\ee
In Eq.(\ref{eq:QC_N6K2}), the perturbative cycle corresponds to $P = C_{(1,5)}$, and $\sqrt{\frac{1+B_{(5,6)}}{1+B_{(1,2)}}}=1$.

\subsubsection*{{Example 2: $(N,K) = (8,3)$}} \label{sec:N8_K6}
Then, we consider the case of $(N,K) = (8,3)$.
Eq.(\ref{eq:D0_evenN_Kgen}) leads to
\be
   {\frak D}^{0} &\propto&  \frac{1}{P^{1/2}}  \left[ 1 + C_{(1,7)} \sqrt{\frac{1+B_{(7,8)}}{1+B_{(1,2)}}} +  \frac{C_{(1,4)}}{\sqrt{1+B_{(1,2)}}\sqrt{1+B_{(3,4)}}}  \right. \nl
     && \qquad \left. + \frac{C_{(1,6)}}{\sqrt{1+B_{(1,2)}}\sqrt{1+B_{(5,6)}}} + \frac{C_{(3,7)} + C_{(3,8)}}{\sqrt{1+B_{(3.4)}}\sqrt{1+B_{(7,8)}}} + \frac{C_{(5,7)} + C_{(5,8)}}{\sqrt{1+B_{(5,6)}}\sqrt{1+B_{(7,8)}}} \right. \nl
     && \qquad \left. + \frac{C_{(3,6)}}{\sqrt{1+B_{(3,4)}} \sqrt{1+B_{(5,6)}}} + \frac{C_{(1,4)} C_{(5,7)} \sqrt{1+B_{(7,8)}}}{\sqrt{1+B_{(1,2)}}\sqrt{1+B_{(3,4)}}\sqrt{1+B_{(5,6)}}} \right]. \label{eq:QC_N8K3}
\ee
It can be reexpressed by
\be
   {\frak D}^{0}
   &\propto&   P^{-1/2} + P^{+1/2} + \frac{B_1 + 2 B_1 B_2}{\sqrt{1+B_{2}}\sqrt{1+B_1^2 B_{2}}} \left( A^{-1/2}_{(3,5)} + A^{+1/2}_{(3,5)}\right)  \nl
   && + \frac{B_1^2 B_2}{1 + B_1^2 B_2} \left( P^{-1/2}A_{(3,5)} + P^{+1/2} A^{-1}_{(3,5)} \right), \label{eq:QC_N8K3_2}
\ee
where $A_{(3,5)} := C_{(3,5)}$ is a cycle with a pure oscillation.
In Eq.(\ref{eq:QC_N8K3}), the perturbative cycle corresponds to $P = C_{(1,7)}$, and $\sqrt{\frac{1+B_{(7,8)}}{1+B_{(1,2)}}}=1$.

\subsubsection*{{Example 3: $(N,K) = (5,2)$}} \label{sec:N5_K2}
We show the case of $(N,K) = (5,2)$.
Eq.(\ref{eq:D0_oddN_Kgen}) gives
\be
   {\frak D} \propto \frac{1}{P^{1/2}} && \left[ 1 +  C_{(1,2)} + C_{(1,4)} + C_{(1,5)} + C_{(3,4)} + C_{(3,5)} \right. \nl
&& \left.  + C_{(1,2)} C_{(3,4)} + C_{(1,2)} C_{(3,5)} \right], \label{eq:QC_N5K2}
\ee
it can be expressed by
\be
   {\frak D} &\propto& P^{-1/2} + P^{+1/2} + B_1 \left( A^{-1}_{(2,3)} + A_{(2,3)} \right) + B_2 \nl
   && +  B_1 B_2 \left( A^{-1}_{(1,2)} + A_{(1,2)} \right) + B_1^2 B_2, \label{eq:QC_N5K2_2}
\ee
where $A_{(n_1,n_2)}:= \exp \left[ i \cdot {\rm Im}[\log C_{(n_1,n_2)}] \right]$.
In Eq.(\ref{eq:QC_N5K2}), the perturbative cycle  corresponds to $P = C_{(1,5)}$.

\subsubsection*{{Example 4: $(N,K) = (7,3)$}} \label{sec:N7_K3}
Then, we consider the case of $(N,K) = (7,3)$.
Eq.(\ref{eq:D0_oddN_Kgen}) leads to
\be
   {\frak D} \propto \frac{1}{P^{1/2}} && \left[ 1 + C_{(1,2)}+ C_{(1,4)} + C_{(1,6)} + C_{(1,7)} \right. \nl
     && \left. + C_{(3,4)} + C_{(3,6)} + C_{(3,7)} + C_{(5,6)} + C_{(5,7)} \right. \nl
     && \left. + C_{(1,2)} C_{(3,4)} + C_{(1,2)} C_{(3,6)} + C_{(1,2)} C_{(3,7)} + C_{(1,2)} C_{(5,6)} + C_{(1,2)} C_{(5,7)} \right. \nl
     &&  \left.  + C_{(1,4)} C_{(5,6)} + C_{(1,4)} C_{(5,7)} + C_{(3,4)} C_{(5,6)} + C_{(3,4)} C_{(5,7)}  \right. \nl
   &&\left. + C_{(1,2)} C_{(3,4)} C_{(5,6)} + C_{(1,2)} C_{(3,4)} C_{(5,7)} \right], \label{eq:QC_N7K3}
\ee
and it can be written as
\be
   {\frak D} &\propto& P^{-1/2} + P^{+1/2} + B_1 \left( A_{(2,4)}^{-1} + A_{(2,4)} \right) + B_2 \left( A_{(3,4)}^{-1} + A_{(3,4)} \right) + B_3 \nl
   && + B_1 B_2 \left( A_{(1,2)}^{-1} + A_{(1,2)} \right) \left( A_{(3,4)}^{-1} + A_{(3,4)} \right)  + B_2 B_3 \left( A_{(1,3)}^{-1}  + A_{(1,3)} \right) \nl
   && + B_1 B_2 B_3 \left( A_{(2,3)}^{-1} + A_{(2,3)} \right)  + B_1^2 B_2 \left( A_{(3,4)}^{-1}  + A_{(3,4)} \right) + B_2^2 B_3 \nl
   && + B_1 B_2^2 B_3 \left( A_{(1,2)}^{-1} + A_{(1,2)} \right)  + B_1^2 B_2^2 B_3, \label{eq:QC_N7K3_2}
\ee
where $A_{(n_1,n_2)} := \exp \left[ i \cdot {\rm Im}[\log C_{(n_1,n_2)}] \right]$.
Notice that $A_{(n_1,n_2)} \cdot A_{(n_2,n_3)} = A_{(n_1,n_3)}$.
In Eq.(\ref{eq:QC_N7K3}), the perturbative cycle corresponds to $P = C_{(1,7)}$.

\section{Formulas from the exact quantization conditions} \label{sec:varous_formula}
In this section, we derive formulas in Fig.~\ref{fig:flowchart} using the exact QCs constructed in Sec.~\ref{sec:quant_cond}.
In Sec.~\ref{sec:gutzwiller}, we consider the Gutzwiller trace formula.
In Secs.~\ref{sec:spec_sum} and \ref{sec:path_int}, we derive the spectral summation form and the Euclidean path-integral, respectively.
We would like to notice that, in the below discussions, we use cycles and energy spectra expressed by transseries, but those can be formally replaced with their Borel resummed forms, which are analytic functions, by acting the median resummation, ${\cal S}_{{\rm med},0}$.

\subsection{Gutzwiller trace formula} \label{sec:gutzwiller}
We derive the Gutzwiller trace formula (GTF) using the resolvent method~\cite{Gutzwiller1971}.
Roughly speaking, the GTF describes particle's periodic orbits on a constant energy plane in a given potential.
The GTF is usually formulated in the semi-classical level (sub-leading of the stationary phase approximation) and has the form that
\be
G(E) 
&=& \sum_{n \in {\mathbb N}} \sum_{\rm p.p.o.} i T(E) (-1)^n e^{n \cdot i \oint_{\rm p.p.o.} p dx} \left| \det \frac{\delta^2 S}{\delta x \delta x} \right|^{-1/2}, \label{eq:gutz_gen_oig}
\ee
where ``p.p.o.'' denotes primal periodic orbits, and $T(E)$ is a period with a fixed energy, $E$.
Especially, the sign $(-1)^n$ known as the Maslov index plays a key role of this formula, and this originates from the number of negative eigenvalues of the Hessian matrix, $\frac{\delta^2 S}{\delta x \delta x}$, expanded around the corresponding classical solutions.
See, for example, Refs.~\cite{Brack:1995zz,Muratore-Ginanneschi:2002sjs,Dunne:2008zza,Dietrich:2007vw,Friedrich:1996zza,Sugita:2000ym,Vicedo:2008jy,Esterlis:2014zoa,Behtash:2017rqj,Wang:2021wtp,Arranz:2023bcx,Joseph:2023gmg} for applications of the GTF and the Maslov index.

One of the ways to construct the GTF is to use the partition function $Z(\beta)$.
By denoting a (Hermitian) Hamiltonian operator as $\widehat{H}$, (trace of) the resolvent, $G(E)$, is defined as~\cite{Zinn-Justin:2004vcw}
\be
 G(E) &:=& \int_0^{+\infty} d\beta Z(\beta) e^{\beta E} = \Tr \, \frac{1}{\widehat{H} - E}, \label{eq:resolvent} \\
 Z(\beta) &=& \frac{1}{2 \pi i} \int^{\delta + i \infty}_{\delta - i \infty} G(E) e^{- \beta E} dE, \label{eq:ZbetaG}
\ee
where $Z(\beta)$ is the partition function, $\Tr$ denotes trace over the (Hermitian) Hilbert space, and $0< \delta \ll 1$ is a regularization parameter to avoid  $E=0$.
The resolvent $G(E)$ can be also expressed by the Fredholm determinant, ${\frak D}_{\rm FH}(E)$, as
\be
&& {\frak D}_{\rm FH}(E) = \det \left( \widehat{H} - E \right) = 0, \label{eq:D_det} \\
&& G(E) = - \pd_E \log {\frak D}_{\rm FH}(E). \label{eq:G_pdE_logD}
\ee
The important point is that Eq.(\ref{eq:D_det}) is essentially the same as our exact QCs, ${\frak D}(E)$, and can be generalized by replacing ${\frak D}_{\rm FH}(E)$ with ${\frak D}(E)$~\cite{Sueishi:2020rug}.
Furthermore, as the greatest benefit, translating from the exact QCs into the GTF is technically and intuitively simple thanks to the cycle-representations of of the exact QCs.
\\ \par
\begin{figure}[tbp]
  \begin{center}
    \begin{tabular}{cc}
      \begin{minipage}{0.5\hsize}
        \begin{center}
          \includegraphics[clip, width=75mm]{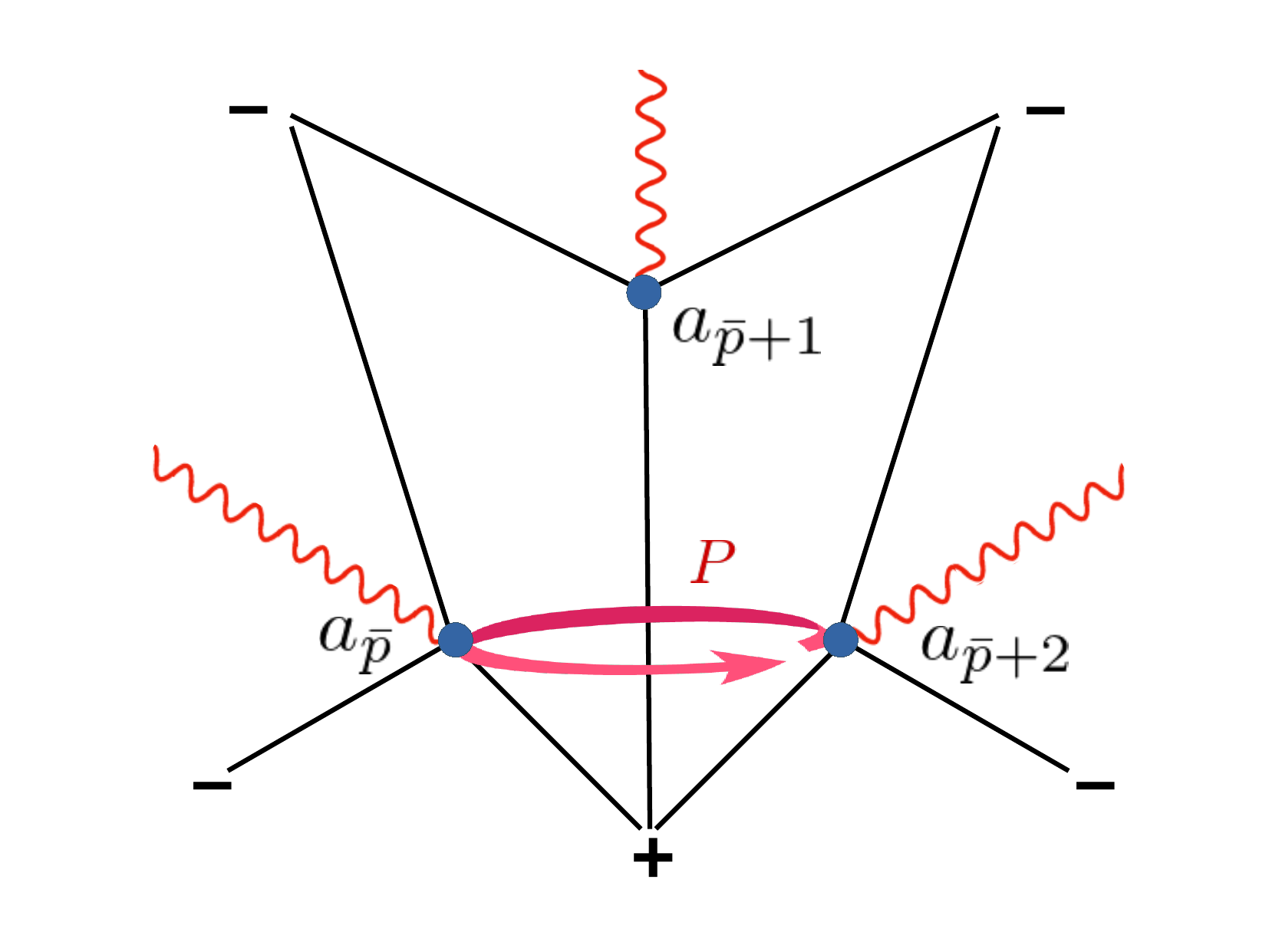}
          \hspace{1.6cm} (a) $P = C_{(\bar{p},\bar{p}+2)}$
        \end{center}
      \end{minipage}
      \begin{minipage}{0.5\hsize}
        \begin{center}
          \includegraphics[clip, width=75mm]{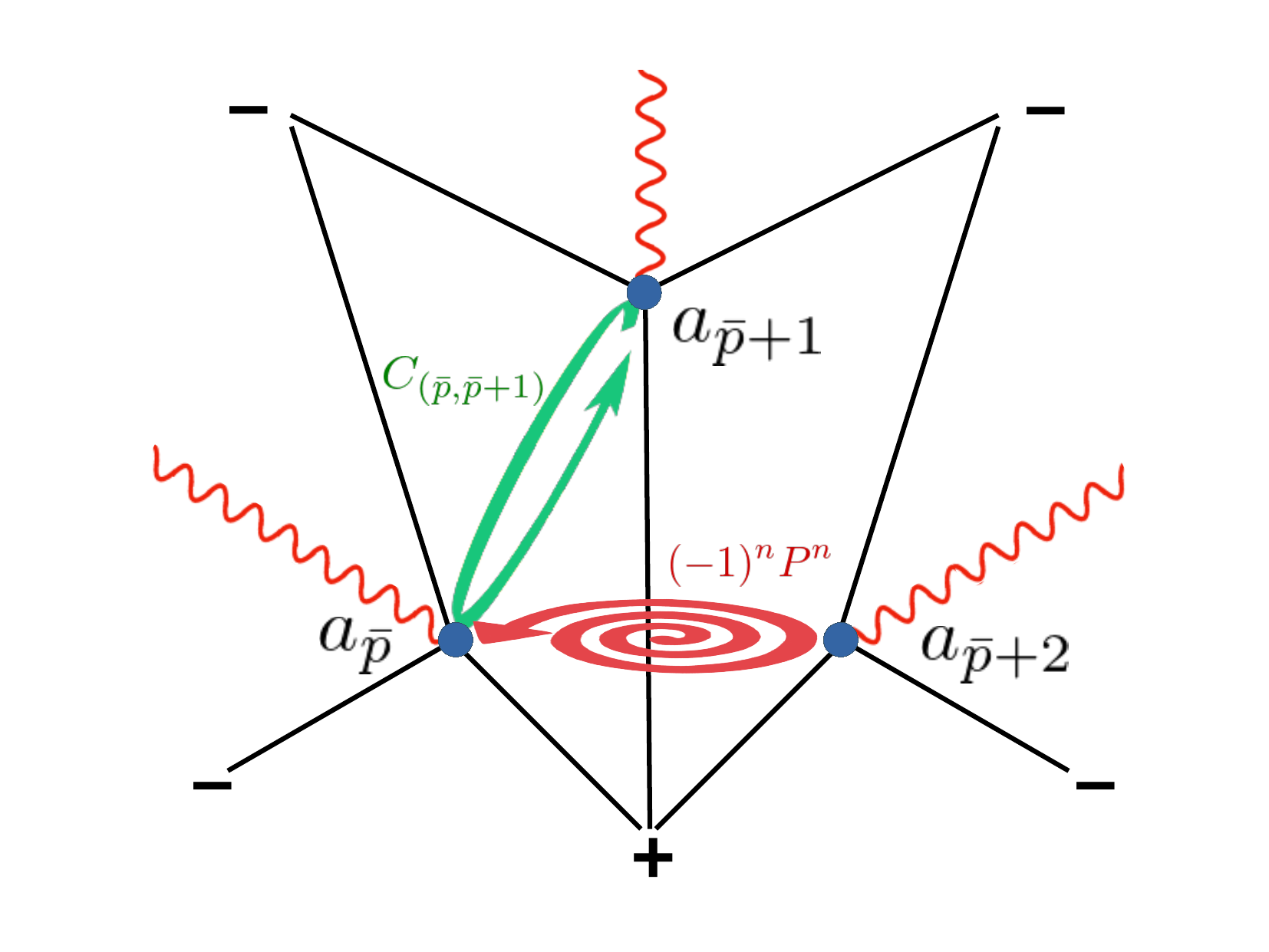}
          \hspace{1.6cm} (b) $L = C_{(\bar{p},\bar{p}+1)}\sum_{n \in {\mathbb N}_0} (-1)^n P^n$
        \end{center}
      \end{minipage} 
    \end{tabular} 
    \caption{Schematic figure of p.p.o.s for $(N,1)$ with odd $N$.
    }
    \label{fig:ppo_N3K1}
  \end{center}
\end{figure}

The GTF for $(N,K)$ with $K=1$ for even $N$ is almost trivial because of vanishing non-perturbative effects in the exact QC.
So that, we firstly show a slightly non-trivial example, $K=1$ for odd $N$, by using Eq.(\ref{eq:D0_odd_K1}).
In order to see the non-perturbative effects more clearly, we factorize the QC into the perturbative and the non-perturbative parts, denoted by ${\frak D}_{\rm P}$ and ${\frak D}_{\rm NP}$, as 
\be
&& {\frak D} \propto 1 + P + C_{(\bar{p}, {\bar{p}+1})} = {\frak D}_{\rm P} \cdot {\frak D}_{\rm NP}, \nl
&& {\frak D}_{\rm P} = 1 + P, \qquad  {\frak D}_{\rm NP} =1 +  \frac{C_{(\bar{p}, {\bar{p}+1})}}{1 + P},
\ee
where $P := C_{(\bar{p},\bar{p}+2)}$ is the perturbative cycle.
From Eq.(\ref{eq:G_pdE_logD}), the resolvent $G(E)$ is written by the cycles as
\be
&& G(E) =  G_{\rm P}(E)  + G_{\rm NP}(E), \\
&& G_{\rm P} = - \frac{\pd_E P}{1+P} = - \pd_E P \sum_{n \in {\mathbb N}_0} (-1)^n P^n = \frac{i}{\hbar} T_{P}(E) P  \sum_{n \in {\mathbb N}_0} (-1)^n P^n, \label{eq:resol_Gp_N3K1} \\
&& G_{\rm NP} = - \frac{\pd_E L}{1+L} =  - \pd_E L \sum_{n \in {\mathbb N}_0} (-1)^n L^n = \frac{i}{\hbar} T_{L}(E) L  \sum_{n \in {\mathbb N}_0} (-1)^n L^n, \label{eq:resol_Gnp_N3K1} \\
&& L =  \frac{C_{(\bar{p}, {\bar{p}+1})}}{1 + P} =  C_{(\bar{p}, {\bar{p}+1})}  \sum_{n \in {\mathbb N}_0} (-1)^{n} P^{n}, \label{eq:Gutz_L_N3K1}
\ee
where $G_{\rm P}$ and $G_{\rm NP}$ are the perturbative and non-perturbative parts, respectively, and $L$ corresponds to the non-perturbative p.p.o. in $G_{\rm NP}$.
The Maslov index, $(-1)^{n}$, naturally arises from the $P$- and the $L$-cycles in Eqs.(\ref{eq:resol_Gp_N3K1})(\ref{eq:resol_Gnp_N3K1}), and the same index also appears from $P$ in the $L$-cycle as Eq.(\ref{eq:Gutz_L_N3K1}).
The schematic figure of the p.p.o.s is shown in Fig.~\ref{fig:ppo_N3K1}.
Moreover, by identification with Eq.(\ref{eq:gutz_gen_oig}), the \textit{quantum} periods including all $\hbar$-orders in $G_{{\rm P}/{\rm NP}}$  are identified from the derivative parts as, $T_{P}(E) := i \hbar\pd_E \log P$ and $T_{L} (E) := i \hbar \pd_E \log  L$, respectively.
Specifically (cf. Ref.~\cite{Bender:1998ke}),
\be
\frac{T_{P}(E)}{\hbar} &=& \frac{(N+2) g^{1/N}}{N E} \sum_{n \in 2{\mathbb N}_0-1} n v_{n} \sin \frac{\pi n}{N} \cdot \eta^{n}, \\
\frac{T_{C}(E)}{\hbar} &=& \frac{i (N+2) g^{1/N}}{2 N E}  \sum_{n \in 2{\mathbb N}_0-1} n v_{n} \left( e^{-i \frac{\pi n}{N}} + 1 \right) \cdot \eta^{n}, \\
T_{L}(E)  &=& T_{C}(E) - \frac{P}{1+P} T_{\rm P}(E),
\ee
where $T_{C} (E) := i \hbar \pd_E \log  C_{(\bar{p},\bar{p}+1)}$, and the coefficients, $v_{n \in 2{\mathbb N}_0-1}$, are defined in Eq.(\ref{eq:cof_v}).
Notice that $\eta \propto \hbar$ as is shown in Eq.(\ref{eq:def_eta_E}).
One can easily see that $T_{\rm P}(E)$ is real, but $T_{C/L}(E)$ are complex values.
It is because the period is defined to be real if the associated cycle is a pure oscillation.
Hence, the non-perturbative cycle with a damping factor generates a complex period~\cite{Sueishi:2020rug}.

It is notable that Eqs.(\ref{eq:resol_Gp_N3K1})(\ref{eq:resol_Gnp_N3K1}) do not contain cycles with negative oscillations, i.e., ${\rm Im}[\log P]>0$ and ${\rm Im}[\log C_{(\bar{p},\bar{p}+1)}]>0$, but the GTF including negative oscillations is available by using complex conjugation of the exact QC, $\bar{\frak D} = {\cal K}[{\frak D}]$.
Since ${\frak D} \propto \bar{\frak D}$, replacing ${\frak D}$ with $({\frak D} \bar{\frak D})^{1/2}$ in Eq.(\ref{eq:G_pdE_logD}) can generate all p.p.o.s with positive and negative oscillations.

\begin{figure}[tbp]
  \begin{center}
    \begin{tabular}{cc}
      \begin{minipage}{0.5\hsize}
        \begin{center}
          \includegraphics[clip, width=75mm]{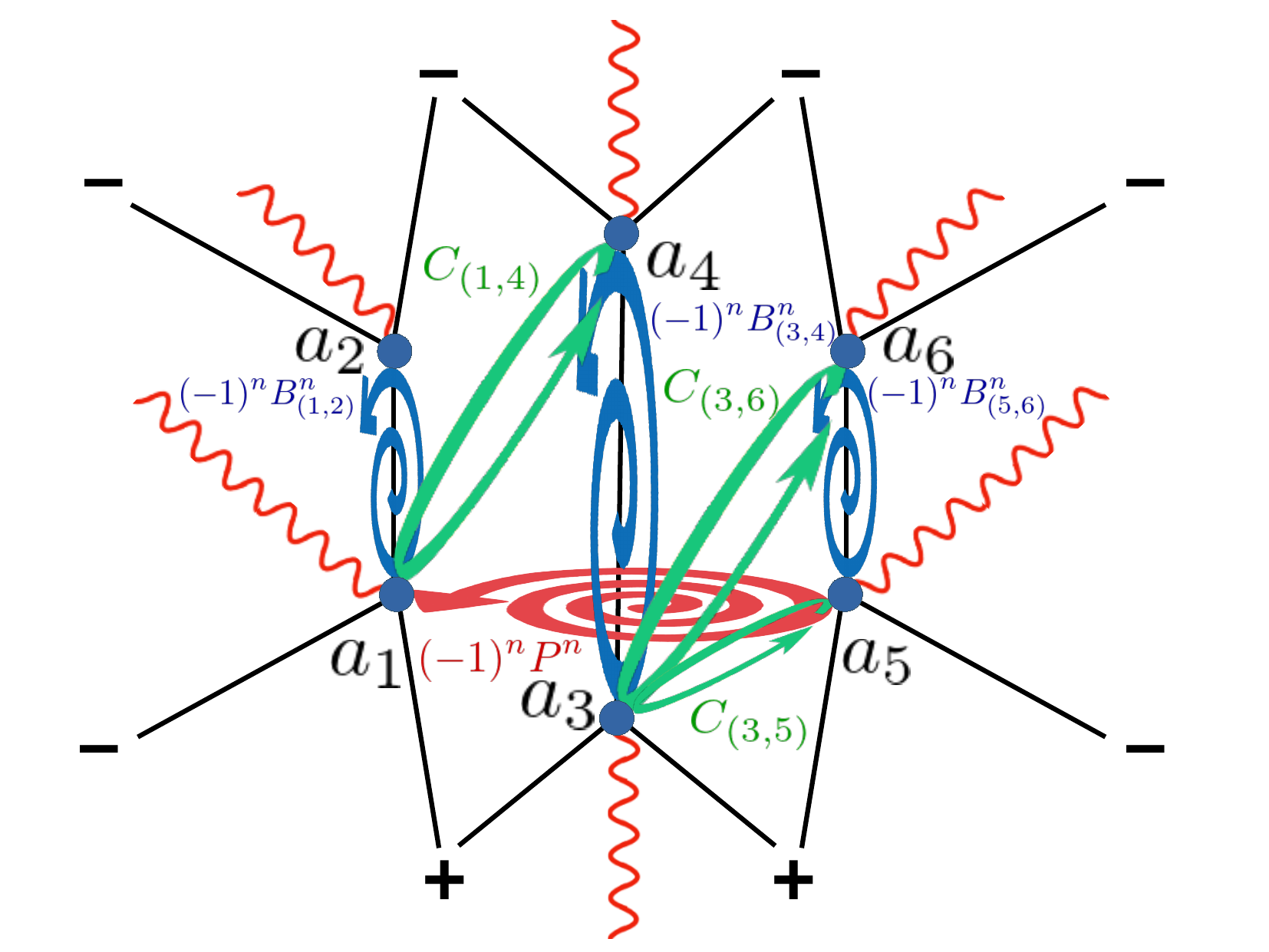}
          \hspace{1.6cm} (a) $(N,K) = (6,2)$
        \end{center}
      \end{minipage}
      \begin{minipage}{0.5\hsize}
        \begin{center}
          \includegraphics[clip, width=75mm]{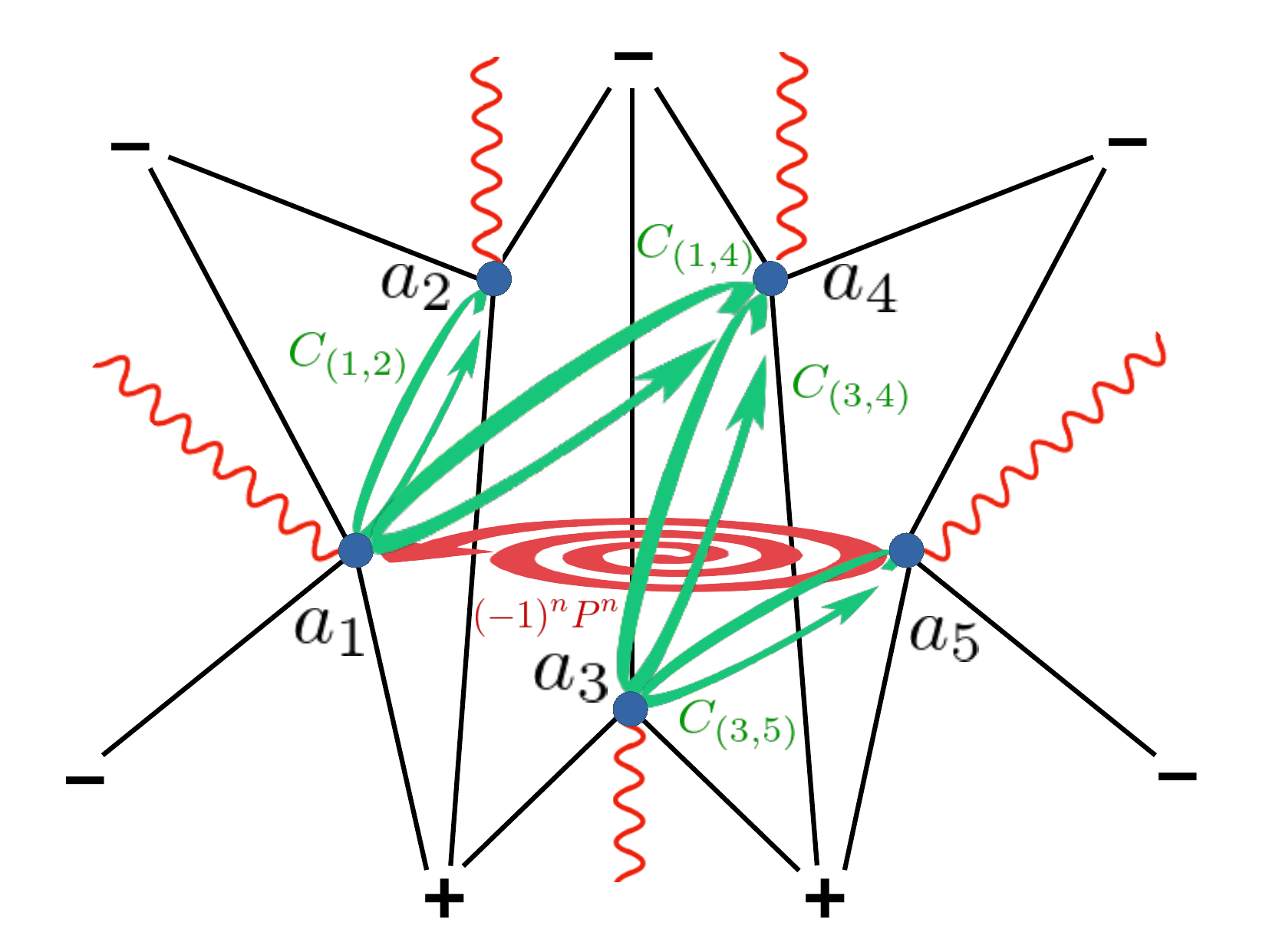}
          \hspace{1.6cm} (b) $(N,K) = (5,2)$
        \end{center}
      \end{minipage} 
    \end{tabular} 
    \caption{Schematic figure of non-perturbative p.p.o.s for $(N,K) = (6,2)$ and $(N,K) = (5,2)$.
      A non-perturbative p.p.o. can be generally constructed by certain combinations of cycles in its QC.
    }
    \label{fig:ppo_Khigh}
  \end{center}
\end{figure}

One can also formulate the generalization of the GTF for  $K > 1$ in the similar way.
Since the exact QCs can be generally expressed as
\be
{\frak D} \propto 1 + P + \delta {\frak D}, \label{eq:D_1_P_dD}
\ee
where $\delta {\frak D}$ is the non-perturbative part, substituting Eq.(\ref{eq:D_1_P_dD}) into Eq.(\ref{eq:G_pdE_logD}) leads to the GTF for $K>1$.
As a result, one can find the same forms to Eqs.(\ref{eq:resol_Gp_N3K1})(\ref{eq:resol_Gnp_N3K1}), but the non-perturbative p.p.o. in Eq.(\ref{eq:Gutz_L_N3K1}) is modified as
\be
L = \frac{\delta {\frak D}}{1+P}.
\ee
For example, from Eqs.(\ref{eq:QC_N6K2})(\ref{eq:QC_N5K2}),
the specific forms of $L$ for $(N,K) = (6,2)$ and $(5,2)$ can be expressed as
\be
L_{(6,2)}
&=& \sum_{n,m_1,m_2 \in {\mathbb N}_0} \left[ \prod_{j=1}^2 \frac{(2 m_j)!}{(2^{m_j} m_j !)^2} \right] \cdot \left[ C_{(1,4)}  (-1)^{n+m_1+m_2} B_{(1,2)}^{m_1} B_{(3,4)}^{m_2}  P^n \right. \nl
  && \left.   + \left( C_{(3,5)} + C_{(3,6)} \right) (-1)^{n+m_1+m_2} B^{m_1}_{(3,4)} B^{m_2}_{(5,6)} P^n \right], \label{eq:L_N6K2} \\ 
L_{(5,2)} &=& \sum_{n \in {\mathbb N}_0}\left[ C_{(1,2)}  + C_{(1,4)}  + C_{(3,4)} + C_{(3,5)} + C_{(1,2)} C_{(3,4)} + C_{(1,2)} C_{(3,5)} \right] (-1)^n P^n. \label{eq:L_N5K2} 
\ee
The schematic figure of their non-perturbative p.p.o.s are shown in Fig.~\ref{fig:ppo_Khigh}.

It is notable that the main difference between the cases of odd and even $N$ is non-perturbative contributions from $B$-cycles constituting ${\bf C}_{{\rm NP}, \theta=0}$, which is a consequence of the  ${\mathbb Z}_2$ symmetry in Eq.(\ref{eq:Z2_symm}).

\subsection{Spectral summation form}
\label{sec:spec_sum}
The spectral summation form (SSF) can be easily derived by replacing the Fredholm determinant in Eqs.(\ref{eq:ZbetaG})(\ref{eq:G_pdE_logD}) with the exact QC.
For construction of non-Hermitian QMs, there generally exists some issues in Hilbert spaces, such as the inner-product and the unitarity condition.
In ${\cal PT}$ symmetric QMs, while the unitarity condition holds, the ${\cal PT}$ symmetric inner-product is nevertheless indefinite.
There is, however, another inner-product compatible with a ${\cal PT}$ symmetric Hamiltonian and holding both positive definiteness and the unitarity condition.
This is called as \textit{${\cal CPT}$ inner-product}.
We summarize the construction in App.~\ref{sec:pseudo_Herm}.
See Refs.~\cite{Mostafazadeh:2001jk,Mostafazadeh:2002pd,Mostafazadeh:2003gz,Weigert:2003py,Bender:2002vv,Bender:2004zz} and references therein in more detail.

We denote the ${\cal CPT}$ inner-product as $\lcurvyangle \psi | \phi \rcurvyangle$, which is defined in Eq.(\ref{eq:CPT_inner_prod}).
The ${\cal PT}$ symmetric Hamiltonian, $\widehat{H}_{\cal PT}$, satisfies 
\be 
\lcurvyangle E_{k_1} | \widehat{H}_{\cal PT} | E_{k_2} \rcurvyangle = \lcurvyangle E_{k_1} | \widehat{H}^\dagger_{\cal PT} | E_{k_2} \rcurvyangle = E_{k_1} \delta_{k_1,k_2}. \qquad (k_{1},k_2 \in {\mathbb N}_0)
\ee
We also define trace over the Hilbert space using the ${\cal CPT}$ inner-product as $\Tr_{\cal CPT} [\widehat{A}] := \sum_{k \in {\mathbb N}_0} 
\lcurvyangle E_k | \widehat{A} | E_k \rcurvyangle$, where $\widehat{A}$ is a ${\cal P}$ (and $\chi$)-pseudo-Hermitian operator satisfying $\widehat{A} = {\cal P} \widehat{A}^\dagger {\cal P} = \chi \widehat{A}^\dagger \chi^{-1}$ with $\chi := {\cal PC}$.
By the ${\cal CPT}$ inner-product, Eqs.(\ref{eq:resolvent})(\ref{eq:ZbetaG}) are modified as
\be
&& G(E) = \Tr_{\cal CPT} \left[ \frac{1}{\widehat{H}_{\cal PT} - E} \right], \qquad {\cal Z}(\beta) := \Tr_{\cal CPT} \, \left[ e^{- \beta \widehat{H}_{\cal PT}} \right] = \sum_{k \in {\mathbb N}_0} e^{- \beta E_k}, \label{eq:Z_E_sum_PT}
\ee
and thus, the resulting forms are essentially the same to the Hermitian cases.

Let us reproduce the SSF in Eq.(\ref{eq:Z_E_sum_PT}) from Eqs.(\ref{eq:resolvent})(\ref{eq:ZbetaG}) using our exact QCs.
For simplicity, we firstly suppose that the QCs include only the perturbative cycle, i.e., ${\frak D} \propto 1+P$.
By taking $P(E) = e^{i a_{\rm P}(E)}$ with $a_{\rm P}(E) \in {\mathbb R}_{>0}$, Eqs.(\ref{eq:ZbetaG})(\ref{eq:G_pdE_logD}) lead to
\be
{\cal Z}(\beta) &=& - \frac{1}{2 \pi i} \int^{\delta + i \infty}_{\delta - i \infty} \pd_E \log (1 + P) e^{- \beta E} dE \nl
&=& - \frac{1}{2 \pi} \oint^{\infty+i 0_+}_{\infty+0_-} \frac{P \pd_E a_{\rm P}}{1+P} e^{- \beta E} dE \, = \, - \frac{1}{2 \pi} \oint^{\infty + i 0_+}_{\infty+ i 0_-}
\frac{e^{- \beta E(a_{\rm P})}}{1+e^{-i a_{\rm P}}} da_{\rm P}, \label{eq:Zb_pt}
\ee
where $\oint^{\infty + i 0_+}_{\infty+ i 0_-} d E$ is the Hankel contour going around $E = \delta$ with $0<\delta \ll 1$.
This has simple poles at $a_{\rm P} = 2 \kappa = \pi (2 k + 1)$ with $k \in {\mathbb N}_0$, that is the same condition to Eq.(\ref{eq:D0_even_K1_2}).
Therefore, it can be evaluated by the residue integration, and $E(a_{\rm P})$ has to be the energy solution of the exact QC.
This means that Eq.(\ref{eq:Zb_pt}) is identical to the perturbative part of the SSF in Eq.(\ref{eq:Z_E_sum_PT}).

One can also obtain the generalization including non-perturbative sectors by replacing $P$ with $P + \delta {\frak D}$ in Eq.(\ref{eq:D_1_P_dD}), where $\delta {\frak D}$ denotes all non-perturbative parts in the exact QC.
By defining
\be
R(E):= e^{i \widetilde{a}(E)}, \qquad \widetilde{a}(E):=  - i \log (P(E) + \delta {\frak D}(E)), \label{eq:R_atil} 
\ee
Eq.(\ref{eq:Zb_pt}) is generalized by replacing $a_{\rm P}$ with $\widetilde{a}$, i.e.
\be
{\cal Z}(\beta) &=& - \frac{1}{2 \pi i} \int^{\delta + i \infty}_{\delta - i \infty} \pd_E \log (1 + R) e^{- \beta E} dE \nl
&=& - \frac{1}{2 \pi} \oint^{\infty+i 0_+}_{\infty+0_-} \frac{R \pd_E \widetilde{a}}{1+R} e^{- \beta E} dE \, = \, - \frac{1}{2 \pi} \oint^{\infty + i 0_+}_{\infty+ i 0_-}
\frac{e^{- \beta E(\widetilde{a})}}{1+e^{-i \widetilde{a}}} d\widetilde{a}, \label{eq:Zb_np}
\ee
where $E(\widetilde{a})$ is the energy solution of the exact QC depending on the energy level, $\widetilde{a} = 2 \kappa = \pi(2k+1)$ with $k \in {\mathbb N}_0$.

\subsection{Euclidean path-integral} \label{sec:path_int}
We formulate the Euclidean path-integral (EPI) under the periodic boundary condition that $x(0) = x(\tau) =: x_{\tau}$ using the exact QCs\footnote{The Minkowskian path-integral can be also formulated in the similar way.}.
In the similar way to the Hermitian cases, the EPI is defined by introducing the complete set, ${\mathbb I} = \int_{\gamma_{\cal PT}} dx_{\tau} | x_{\tau} \rangle \langle x_{\tau} |$ with $\langle x_\tau | := | x_\tau \rangle^\dagger$, as\footnote{
The states $| x \rangle$ and $\langle x |$ are consistent with the inner-product with the ${\cal CPT}$ states, $| E_k \rcurvyangle$ and $ \lcurvyangle E_k |$, as
\be
\langle x | E_k \rcurvyangle &=& \langle x | E_k \rangle = \phi_k(x), \\
\lcurvyangle E_k | x \rangle &=& \langle E_k | {\cal PC}| x \rangle = \overline{\langle x | {\cal CP}| E_k \rangle} =   {\cal CPT} [\phi_k(x)] =  \zeta_k \overline{\phi_k(-x)}, \qquad (\zeta_k^2 = 1)
\ee
where $\phi_k(x) = \overline{\phi_k(-x)}$ is the ${\cal PT}$ symmetric energy eigenfunction with $\int_{\gamma_{\cal PT}} dx \,\phi_{k_1}(x) \phi_{k_2}(x) = \zeta_{k_1} \delta_{k_1,k_2}$.
}
\be
Z(\beta = \tau/\hbar) &:=& \int_{\gamma_{\cal PT}} dx_{\tau} \, \langle x_\tau | e^{- \beta \widehat{H}_{\cal PT}}  | x_{\tau} \rangle, 
\label{eq:Z_beta}
\ee
where $\gamma_{\cal PT}$ is the domain of $x$ given by Eq.(\ref{eq:gam_PT}).
Thanks to the ${\cal CPT}$ inner-product, one can construct its familiar form expressed by the Euclidean action, $S_{\cal PT} = \int_0^\tau d t \, L_{\cal PT}$, with the Lagrangian, $L_{\cal PT}$ in the standard way.
Defining the complete set of momentum, ${\mathbb I} = \int \frac{dp}{2 \pi \hbar} |p \rangle \langle p|$ with $\langle p|x \rangle = e^{i p  x/\hbar}$, and using the Legendre transform, one can obtain
\be
Z(\beta) &=& \int_{\gamma_{\cal PT}} dx_{\tau} \, \langle x_{\tau}| e^{-\beta \widehat{H}_{\cal PT}} |x_{\tau} \rangle =  \int_{\gamma_{\cal PT}} {\cal D} x \int {\cal D} p\, e^{  \int^{\beta}_{0} dt \, \left[ i \frac{p}{\hbar} \frac{dq}{d t} - {H}_{\cal PT} \right]} \nl
&=& {\cal N} \int_{\gamma_{\cal PT}} {\cal D} x \,  e^{ - \frac{1}{\hbar} S_{\cal PT}}, \qquad {\cal D} x :=  \prod_{t \in [0,\tau)} dx (t), \ \  {\cal D} p := \prod_{t \in [0,\tau)} \frac{dp (t)}{2 \pi},
\ee
where ${\cal N}$ is a normalization factor.

As is well-known in the Hermitian cases, the EPI is identical to the SSF in Eq.(\ref{eq:Z_E_sum_PT}), i.e., ${Z}(\beta) = {\cal Z}(\beta)$.
The same argument works for the ${\cal PT}$ symmetric Hamiltonian, due to the ${\cal CPT}$ inner-product.
Hence, the simplest way to find a transseries of the EPI is expanding the energy solution in the SSF.
Since the energy for the massless cases is a monomial of $\hbar^{\frac{2N}{N+2}}$, one can write down its explicit form as
\be
Z(\beta) &=& \sum_{k \in {\mathbb N}_0} e^{-\beta E_k} 
= \sum_{k \in {\mathbb N}_0} \sum_{n \in {\mathbb N}_0} \frac{(- \tau e(k))^n}{n!} \left[ g^2 \hbar^{N-2} \right]^{\frac{n}{N+2}}, \label{eq:pathint_spec}
\ee
where the energy is given by Eq.(\ref{eq:Ek_c}) with a transseries $e(k)$ of the energy level, $k$.
Notice that $e(k)$ in Eq.(\ref{eq:pathint_spec}) is a divergent series of $k^{-1}$ or $\kappa^{-1} = [\pi (k + \frac{1}{2})]^{-1}$.
Using the property of the Stokes automorphism which is a homomorphism, the Borel resummed form is formally obtained by replacing $e(k)$ in $E_k$ with $\widehat{e}(k)$, where $\widehat{f} = {\cal S}_{{\rm med},0}[f]$.

As is shown in Fig.~\ref{fig:flowchart}, the same result can be derived by the integration-by-part as
\be
Z(\beta) &=& - \frac{\beta}{2 \pi i} \int^{\delta + i \infty}_{\delta - i \infty} \log {\frak D} \cdot e^{-\beta E} dE 
= - \frac{\beta}{2 \pi i} \int^{\delta + i \infty}_{\delta - i \infty} \log \left[ 2 \cos \frac{\widetilde{a}(E)}{2} \right]  e^{-\beta E} dE \nl
&=&  \frac{\beta}{2 \pi i} \int^{\delta + i \infty}_{\delta - i \infty} \log \left[ \Gamma \left(\frac{1}{2} - \frac{\widetilde{a}(E)}{2\pi} \right)    \right] e^{-\beta E} dE,
\ee
where $\widetilde{a}(E)$ is defined in Eq.(\ref{eq:R_atil}).
Singularities appear from the gamma function at $\widetilde{a}(E) = 2 \kappa = \pi (2k + 1)$ with $k \in {\mathbb N}_0$  and consequently leads to the same form to the SSF in Eq.(\ref{eq:Zb_np}).

\section{The massive cases} \label{sec:mass_cases}
We briefly describe the massive cases defined by the potential in Eq.(\ref{eq:V_PT}) with $\omega > 0$.
In contrast to the massless cases, the perturbative expansion using $\hbar$ naively works, and the dependence of the energy level appears as a polynomial in each the coefficients.
Application of EWKB to the $(N,K) = (4,1)$ case was investigated in Ref.~\cite{Kamata:2023opn}, and its exact QC has the form that
\be
   {\frak D}^{0} \propto 1 + {\frak A} \sqrt{\frac{1+{\frak B}_1}{1+{\frak B}_2}} =  1 + {\frak A}, \qquad {\frak B}_1 = {\frak B}_2, \label{eq:QC_mass}
\ee
where ${\frak A}$ is a perturbative cycle and is given by a residue integration of $S_{\rm od}$ around $x = 0$.
The double turning point at $x=0$ is connected to two single turning points by Stokes lines which corresponds to ${\frak B}_{1,2}$, but those contributions canceled to each other due to the ${\mathbb Z}_2$ symmetry in Eq.(\ref{eq:Z2_symm}).
Hence, the energy spectrum contains the perturbative sector only.
When considering the potential in Eq.(\ref{eq:V_PT}) with $\omega > 0$ for an arbitrary $N$, the above situation is unchanged due to the following fact:
\begin{fact*}[Uniqueness of $\gamma_{\cal PT}$ and Borel summability of energy spectra]
Consider the potential in Eq.(\ref{eq:V_PT}) with $\omega > 0$, and suppose that the path of analytic continuation $\gamma_{\cal PT}$ is defined by Eq.(\ref{eq:gam_PT}).
Then, for any $N \in {\mathbb N}+2$, the path providing a solution of the exact QC is uniquely determined as the nearest path to the real axis, i.e., $K=\lfloor (N-1)/2 \rfloor$ ($\varepsilon = 1$ and $2$ for odd and even N, respectively).
The resulting exact QC contains only the perturbative cycle around $x=0$, ${\frak A}$, as ${\frak D} = 1 + {\frak A}$.
The ${\frak A}$-cycle is Borel non-summable and summable for even and odd $N$, respectively.
Borel summability of the energy spectrum is also the same.
\end{fact*} \noindent
The derivation is summarized in App.~\ref{sec:mass_borel}.
As a result, the resulting energy spectrum, the GTF, the SSF, and the EPI are all purely perturbative.

\begin{figure}[t]
  \begin{center}
    \begin{tabular}{cc}
      \begin{minipage}{0.5\hsize}
        \begin{center}
          \includegraphics[clip, width=75mm]{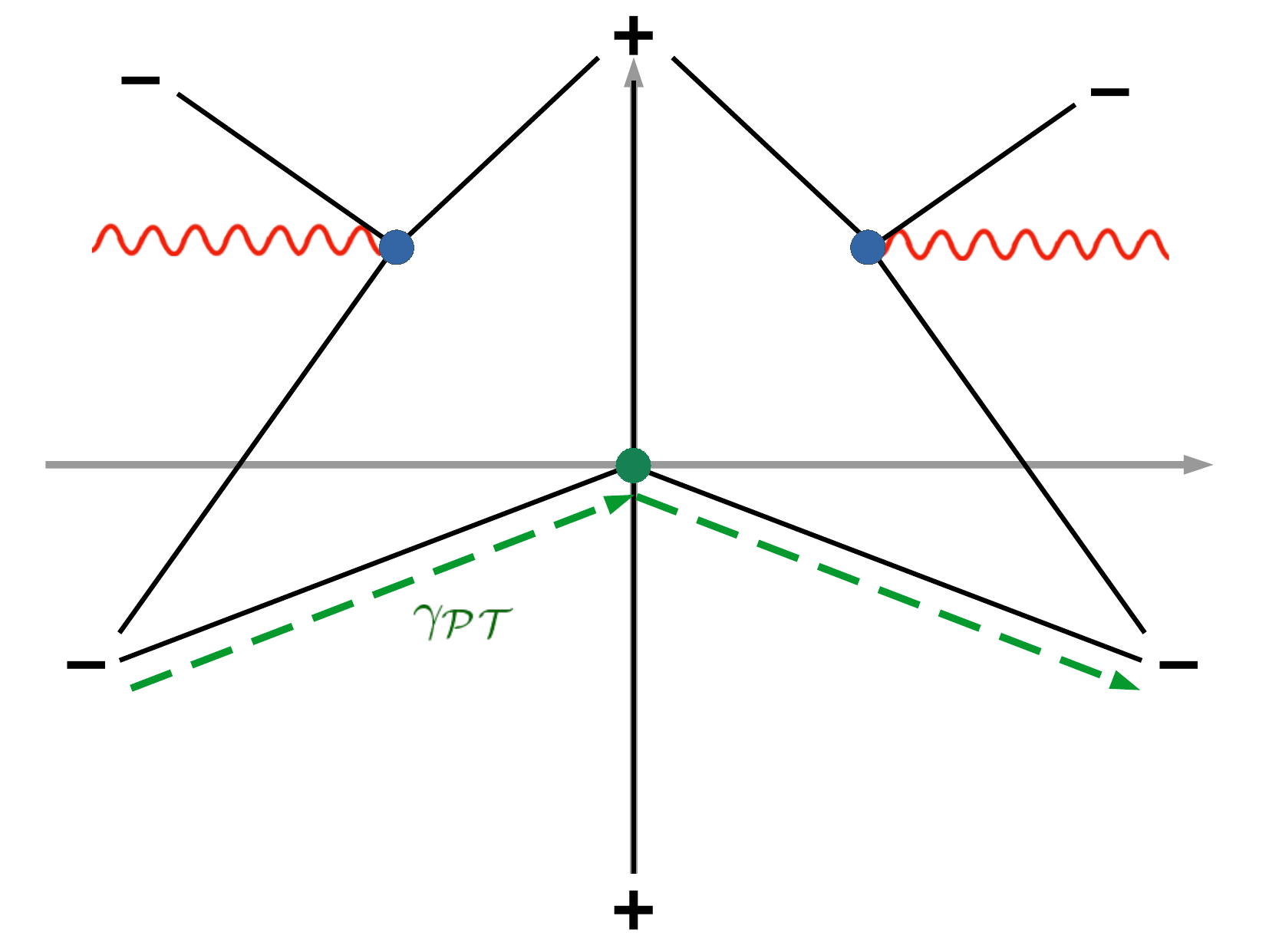}
          \hspace{1.6cm} (a) $\delta V_{\cal PT} =  g_3 x^2 (ix) + g_4 x^2 (ix)^2$
        \end{center}
      \end{minipage}
      \begin{minipage}{0.5\hsize}
        \begin{center}
          \includegraphics[clip, width=75mm]{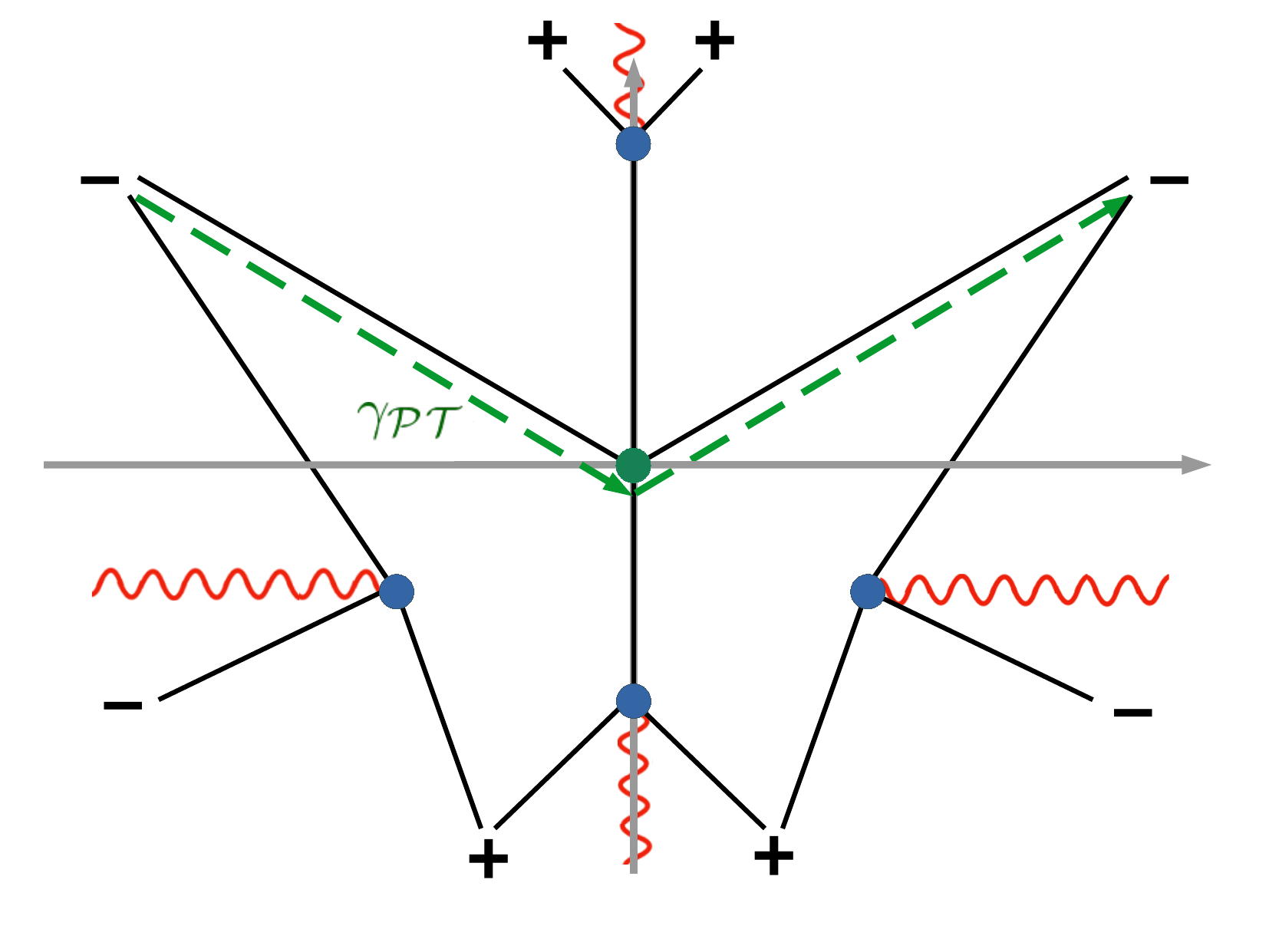} 
          \hspace{1.6cm} (b) $\delta V_{\cal PT} = g_3 (ix)^3 + g_4 x^2 (ix)^2 + g_6 x^4 (ix)^2$
        \end{center}
      \end{minipage} \\ \\
            \begin{minipage}{0.5\hsize}
        \begin{center}
          \includegraphics[clip, width=75mm]{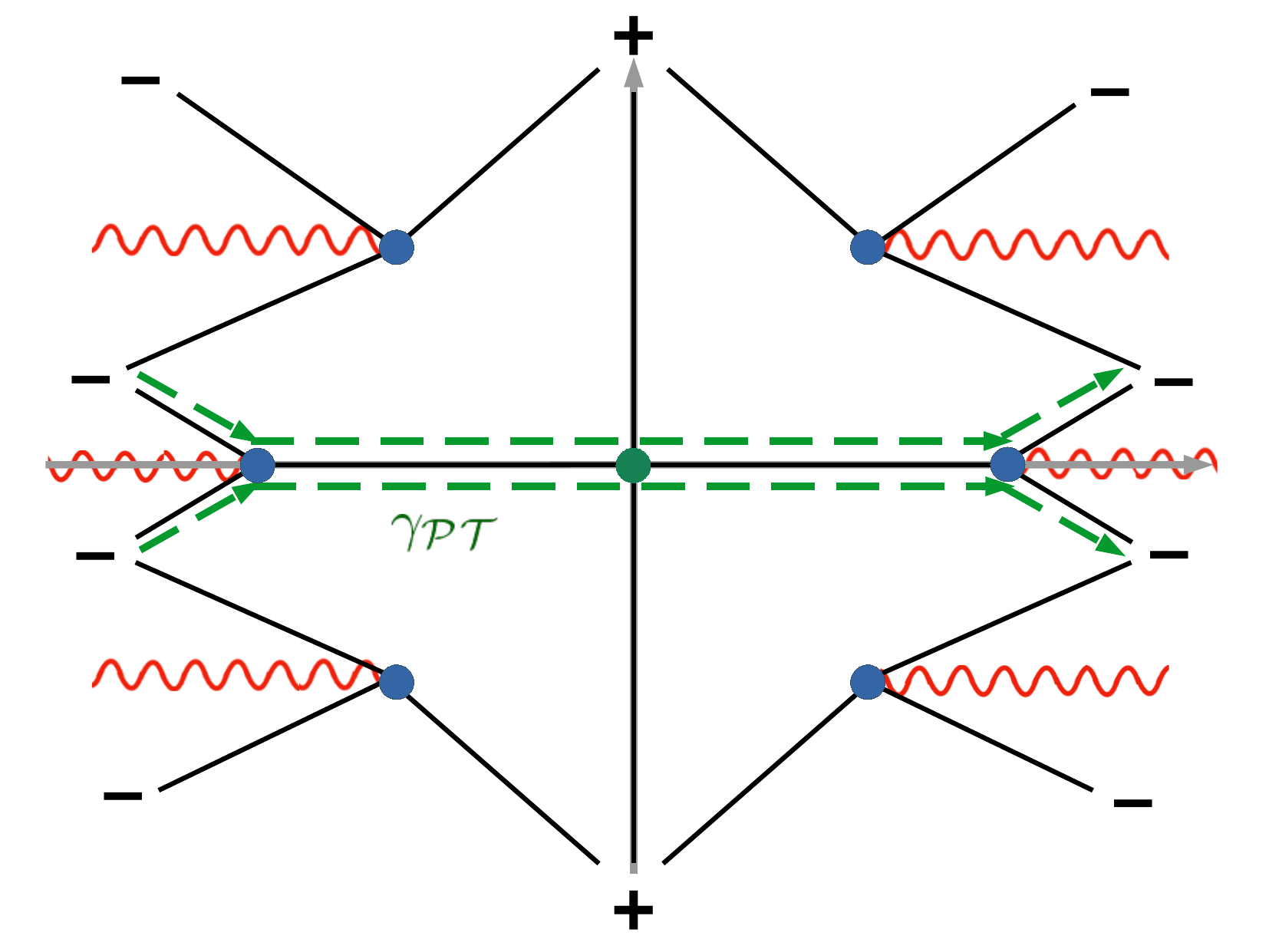} 
          \hspace{1.6cm} (c) $\delta V_{\cal PT} = g_4 x^2 (ix)^2 + g_6 x^6 + g_8 x^6 (ix)^2$
        \end{center}
      \end{minipage} 
    \end{tabular} 
    \caption{
Examples of the Stokes graph for the massive cases with a polynomial potential, $V_{\cal PT} = \omega^2 x^2 + \delta V_{\cal PT}$ for $ \omega, g_i \in {\mathbb R}_{>0}$ and the zero-classical energy, $E_0 = 0$, with $\arg(\hbar)=0$.
The green dot at the origin and the blue ones denote the double and the simple turning points, respectively.
In order to find a quantized energy spectrum by $\hbar$, the path of analytic continuation, $\gamma_{\cal PT}$, has to be taken as the green dashed line.
In (c), there exist two suitable path as a complex conjugation pair giving the same result because of the ${\mathbb Z}_2$ symmetry (\ref{eq:Z2_symm}) in $V_{\cal PT}$.
    }
    \label{fig:massive_Stokes_ex}
  \end{center}
\end{figure}

We would emphasize that the same statement to the above {\bf Fact} holds for a wide class of classical ${\cal PT}$ symmetric polynomial potentials with a single quadratic vacuum.
Here, let us remove choice of the real axis for a path of analytic continuation even when the wavefunction is normalizable at $x = \pm \infty$, such as (E-2) and (E-3) of Fig.~\ref{fig:Stokes_shape_even}, because it is a Hermitian QM.
In such a case, once a Stokes graph is drawn, a suitable path of analytic continuation is automatically determined to give a quantized energy by $\hbar$ without introducing $\varepsilon$ as a deformation parameter from the real axis.
The uniqueness of $\gamma_{\cal PT}$ is broken when the set of turning points is invariant under complex conjugation, which arises from the ${\mathbb Z}_2$ symmetry (\ref{eq:Z2_symm}) in the potential.
However, the reasonable paths are given as a complex conjugate pair, and either choice from the pair gives the same result\footnote{
  In this sense, the uniqueness of $\gamma_{\cal PT}$ in {\bf Fact} for even $N$ is guaranteed by positivity of $\varepsilon$.
  When $\varepsilon \in {\mathbb Z}$ for $N \in 2 {\mathbb N}+2$, those paths of analytic continuation are given by $(N,K) = (N, \frac{N}{2}-1)$ and $(N,K) = (N,\frac{N}{2}+1)$ which correspond to $\varepsilon = +2$ and  $\varepsilon = -2$, respectively, and these give the same exact QC.
}.
Examples of the Stokes graphs given by polynomial potentials with a mass term are depicted in Fig.~\ref{fig:massive_Stokes_ex}.
The energy spectrum is Borel non-summable only if a Stokes phenomenon occurs by degeneracies of Stokes lines, which are parallelly flowing to the real axis, from the double turning point to simple turning points.
See discussions in App.~\ref{sec:mass_borel}.

In summary, this ${\bf Fact}$ tells us a nice observation;
even for more generic polynomial potentials, exact QCs for the massive ${\cal PT}$ symmetric QMs with a single quadratic vacuum have the same form to Eq.(\ref{eq:QC_mass}) and can be calculated by performing \textit{only} the residue integration of $S_{\rm od}(x, \widetilde{E}; \hbar)$ around the vacuum in Eq.(\ref{eq:F_Sod}).
These results lead to the energy spectra and the three formulas in Fig.~\ref{fig:flowchart} as simple forms without non-perturbative sectors.

\section{Additional remarks} \label{sec:comments}
In this section, we make some additional remarks related to our analysis.
We briefly discuss similarities to the Hermitian cases in Sec.~\ref{sec:CPT_cases}, and then comment on resurgence in Sec.~\ref{sec:resurgence_analysis}.

\subsection{Similarities to Hermitian QMs} \label{sec:CPT_cases}
We discuss similarities of transseries structure to Hermitian QMs.
Here, we consider the Hermitian potential defined by
\be
V_{\cal H}(x) := \omega^2 x^2 + \lambda x^{N}, \qquad \omega \in {\mathbb R}_{\ge 0}, \  \lambda \in {\mathbb R}_{>0}, \  N \in 2{\mathbb N}+2, \label{eq:V_herm}
\ee
and take $\omega = 0$ for a while.
Stokes graphs of the Hermitian potential for $N \in 4{\mathbb N}+2$ and $N \in 4{\mathbb N}$ are the same to (E-2) or (E-3) in Fig.~\ref{fig:Stokes_shape_even}, respectively, and a path of analytic continuation is taken along a line slightly below the real axis.
As we can see below, transseries structure of the Hermitian QM is quite similar to the ${\cal PT}$ symmetric QM.

In this part, we only consider exact QCs of the Hermitian QM because the procedure for each the formula is parallel to analyses in the above sections.
In the Hermitian cases, turning points consisting of a perturbative cycle are $a_1$ and $a_N$, and the cycle $P$ can be evaluated as
\be
P &=& \exp \left[ 2  i \sum_{n \in 2 {\mathbb N}_0 -1} (-1)^{\frac{n-1}{2}} v_{n} \cdot \eta^{n} \right], \label{eq:C_pp_Herm}
\ee
where the coefficients, $v_{n \in 2 {\mathbb N}_0-1}$, are given in Eq.(\ref{eq:cof_v}).
Since the Stokes graphs (E-2)(E-3) in Fig.~\ref{fig:Stokes_shape_even} have a Stokes phenomenon at $\arg(\eta)=0$, one has to treat discontinuities in the QCs.
In the similar way to Sec.~\ref{sec:quant_cond}, the exact QCs take the forms that
\be
&& {\frak D}^{0} \propto  \frac{1}{P^{1/2}} \sum_{(n_1,\cdots, n_{N/2}) \in \{0,1\}^{N/2}}  \left[ \prod_{\ell=1}^{N/2-1}  {\frak D}^{(\ell)}_{n_\ell, n_{\ell + 1}}  \right] {\frak D}^{(N/2)}_{n_{N/2}}, \nl
&& {\frak D}^{(1)}_{n_1,n_{2}} := \widetilde{C}_{(1,3)} ^{n_1} \left( \delta_{n_{2},0} + \widetilde{B}^{-1}_{(3,2)} \cdot \delta_{n_{2},1} \right)^{n_1}, \nl
&& {\frak D}^{(\ell \in \{2,3,\cdots, N/2-1\})}_{n_\ell,n_{\ell + 1}} := \widetilde{C}_{(2\ell-2,2\ell+1)}^{n_\ell} \left( \delta_{n_{\ell+1},0} + \widetilde{B}^{-1}_{(2\ell+1,2\ell)} \cdot \delta_{n_{\ell+1},1} \right)^{n_\ell}, \nl
&& {\frak D}^{(N/2)}_{n_{N/2}}:= \widetilde{C}_{(N-2,N)}^{n_{N/2}}, \label{eq:D0_evenN_Herm} 
\ee
where 
\be
&& \widetilde{C}_{(1,3)} :=  \frac{C_{(1,3)}}{\sqrt{1 + B_{(2,3)}}}, \nl
&& \widetilde{C}_{(2 \ell-2,2 \ell+1)} := \frac{C_{(2 \ell-2,2 \ell+1)}}{\prod_{n=0}^1 \sqrt{1 + B_{(2 \ell+2n -2,2 \ell+2n-1)}}}, \qquad (\ell \in \{2,3,\cdots, N/2 \}) \nl
&& \widetilde{C}_{(N-2,N)} :=  \frac{C_{(N-2,N)}}{\sqrt{1 + B_{(N-2,N-1)}}}, \nl
&&  \widetilde{B}^{-1}_{(2 \ell + 1,2 \ell) } := \frac{1+B_{(2 \ell,2 \ell+1)}}{B_{(2 \ell,2 \ell+1)}} \quad \in {\mathbb R}_{>0}.  \label{eq:C_evenN_Herm} 
\ee
For $N=4$ and $6$, those can be explicitly written down as
\be
   {\frak D}^0_{N=4} &\propto& \frac{1}{P^{1/2}} \left[ 1 + P + \frac{C_{(1,3)} + C_{(2,4)}}{\sqrt{1+B_{(2,3)}}}
     \right], \\
   {\frak D}^0_{N=6} &\propto& \frac{1}{P^{1/2}} \left[ 1 + P  +\frac{C_{(1,3)} + C_{(2,6)}}{\sqrt{1+{B_{(2,3)}}}} +\frac{C_{(1,5)} + C_{(4,6)}}{\sqrt{1+{B_{(4,5)}}}}  +\frac{C_{(2,5)} + C_{(1,3)} C_{(4,6)}}{\sqrt{1+{B_{(2,3)}}} \sqrt{1+{B_{(4,5)}}}} \right].
\ee

As one can see from Eq.(\ref{eq:D0_evenN_Kgen}), the Hermitian QCs have a similar feature to the ${\cal PT}$ symmetric QCs for even $N$.
Although details of the energy spectra of the Hermitian QM such as the specific values and the number of non-perturbative sectors differ from the ${\cal PT}$ symmetric QM with the same $N$, the fundamental feature is almost the same because the difference comes only from the paths of analytic continuation on the same Stokes graphs.
In addition, the ${\cal PT}$ symmetry, ${\cal PT}: x \rightarrow - \bar{x}$,  constrains locations of turning points, and the real part of both turning points and paths of analytic continuation are always ${\mathbb Z}_2$ symmetric.
This is also the same in the Hermitian cases and the crucial reason to give the similar feature in the transseries.

As we discussed in Sec.~\ref{sec:mass_cases}, the similarity also holds in the massive cases, i.e., $\omega > 0$, but there is generally a difference in Borel summability from the Hermitian QM.
In other words, although both the exact QCs are purely perturbative, the exact QCs of the Hermitian QM defined by Eq.(\ref{eq:V_herm}) are always Borel summable, but it is not always true for the ${\cal PT}$ symmetric QM defined by Eq.(\ref{eq:V_PT}).

\subsection{Resurgence} \label{sec:resurgence_analysis}
We make some comments on resurgence.
Construction of resurgent relations of the energy spectra for the massless cases is possible by alien calculus, but it does not mean that these resurgent relations perfectly reproduce each the non-perturbative sector.
Indeed, even the first alien derivative does not generate information of all non-perturbative cycles, $B_{(n_1,n_2)}$ and $C_{(n_1,n_2)}$, in Eqs.(\ref{eq:D0_evenN_Kgen})(\ref{eq:D0_oddN_Kgen}).
We summarize the details in App.~\ref{sec:Alien_energy}.

This reason originates from the structure of the Stokes graphs.
In order to extract all the non-perturbative information from the perturbative cycle at once by the alien derivatives at a certain complex phase, $\theta = \arg(\eta)$, all the non-perturbative cycles in the exact QCs need to simultaneously have degeneracies of Stokes lines and intersections with the perturbative cycle.
However, this situation can not be realized by any $\theta$, and only some of the non-perturbative cycles can have them at a certain $\theta$.
In this sense, the resurgent relations can only extract \textit{partial} non-perturbative information from the perturbative part for each $\theta$ causing a Stokes phenomenon.
This situation is also unchanged for the Hermitian cases discussed in Sec.~\ref{sec:CPT_cases}.

In contrast, the situation in the massive cases completely differs from the massless cases.
As we described in Sec.~\ref{sec:mass_cases}, the exact QCs contain only a perturbative cycle even if a Stokes phenomenon happens at $\theta = 0$.
This is a consequence of the fact that Borel non-summability is in general irrelevant to existence of non-perturbative contributions in the exact QCs, i.e., it only concludes performability of Borel resummation due to Borel singularities.
In the Hermitian cases with a single quadratic vacuum, the exact QCs are always not only purely perturbative but also Borel summable.

\section{Summary and conclusion}  \label{sec:summary}
In this paper, we have studied exact WKB analysis (EWKB) for a ${\cal PT}$ symmetric quantum mechanics (QM) defined by the potential that $V_{\cal PT}(x) = \omega^2 x^2 + g x^{2 K} (i x)^{\varepsilon}$ with $\omega \in {\mathbb R}_{\ge 0}$, $g \in {\mathbb R}_{>0}$ and $K, \varepsilon \in {\mathbb N}$ to clarify its perturbative/non-perturbative structure.
In our analysis, we have mainly considered the massless cases, i.e., $\omega = 0$, and obtained:
\begin{enumerate}
\item[I)] the exact quantization conditions (QCs) for arbitrary $(K,\varepsilon)$ including all order non-perturbative corrections (Sec.~\ref{sec:quant_cond}),
\item[II)] clarification of full transseries structure of the energy spectra with respect to the inverse energy level expansion (Sec.~\ref{sec:energy_trans}), and
\item[III)] derivations of the Gutzwiller trace formula (GTF), the spectral summation form (SSF), and the Euclidean path-integral (EPI) using the exact QCs (Sec.~\ref{sec:varous_formula}).
\end{enumerate}
After the investigation of the massless cases, we have then discussed the massive cases, i.e., $\omega > 0$, and shown:
\begin{enumerate}
\item[IV)] uniqueness of the path of analytic continuation for a given $N$, and  non-existence of non-perturbative contributions in the exact QCs, the energy spectra, and all the formulas in III)  (Sec.~\ref{sec:mass_cases}).
\end{enumerate}
We have finally made additional remarks on similarities to the Hermitian cases for even $N$ and resurgence (Sec.~\ref{sec:comments}).
\\ \\ \indent
In our EWKB, the exact QCs can be expressed by Voros symbols (periodic cycles), and the cycle-representation of the QCs is quite helpful for the analysis based on  Borel resummation theory.
For the massless cases, the ${\mathbb Z}_2$ symmetry, ${\mathbb Z}_2: x \rightarrow -x$, in the potential crucially affect their transseries and non-perturbative structures through the DDP formula, and the effect appears as (non-)existence of extra non-perturbative cycles without oscillations in the cases of even (odd) $N = 2 K + \varepsilon$. 
The perturbative/non-perturbative structure of the exact QCs directly propagate not only to the energy spectra but also to all the formulas in III).
We should emphasize that, although we have performed those analyses by using transseries, their analytic forms can be formally obtained by taking the median resummation to them.
Thus, our results are formally exact.

For the massive cases, from the requirement of existence of solution of the exact QCs, the path of analytic continuation are uniquely determined, and in consequence the transseries structure become quite simplified because of a constrained $K$ as $K=\lfloor (N-1)/2 \rfloor$.
As a result, the non-perturbative contributions do not appear in the exact QCs, and thus, the energy spectra and all the formulas in III) are perturbative.
However, those are in general Borel non-summable.
This result is extendable to more generic polynomial potentials with a single quadratic vacuum.

Notice that, for constructions of the formulas in Fig.~\ref{fig:flowchart} from the exact QCs, pseudo-Hermiticity and the ${\cal CPT}$ inner-product are quite essential, which guarantee the unitarity condition and positive definiteness.

Since this study addressed a quite simple potential, there are many questions remained even in quantum mechanical level: more generic potentials,  constraint to non-perturbative effects by ${\cal PT}$ symmetry, generalizations of ${\cal PT}$/${\cal CPT}$ duality and their non-perturbative effects, and so on.
Furthermore, a generalization to field theories and study of their non-perturbative structure are interesting problems as a future work.


\acknowledgments
We would thank Naohisa~Sueishi for helpful discussion about the Gutzwiller trace formula.
We would thank Illust AC (https://www.ac-illust.com) for illustrations in our figures.
S.~K is supported by JSPS KAKENHI Grant No.~22H05118.

\appendix

\section{Pseudo-Hermiticity and ${\cal CPT}$ inner-product} \label{sec:pseudo_Herm}
In this part, we briefly review pseudo-Hermiticity and construction of the ${\cal CPT}$ inner-product.
For now, we consider the Minkowski spacetime, but extension to the Euclid spacetime is straightforward by the Wick rotation.
See Refs.~\cite{Mostafazadeh:2001jk,Mostafazadeh:2002pd,Mostafazadeh:2003gz,Weigert:2003py,Bender:2002vv,Bender:2004zz} and references therein in detail.

We denote $\widehat{H}_{\cal PT}$ as a ${\cal PT}$ Hamiltonian operator and define ${\cal C}$, ${\cal P}$, and ${\cal T}$ operators which satisfy
\be
&&   [\widehat{H}_{\cal PT}, {\cal PT}] =[\widehat{H}_{\cal PT}, {\cal C}] =  [{\cal C}, {\cal PT}] = 0, \\
&& {\cal O}^2
= {\cal O} {\cal O}^\dagger = {\cal O}^{\dagger} {\cal O}
= {\mathbb I}, \qquad {\cal O} \in \{ {\cal C}, {\cal P}, {\cal T} \},
\ee
where $[A,B] := A B - B A$.
The parity operator, ${\cal P}$, flips the sign of space as $x \rightarrow - x$, and the time-reversal operator, ${\cal T}$, corresponds to complex conjugation, ${\cal K}$.
The `charge conjugation', ${\cal C}$, is not a usual transform acting to a charged particle and will be determined to find the ${\cal CPT}$ inner-product later.
Notice that the time-reversal, ${\cal T}$, is an anti-unitary operator.

There are a couple of notations of the ${\cal PT}$ symmetric Hilbert space, but the most familiar way might be to start with the Dirac bra-ket of the energy eigenstates:
\be
&& \widehat{H}_{\cal PT} | E_k \rangle = E_k |E_k \rangle, \qquad {\cal PT} | E_k \rangle = | E_k \rangle, \qquad \langle E_k | := | E_k \rangle^\dagger, \label{eq:Ek_states} \\
&& \widehat{H}_{\cal PT}^\dagger | \bar{E}_k \rangle = \bar{E}_k | \bar{E}_k \rangle, \qquad {\cal PT} | \bar{E}_k \rangle = | \bar{E}_k \rangle, \qquad \langle \bar{E}_k | := | \bar{E}_k \rangle^\dagger, \label{eq:Ek_states_dag} 
\ee
where $k$ is a label of the energy level.
Here, we assume that the energy does not have degeneracies and that ${\cal PT}$ symmetry is unbroken, i.e., the energy spectrum is real and ${\cal K}[E_k] = \bar{E}_k$, where ${\cal K}$ is complex conjugation.
The fact that the Hamiltonian is not Hermitian, i.e., $\widehat{H}_{\cal PT} \ne \widehat{H}_{\cal PT}^\dagger$, implies that $\widehat{H}_{\cal PT}^\dagger | E_k \rangle \ne E_k |E_k \rangle$ even if the spectrum is real.
Although the ${\cal PT}$ Hamiltonian is not Hermitian, it satisfies ${\cal P}$-pseudo-Hermiticity condition~\cite{Mostafazadeh:2001jk}:
\be
\widehat{H}_{\cal PT}^\dagger  = {\cal P} \widehat{H}_{\cal PT} {\cal P}. \label{eq:pseudo_Herm}
\ee
By this condition, one finds that
\be
\langle E_k| {\cal P} \widehat{H}_{\cal PT} |E_k \rangle &=& \langle  E_k | \widehat{H}_{\cal PT}^\dagger {\cal P} |E_k \rangle =  E_k \langle E_k| {\cal P} |E_k \rangle, \label{eq:Ek_<Ek|P|Ek>}
\ee
and, according to Ref.~\cite{Weigert:2003py}, $\langle E_k| {\cal P} |E_k \rangle$ in the the last equality can be replaced with 
\be
\zeta_k = \langle E_k| {\cal P} |E_k \rangle, \qquad \zeta_k^2 = 1.
\ee
Eq.(\ref{eq:Ek_<Ek|P|Ek>}) also implies that the ${\cal PT}$ symmetric states, $|E_k)$ and $(E_k|:= {\cal PT}|E_k)$, can be found by the identification that
\be
| E_k \rangle \rightarrow | E_k), \qquad \langle E_k | {\cal P} \rightarrow ( E_k|, \qquad
(E_{k_1} | E_{k_2} ) = \zeta_{k_1} \delta_{k_1,k_2},
\ee
where $\delta_{k_1,k_2}$ is the Kronecker delta.
Hence, the inner-product of the ${\cal PT}$ symmetric states is indefinite, and the complete set is given by
\be
   {\mathbb I} =    \sum_{k} \zeta_k {\cal P} | E_k \rangle \langle E_k| = \sum_{k} \zeta_k  | E_k \rangle \langle E_k| {\cal P}. \label{eq:PTcomp_set_E_l}
\ee
This implies that the $| \bar{E}_k \rangle$ in Eq.(\ref{eq:Ek_states_dag}) can be expressed by $|E_k \rangle$ as
\be
| \bar{E}_{k} \rangle = \zeta_{k} {\cal P} |{E}_{k} \rangle, \qquad \langle \bar{E}_{k} | = \zeta_{k} \langle {E}_{k} | {\cal P}, \qquad \langle E_{k_1} | \bar{E}_{{k_2}} \rangle = \langle \bar{E}_{k_1} | E_{{k_2}} \rangle = \delta_{k_1,k_2}.
\ee
However, for the time-dependent ${\cal PT}$ symmetric states, the inner-product is invariant under the time-evolution:
\be
\frac{\pd}{\pd t} (E_k (t)| E_k (t)) = \frac{1}{i \hbar} \langle E_k (t)| \left(- \widehat{H}_{\cal PT}^\dagger {\cal P} + {\cal P} \widehat{H}_{\cal PT} \right) | E_k (t)\rangle = 0. \label{eq:unitary_cond}
\ee
Therefore, the unitarity condition is satisfied due to the pseudo-Hermiticity in Eq.(\ref{eq:pseudo_Herm}).

As we saw above, the inner-product of the ${\cal PT}$ symmetric states is indefinite.
One can solve this problem by introducing ${\cal C}$ operator and construct the ${\cal CPT}$ inner-product.
We define action of the ${\cal C}$ operator to the energy state as
\be
{\cal C} | E_k \rangle = \zeta_k | E_k \rangle \quad \Rightarrow \quad | \bar{E}_k \rangle = \zeta_k {\cal P} | E_k \rangle = {\cal P C} | E_k \rangle. 
\ee
By this operator, one can find the ${\cal CPT}$ inner-product which is positive definite:
\be
\delta_{k_1,k_2} &=& \langle E_{k_1} | \bar{E}_{k_2} \rangle = \langle E_{k_1} | {\cal P} {\cal C}|{E}_{k_2} \rangle \nl
&=& \int_{\gamma_{\cal PT}} dx \,  \langle E_{k_1} | {\cal P C} | x \rangle \langle x |{E}_{k_2} \rangle = \int_{\gamma_{\cal PT}} dx \, {\cal C P}[ \overline{\phi_{k_1}(x)}] \phi_{k_2}(x) \nl
&=& \int_{\gamma_{\cal PT}} dx \, {\cal C P T} [\phi_{k_1}(x)]\phi_{k_2}(x),
\ee
where $\phi_k(x) := \langle x |E_{k} \rangle$ is the ${\cal PT}$ symmetric energy eigenfunction satisfying ${\cal PT}[\phi_k(x)] = \langle x | {\cal PT} | E_{k}\rangle = \overline{\phi_k(-x)} =  \phi_k(x)$.
By denoting $\chi := {\cal PC}$, $\widehat{H}_{\cal PT}$ is $\chi$-pseudo-Hermitian:
\be
{\cal P C} \widehat{H}_{\cal PT} =  {\cal P} \widehat{H}_{\cal PT} {\cal C} =  \widehat{H}_{\cal PT}^\dagger {\cal P C} \quad \Rightarrow \quad \widehat{H}_{\cal PT}^\dagger = \chi \widehat{H}_{\cal PT} \chi^{-1}.
\ee
Finally, we define the ${\cal CPT}$ inner-product, $\lcurvyangle \psi | \phi \rcurvyangle$, as
\be
\lcurvyangle \psi | \phi \rcurvyangle := \langle \psi | \chi \phi\rangle = \langle \chi^{-1} \psi | \phi\rangle. \label{eq:CPT_inner_prod}
\ee
By using $\chi = {\cal PC}$, the complete set (\ref{eq:PTcomp_set_E_l}) is expressed by
\be
{\mathbb I} = \sum_{k} | \chi E_k \rangle \langle E_k | = \sum_{k} | E_k \rangle \langle \chi^{-1} E_k |,
\ee
and one can derive
\be
\delta (x - y) &=& \langle x | y \rangle = \sum_{k} \langle x | E_k \rangle \langle \chi^{-1} E_k | y \rangle \nl
&=&  \sum_{k} \phi_k(x) \chi \overline{\phi_k(y)} =  \sum_{k} {\cal CPT} [\phi_k(y)] \phi_k(x).
\ee
One can easily prove the unitarity condition for the ${\cal CPT}$ inner-product  in the similar way to Eq.(\ref{eq:unitary_cond}).

\section{Derivation of {\bf Fact} in Sec.~\ref{sec:mass_cases}}
\label{sec:mass_borel}.
\begin{figure}[tbp]
  \begin{center}
    \begin{tabular}{cc}
      \begin{minipage}{0.5\hsize}
        \begin{center}
          \includegraphics[clip, width=75mm]{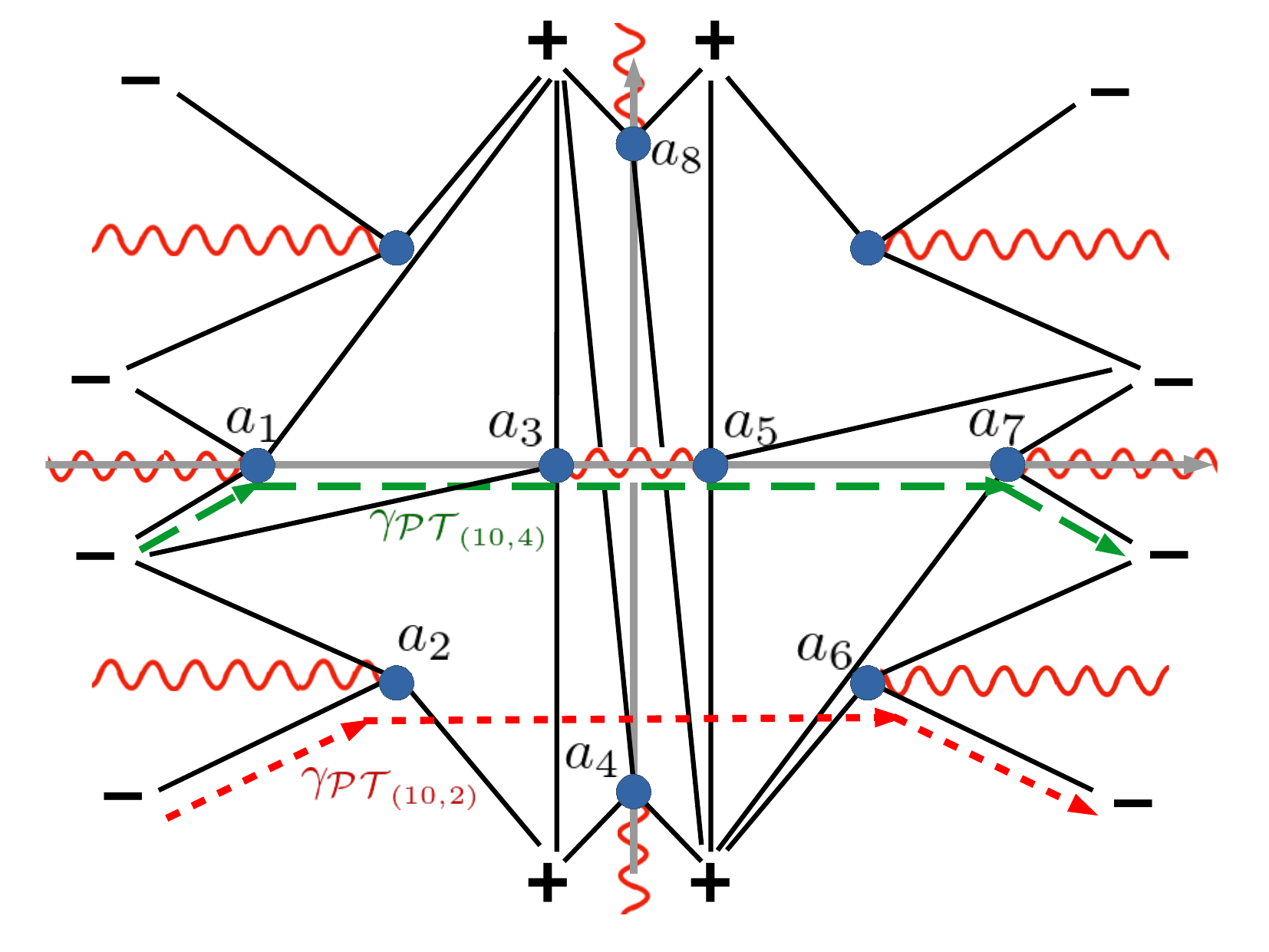}
          \hspace{1.6cm} (a) $\arg(\hbar) = 0_+$
        \end{center}
      \end{minipage}
      \begin{minipage}{0.5\hsize}
        \begin{center}
          \includegraphics[clip, width=75mm]{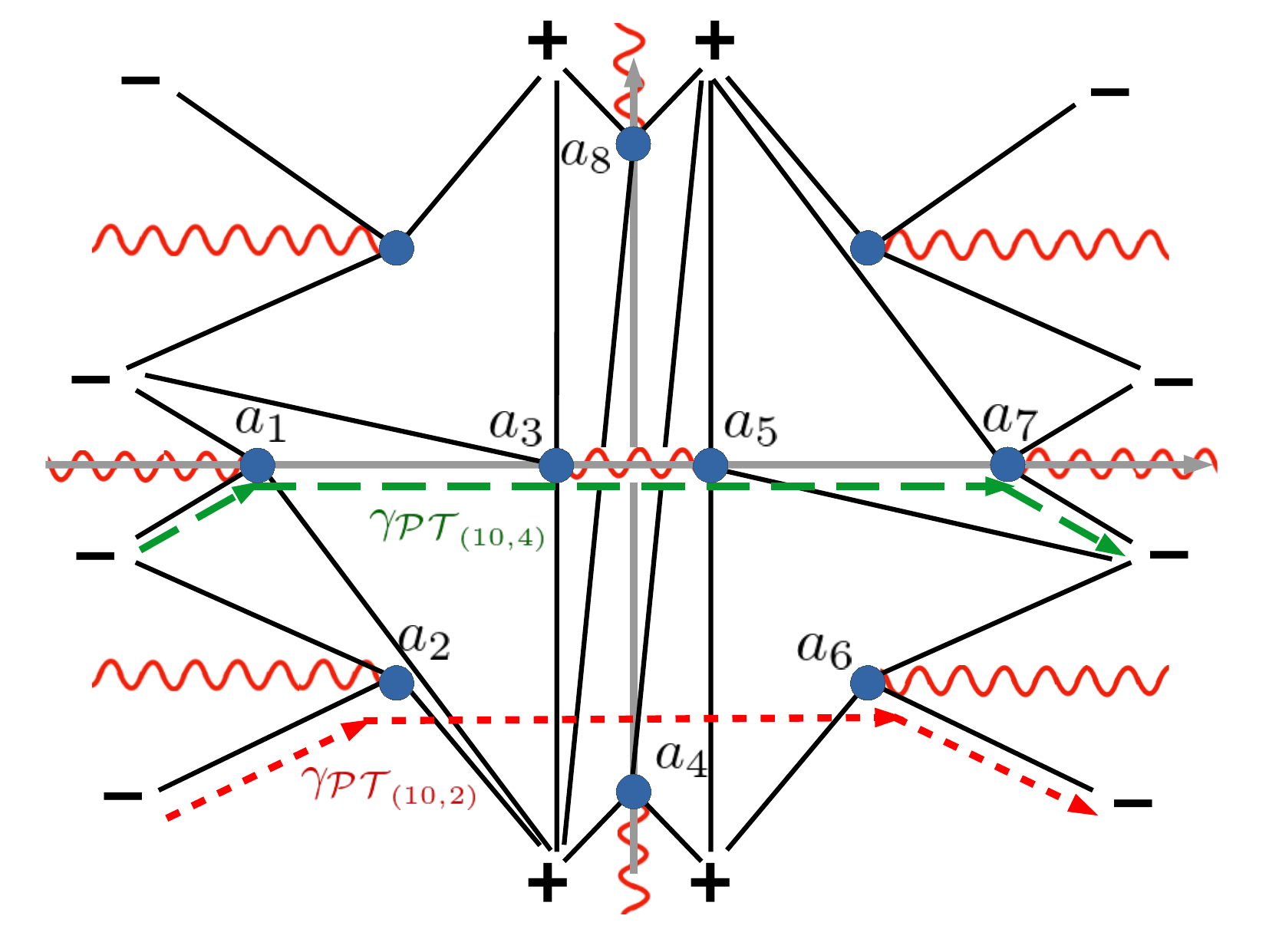}
          \hspace{1.6cm} (b) $\arg(\hbar) = 0_-$
        \end{center}
      \end{minipage} 
    \end{tabular} 
    \caption{Stokes graphs in the massive cases with $\arg(\hbar)=0_\pm$ for $(N,K) = (10,2)$ and $(N,K) = (10,4)$.
      The red and green lines denote the path of analytic continuation for $K=2$ and $K=4$, respectively.
      In these figures, we assume that $E = O(1)$ in the Schr\"{o}dinger equation, so that the perturbative cycle $C_{(3,5)}$ consists of two simple turning points.}
    \label{fig:Stokes_mass_Ai_pm}
  \end{center}
\end{figure}
\begin{figure}[tbp]
  \begin{center}
    \begin{tabular}{cc}
      \begin{minipage}{0.5\hsize}
        \begin{center}
          \includegraphics[clip, width=75mm]{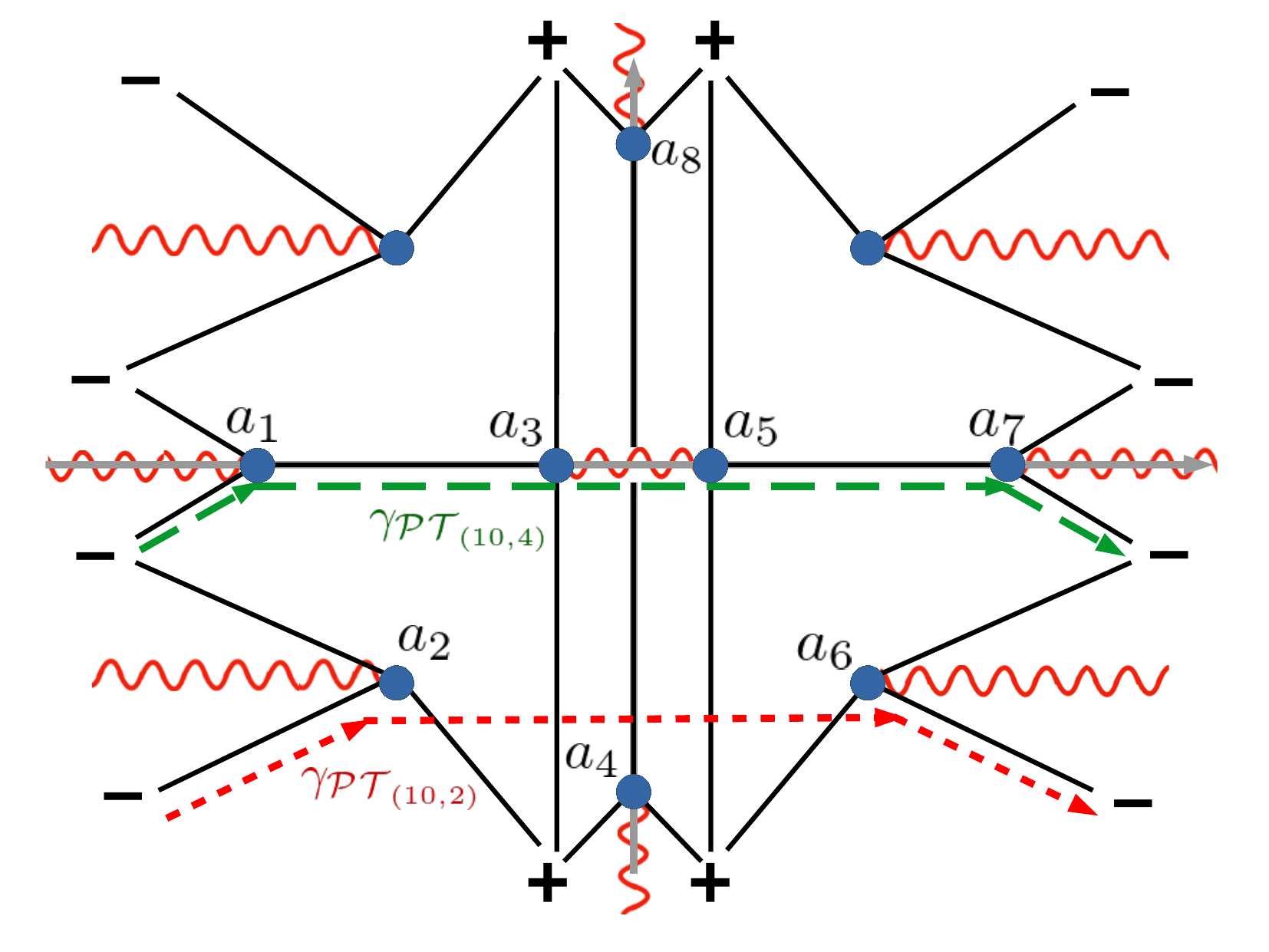}
          \hspace{1.6cm} (a) Airy type ($0 < E_0 \ll 1$)
        \end{center}
      \end{minipage}
      \begin{minipage}{0.5\hsize}
        \begin{center}
          \includegraphics[clip, width=75mm]{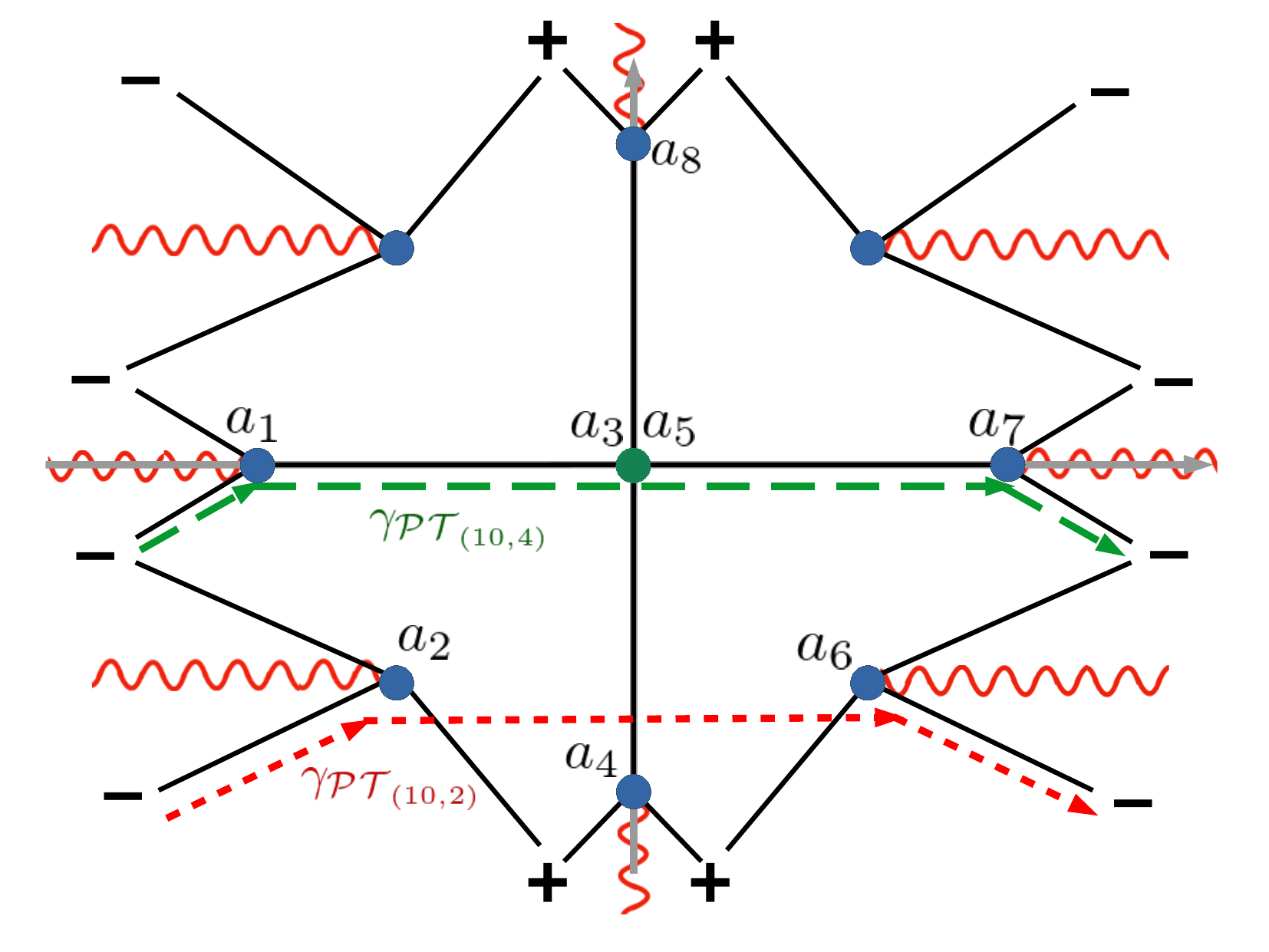}
          \hspace{1.6cm} (b) Degenerate Weber-type ($E_0 = 0$)
        \end{center}
      \end{minipage} 
    \end{tabular} 
    \caption{Stokes graphs in the massive cases with $\arg(\hbar)=0$ for $(N,K) = (10,2)$ and $(N,K) = (10,4)$ using Airy-type and degenerate Weber-type for the quadratic vacuum at the origin.
The two simple turning points blue-colored in (a), $a_3$ and $a_5$, collide to each other as varying $E_0 \rightarrow 0_+$, where $E_0$ is the zero-th order of the energy, and consequently become a double turning point green-colored in (b).
    }
    \label{fig:Stokes_mass_Ai_dw}
  \end{center}
\end{figure}
\begin{figure}[tbp]
 \centering
 \includegraphics[clip, width=75mm]{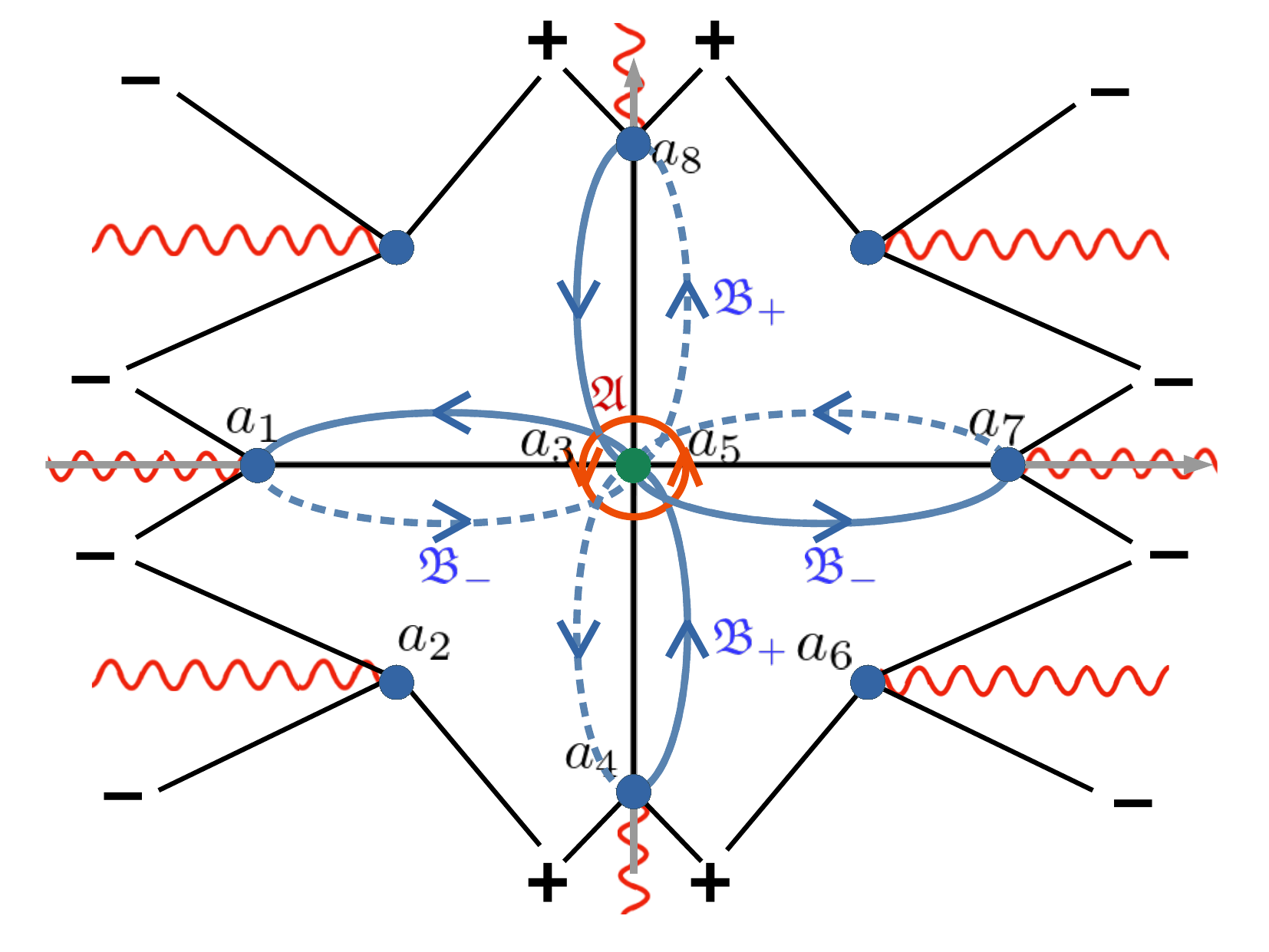}
 \caption{Cycles on the Stokes graph defined by the potential in Eq.(\ref{eq:V_massive_N10}) with $E_0 = 0$.}
    \label{fig:Stokes_mass_dw_cycle}
\end{figure}
We explain the derivation of {\bf Fact} in Sec.~\ref{sec:mass_cases}.
As an example, we consider $N=10$ and take the two paths given by $K=2$ and $K=4$ in Eq.(\ref{eq:gam_PT}) for analytic continuations.
By these choices of $(N,K)$, the potential is given by
\be
V_{\cal PT}(x) = \omega^2 x^2 - g x^{10}, \qquad \omega, g \in {\mathbb R}_{>0}. \label{eq:V_massive_N10}
\ee
The Stokes graph and the two paths, $\gamma_{{\cal PT}_{(10,2)}}$ and $\gamma_{{\cal PT}_{(10,4)}}$, are shown in Fig.~\ref{fig:Stokes_mass_Ai_pm}.
In the below discussion, the labels of turning points are taken in the manner in Fig.~\ref{fig:Stokes_mass_Ai_pm}.

We firstly consider the lower path, $\gamma_{{\cal PT}(10,2)}$, for analytic continuation.
The monodromy matrices are given by
\be
   {\cal M}^{0_+} &=& M_+ N_{a_2,a_3} M_+ N_{a_3,a_4} M_-^{-1} N_{a_4,a_8} M_+ N_{a_8,a_5} M_+ N_{a_5,a_7} M_+ N_{a_7,a_6} M_+ N_{a_6,a_2}, \\
   {\cal M}^{0_-} &=& M_+ N_{a_2,a_1} M_+ N_{a_1,a_3} M_+ N_{a_3,a_8} M_+ N_{a_8,a_4} M_-^{-1} N_{a_4,a_5} M_+ N_{a_5,a_6} M_+ N_{a_6,a_2}, 
\ee
thus, normalizability of the wavefunction, ${\cal M}^{\pm 0}_{12} = 0$, gives
\be
   {\frak D}^{0_+}  &\propto&   C_{(4,2)} + C_{(4,3)} + (1 + C_{(4,2)} + C_{(4,3)})(C_{(4,5)} + C_{(4,6)} + C_{(4,7)} + C_{(4,8)}), \\
   {\frak D}^{0_-}  &\propto&   C_{(4,5)} + C_{(4,6)} + ( 1 + C_{(4,5)} + C_{(4,6)})(C_{(4,1)} + C_{(4,2)} + C_{(4,3)} + C_{(4,8)}).
\ee
Even though the QCs have a discontinuity, a perturbative part of the energy should be derived from a common perturbative part in the QCs, which is 
\be
   {\frak D}^{0_\pm}_{\rm P} \propto 1 + C_{(2,6)}. \label{eq:D_P_low}
\ee 
When the energy in the cycles is replaced as $E \rightarrow \widetilde{E} \hbar$ with $\widetilde{E} = O(\hbar^0)$, the two simple turning points, $a_3$ and $ a_5$, becomes a double turning point, but Eq.(\ref{eq:D_P_low}) is unchanged.
One can see that, by this replacement, the leading order of $\log C_{(2,6)}$ is $ O(\hbar^{-1})$, but the energy parameter, $\widetilde{E}$, is not included in the same order.
This means that the QC in Eq.(\ref{eq:D_P_low}) has no solution of the energy to satisfy ${\frak D}^{0_\pm}_{\rm P}=0$.

Next, let us consider the upper path, $\gamma_{{\cal PT}_{(10,4)}}$.
The monodromy matrices are obtained by
\be
   {\cal M}^{0_+} &=&  M_+ N_{a_3,a_4} M_-^{-1} N_{a_4,a_8} M_+ N_{a_8,a_5} M_+ N_{a_5,a_7} M_+  N_{a_7,a_2}, \\
   {\cal M}^{0_-} &=& N_{a_3,a_1} M_+ N_{a_1,a_3} M_+ N_{a_3,a_8} M_+ N_{a_8,a_4} M_-^{-1} N_{a_4,a_5} M_+ N_{a_5,a_2}.
\ee
By imposing normalizability to the wavefunction, i.e., ${\cal M}^{0_\pm}_{12}=0$, the QCs are obtained as
\be
   {\frak D}^{0_+}  &\propto& 1 + A_{(3,5)} \left( 1 + B_{(5,7)} \right) + C_{(3,8)} + C_{(4,5)} + C_{(4,7)} + B_{(4,8)}, \\
   {\frak D}^{0_-}  &\propto& 1 + A_{(3,5)}^{-1} \left( 1 + B_{(3,1)} \right) + C_{(5,8)} + C_{(4,3)}  + C_{(4,1)} + B_{(4,8)}.
\ee
Here, we used the symbols, $A_{(3,5)} = C_{(3,5)}$ and $B_{(\bullet,\bullet)} = C_{(\bullet, \bullet)}$, to emphasize to be purely perturbative and non-perturbative cycles, respectively.
The DDP formula for the cycles is available by counting the intersections and given by
\be
&& {\frak S}_{0}^\nu[A_{(3,5)}] = A_{(3,5)} (1+B_{(3,1)})^{-\nu}(1+B_{(5,7)})^{-\nu}(1+B_{(4,8)})^{+2\nu}, \\
&& {\frak S}_{0}^\nu[B_{(\bullet,\bullet)}] = B_{(\bullet,\bullet)}, \\
&& {\frak S}_{0}^\nu[C_{(3,8)}] = C_{(3,8)} (1+B_{(3,1)})^{-\nu}(1+B_{(4,8)})^{+\nu},  \nl
&&{\frak S}_{0}^\nu[C_{(4,5)}] = C_{(4,5)} (1+B_{(5,7)})^{-\nu}(1+B_{(4,8)})^{+\nu}, \nl
&&{\frak S}_{0}^\nu[C_{(4,7)}] = C_{(4,7)} (1+B_{(5,7)})^{-\nu}(1+B_{(4,8)})^{+\nu}, \nl
&& {\frak S}_{0}^\nu[C_{(5,8)}] = C_{(5,8)} (1+B_{(5,7)})^{+\nu}(1+B_{(4,8)})^{-\nu},  \nl
&&{\frak S}_{0}^\nu[C_{(4,3)}] = C_{(4,3)} (1+B_{(3,1)})^{+\nu}(1+B_{(4,8)})^{-\nu}, \nl
&&{\frak S}_{0}^\nu[C_{(4,1)}] = C_{(4,1)} (1+B_{(3,1)})^{+\nu}(1+B_{(4,8)})^{-\nu}.
\ee
From these, one can obtain the exact QC, ${\frak D}^0 := {\frak S}^{\pm 1/2}_0[{\frak D}^{0_\pm}]$, as
\be
   {\frak D}^0 &\propto& 1 + \sqrt{\frac{1 + B_{(5,7)}}{1 + B_{(3,1)}}} A_{(3,5)} + \sqrt{\frac{1 + B_{(5,7)}}{1 + B_{(4,8)}}} C_{(4,5)} + \frac{C_{(3,8)}}{\sqrt{1 + B_{(3,1)}}\sqrt{1 + B_{(4,8)}}}. \label{eq:D0_high}
\ee
By replacing $E \rightarrow \widetilde{E} \hbar$, the quadratic vacuum at the origin becomes a double turning point, as is shown in Fig.~\ref{fig:Stokes_mass_Ai_dw}.
Using the ${\mathbb Z}_2$ symmetry in Eq.(\ref{eq:Z2_symm}), the exact QC becomes
\be
   {\frak D}^0 &\propto& 1 +  {\frak A} + \left( \sqrt{1 + {\frak B}_-} + \frac{1}{\sqrt{1 + {\frak B}_-}} \right)  \frac{{\frak B}_+}{\sqrt{1 + {\frak B}_+^2}}, \label{eq:D0_high_dw}
\ee
where $A_{(3,5)} \rightarrow {\frak A}$ is a perturbative cycle going around the double turning point corresponding to the quadratic vacuum, and ${\frak B}_\mp$ are introduced through $E \rightarrow \widetilde{E} \hbar$ as
\be
B_{(3,1)} = B_{(5,7)} \rightarrow {\frak B}_-, \qquad C_{(4,5)} = C_{(3,8)}\rightarrow {\frak B}_+, \qquad B_{(4,8)} \rightarrow {\frak B}_+^2. 
\ee
Fig.~\ref{fig:Stokes_mass_dw_cycle} shows the cycles on the Stokes graph with $E = O(\hbar)$.
These cycles are expressed by $F$ as~\cite{Sueishi:2021xti,Kamata:2021jr,Kamata:2023opn,Bucciotti:2023trp},
\be
&& {\frak A} = e^{-2 \pi i F}, \\
&& {\frak B}_\mp = C_\mp^2\frac{\sqrt{2 \pi} {\frak B}_0 e^{\pm \pi i F} \hbar^{\pm F}}{\Gamma(1/2 \mp F)}, \qquad {\frak B}_0 := e^{-\frac{S_{\frak B}}{\hbar}}, \quad (S_{\frak B} \in {\mathbb R}_{>0}) \label{eq:B_CC}
\ee
where $F$ and $C_{\mp}$ are formal power series of $\widetilde{E}$ and $\hbar$, and $F$ is obtained by a residue integration of $S_{\rm od}$ around $x=0$ as
\be
F(\widetilde{E};\hbar) = - {\rm Res}_{x=0}\, S_{\rm od}(x,\widetilde{E};\hbar) = - c \widetilde{E} + O(\hbar), \label{eq:F_Sod}
\ee
with $c \in {\mathbb R}_{>0}$.
The perturbative energy solution is given by
\be
{\frak D}^0_{\rm P} \propto 1 + {\frak A} = 0, 
\ee
and the positive energy condition leads to
\be
F = - k \quad \Rightarrow \quad \widetilde{E} = \frac{k}{c} + O(\hbar) , \qquad k \in {\mathbb N}_{0} + \frac{1}{2}.
\ee
Substituting $F$ into Eq.(\ref{eq:B_CC}) gives
\be
{\frak B}_+ = 0,
\ee
because of the gamma function in the denominator.
Thus, the exact QC (\ref{eq:D0_high_dw}) becomes
\be
   {\frak D}^0 &\propto& 1 +  {\frak A},
\ee
which contains the perturbative cycle only, i.e., the energy solution contains no non-perturbative part.
Notice that the DDP formula of ${\frak A}$ is still non-trivial because ${\frak B}_- \ne 0$, which means that the energy solution is Borel non-summable. 
 \\ \par
\begin{figure}[tbp]
 \centering
 \includegraphics[width=80mm]{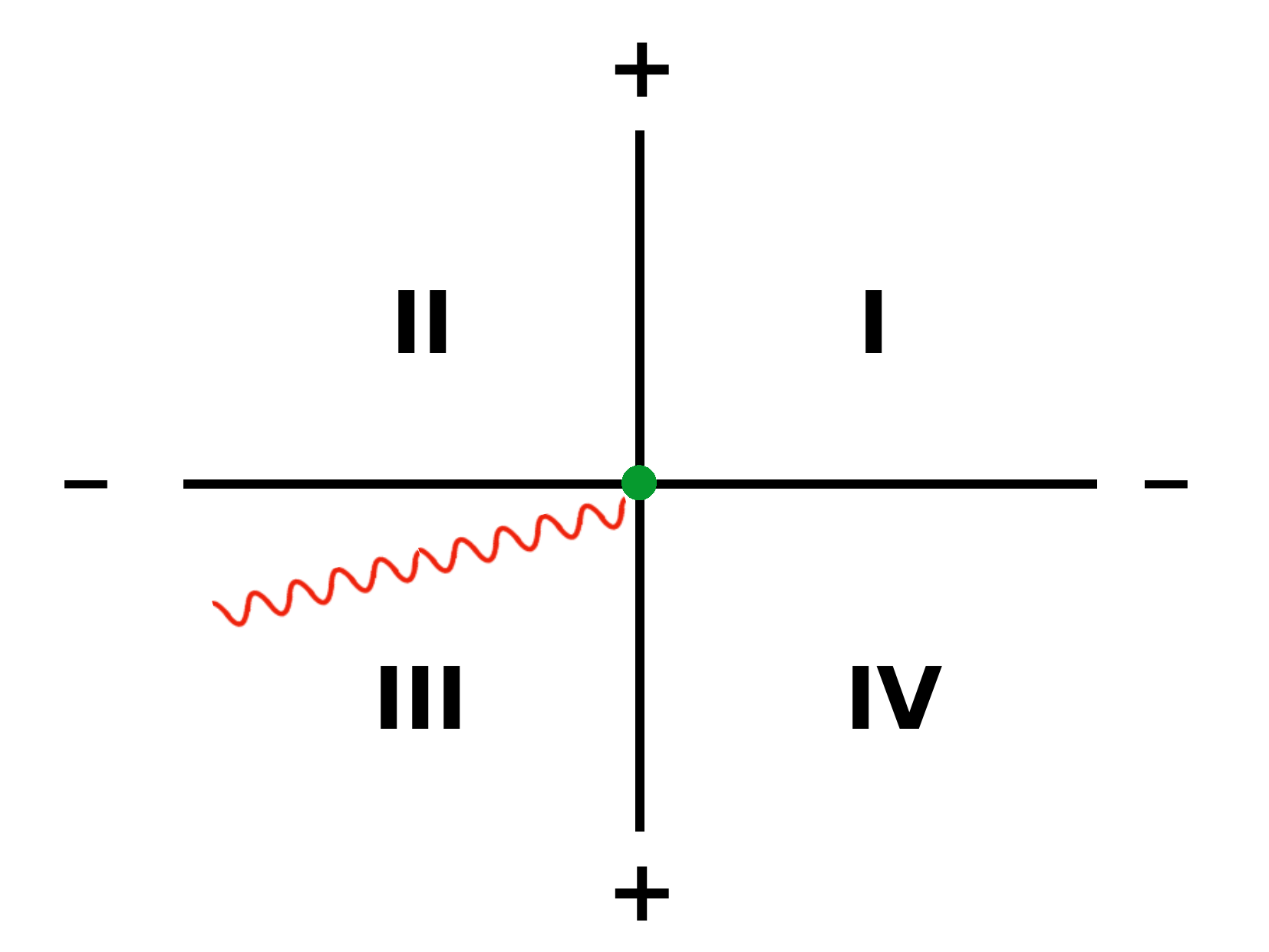}
 \caption{Stokes graph for the degenerate Weber equation.
   The green dot is a double turning point.
   The black solid and red wave lines denote Stokes lines and a branch-cut, respectively.
   This figure is brought from Ref.~\cite{Kamata:2023opn}.
   See Refs.~\cite{Sueishi:2021xti,Kamata:2021jr,Kamata:2023opn,Bucciotti:2023trp} in detail, for example.
}
\label{fig:DW_connection}
\end{figure}

The same discussions are applicable to any other $(N,K)$.
Here are observations from the above analysis:
\begin{enumerate}
\item[(1)] The perturbative part of the energy solution is given by a cycle with purely oscillating.
  When replacing the energy as $E \rightarrow \widetilde{E} \hbar$ with $\widetilde{E} = O(\hbar^0)$, if the perturbative cycle consists of two simple turning points in the exact QC, no appropriate solution can be found from the exact QC.
  This means that the exact QC must contain a cycle going around a double turning point as a reasonable perturbative cycle.
  It is possible only when the nearest path to the real axis is taken as the path of analytic continuation.
  As a result, the exact QCs for arbitrary $(N,K)$ consist of a perturbative cycle defined by the double turning point and non-perturbative cycles having non-trivial intersection numbers with the perturbative cycle.
  \item[(2)] By taking $E \rightarrow \widetilde{E} \hbar$, the quadratic vacuum is expressed by the degenerate Weber-type Stokes graph shown in Fig.~\ref{fig:DW_connection}.
  Suppose that there exists a single quadratic vacuum and that the asymptotic behavior of the local Stokes graph is taken in the similar way to Fig.~\ref{fig:DW_connection} by appropriately taking branch-cuts.
  When there exist non-perturbative cycles along the ``$+$''-directions in Fig.~\ref{fig:DW_connection}, which correspond to ${\frak B}_+$ in Eq.(\ref{eq:D0_high_dw}),   their contributions do not exist because ${\frak B}_+ = 0$.
\item[(3)]   When there exist non-perturbative cycles along the ``$-$''-directions in Fig.~\ref{fig:DW_connection}, which correspond to ${\frak B}_-$ in Eq.(\ref{eq:D0_high_dw}), 
  their contributions are canceled in the exact QC.
  In such a case, the energy solution is Borel non-summable.
  It is notable that these ${\frak B}_-$-type non-perturbative cycles coupled to ${\frak A}$, always appear as a pair in the numerator and denominator as, e.g., $(B_{(3,1)}, B_{(5,7)})$ in Eq.(\ref{eq:D0_high}).
  Such a pair appears when the ${\mathbb Z}_2$ symmetry in Eq.(\ref{eq:Z2_symm}) is preserved in the potential, and these two contributions are equivalent to each other.  
\end{enumerate}
\noindent
One can generate other double turning points from the simple turning points by changing $E_0 \in {\mathbb R}$ as a control parameter and inducing bifurcations.
In the potential (\ref{eq:V_PT}), however, the solution can be obtained only when a double turning point exists at the origin due to the above (1)-(3) and topology of those Stokes graphs.
By using these observations, {\bf Fact} in Sec.~\ref{sec:mass_cases} can be proved for any $(N,K)$.

Notice that the locations of the branch-cuts and the asymptotic behaviors of the Stokes graph, $``\pm"$, are arbitrary as far as being consistent with each other, and thus, the result must be unchanged by changing them.

\section{Alien calculus for energy spectra} \label{sec:Alien_energy} 
In this part, we describe alien calculus for the energy spectra.
We begin with the exact QCs given by Eqs.(\ref{eq:D0_evenN_Kgen})(\ref{eq:D0_oddN_Kgen}).
Suppose that we have already found a transseries solution of the energy, denoted by $E(\kappa)$, which satisfies\footnote{
In the below discussions such as Sec.~\ref{sec:alien_evenN}, we mainly address action of the alien derivative to a perturbative sector for simplicity, but the solution does not need to only contain a perturbative cycle.
It generally contains non-perturbative contributions, and action of the alien derivative to non-perturbative sectors is non-trivial.
}
\be
\left. {\frak D}^0 \right|_{E=E(\kappa)} = 0.
\ee
The Stokes automorphism ${\frak S}_{\theta}^{\nu \in {\mathbb R}}$ can be generally expressed by the alien derivative $\bul{\Delta}_{\theta}$ as
\be
{\frak S}_{\theta}^{\nu \in {\mathbb R}} = \exp \left[ \nu \bul{\Delta}_{\theta} \right] \sim 1 + \nu \bul{\Delta}_{\theta},
\ee
where $\theta$ is the angle of the integration-ray in the Laplace integral, ${\cal L}_{\theta}$.
Below, we consider the case of $\theta := \arg(\eta)= 0$ for simplicity, but a generalization to non-zero $\theta$ is straightforward.

Instead of seeing the energy solutions, it is convenient to deal with $\eta$ because our exact QCs for the massless cases are transseries of $\eta$.
We should remind that the DDP formula of the cycles in Eq.(\ref{eq:DDP_evenN_Kgen}) is related to $\arg(\eta)$, but $\eta$ (or the energy $E$) is a free parameter until solving the exact QCs.
Hence, for the transseries of $\eta$ in the cycles, $\eta(\kappa)$, the alien derivative can be split into two parts as
\be
&& \bul{\Delta}_0 [f(\eta(\kappa))] :=  \left. \pd_{\eta^{-1}} f(\eta) \right|_{\eta=\eta(\kappa)} \cdot \bul{\Delta}_0[\eta^{-1}(\kappa)] + \bul{\Delta}_{\eta,0} [f(\eta(\kappa))], \label{eq:alien_split} \\
&& \bul{\Delta}_{\eta,0} [f(\eta(\kappa))] := \bul{\Delta}_{\eta,0} [f(\eta)]|_{\eta = \eta(\kappa)}, 
\ee
where $\bul{\Delta}_{\eta,0}$ is the transformation law of the DDP formula with a constant $\eta$ in Eq.(\ref{eq:DDP_evenN_Kgen}).
If one acts $\bul{\Delta}_{\eta,0}$ to ${\frak D}^0(\eta(\kappa))$, the result is in general non-zero.
Thus, from Eq.(\ref{eq:alien_split}), we determine $\bul{\Delta}_0[\eta]$ to keep zero under the action of $\bul{\Delta}_{0}$ to ${\frak D}^0(\eta(\kappa))$.
From Eq.(\ref{eq:alien_split}), one can readily find that
\be
&& \bul{\Delta}_0 [{\frak D}^0(\eta(\kappa))] = 0 \quad \Rightarrow \quad \bul{\Delta}_0[\eta^{-1}(\kappa)] = - \left. \frac{\bul{\Delta}_{\eta,0} [{\frak D}^0(\eta)]}{\pd_{\eta^{-1}} {\frak D}^0(\eta)}\right|_{\eta = \eta(\kappa)}. \label{eq:del_eta_evenN}
\ee
After obtaining $\bul{\Delta}_0[\eta^{-1}(\kappa)]$, the result can be translated into $\bul{\Delta}_0[E(\kappa)]$ by using Eq.(\ref{eq:def_eta_E}), that is
\be
\bul{\Delta}_0[E(\kappa)] = \frac{2 N}{N+2} \cdot \frac{E(\kappa)}{\eta^{-1}(\kappa)} \cdot \bul{\Delta}_0[\eta^{-1}(\kappa)]. \label{eq:Del_E_Del_eta}
\ee
By repeating the same procedure, one can recursively obtain the higher order alien derivatives, $(\bul{\Delta}_0)^n[\eta^{-1}(\kappa)]$ (or $(\bul{\Delta}_0)^n[E(\kappa)]$), by
\be
&& \sum_{s=1}^{n} \sum_{{\bf k} \in {\mathbb N}_0^s}^{\sum_{t=1}^s t  k_t \le n } \frac{\nu^{n}}{(n - \sum_{t=1}^s  t k_t)!} \nl
&& \cdot \left[ \prod_{t=1}^s \frac{((\bul{\Delta}_0)^t[\eta^{-1}(\kappa)] )^{k_t}}{(t !)^{k_t}k_t !} (\bul{\Delta}_{\eta,0})^{ - t k_t} \right] (\bul{\Delta}_{\eta,0})^{n} [ \pd^{|{\bf k}|}_{\eta^{-1}}{\frak D}^{0}(\eta(\kappa))] = 0,
\ee
where $| {\bf v} |$ denotes the $L_1$-norm.

In Sec.~\ref{sec:alien_evenN}, we demonstrate the specific calculations using these formulas.
In this paper, since it is sufficient for our purpose, we only argue the first order alien derivative.

\subsection{Even $N$} \label{sec:alien_evenN}
We consider resurgent relations for the even $N$ cases.
From the DDP formula given in Eq.(\ref{eq:DDP_evenN_Kgen}), the alien derivative to the cycles are obtained as
\be
&& \bul{\Delta}_{\eta,0}  [C_{(\bar{p}+2\ell-2,\bar{p}+2 \ell+1)}] = C_{(\bar{p}+2\ell-2,\bar{p}+2 \ell+1)} \cdot \log \left[ \prod_{n=0}^1 (1 + B_{(\bar{p}+2\ell+2n-2,\bar{p}+2 \ell + 2n-1)}) \right], \nl
&& \bul{\Delta}_{\eta,0} [B_{(\bar{p}+2 \ell+2n-2,\bar{p}+2 \ell + 2n-1)}] = 0 \quad \mbox{with} \quad n = 0, 1, \label{eq:aliden_evenN_Kgen}
\ee
with $\ell \in \{1,2,\cdots, K\}$.
Here, let us consider the perturbative sector.
This sector can be extracted by picking up the part corresponding to $n_{\ell} = 0$ and $n_{\ell} = 1$ for all $\ell$ from the summation in Eq.(\ref{eq:D0_evenN_Kgen}), and it is given by
\be
 {\frak D}_{\rm P}^{0} \propto 1 + C_{(\bar{p},\bar{p}+2K)} = 1 + P, \label{eq:D0_p}
\ee
where ${\frak D}_{\rm P}^{0}$ denotes the perturbative sector in the exact QC.
Since $\bul{\Delta}_{\theta}[E(\kappa)]$ is given from $\bul{\Delta}_{\theta}[\eta^{-1}(\kappa)]$ in Eq.(\ref{eq:Del_E_Del_eta}), we below argue $\bul{\Delta}_{\theta}[\eta^{-1}(\kappa)]$.

We compute Eq.(\ref{eq:del_eta_evenN}) using the definition of $P$ in Eq.(\ref{eq:C_pp_Kgen}).
We suppose that the specific form of $\eta^{-1}(\kappa)$ (or $\eta(\kappa)$) has been known by solving Eq.(\ref{eq:D0_p}).
Since
\be
\pd_{\eta^{-1}} {\frak D}_{\rm P}^0 (\eta) &=& - 2 i P \cdot \sum_{n \in 2 {\mathbb N}_0 -1} n v_{n} \sin \frac{\pi K n}{N} \cdot \eta^{n+1}, \label{eq:deta_C_pp_Kgen} \\
\bul{\Delta}_{\eta,0}  [{\frak D}_{\rm P}^0 (\eta)] &=& P \cdot \log \left[ \prod_{\ell=1}^{K-1} \prod_{n=0}^1 (1 + B_{(\bar{p}+2\ell+2n-2,\bar{p}+2\ell+2n-1)}) \right] \nl
&=& P \cdot \log \left[ (1 + B_{(\bar{p},\bar{p}+1)})(1 + B_{(\bar{p}+ 2 K,\bar{p}+2 K + 1)}) \prod_{\ell=1}^{K-1} (1 + B_{(\bar{p} + 2\ell,\bar{p} + 2\ell + 1)})^2 \right] \nl
&=& 2 P \cdot \left[ \log \left( 1 + B_{(\bar{p},\bar{p}+1)} \right) + \log \left( 1 + B_{(\bar{p}+K,\bar{p}+K+1)} \right) \delta_{K \, {\rm mod} \, 2, 0}  \right] \nl
&& + 4 P \cdot \sum_{\ell=1}^{\lfloor (K-1)/2 \rfloor} \log \left( 1 + B_{(\bar{p}+2\ell,\bar{p}+2\ell+1)} \right),
\label{eq:alieneta_C_pp_Kgen} 
\ee
where we used $B_{(\bar{p}+2\ell-2,\bar{p}+2 \ell - 1)} = B_{(\bar{p}+ 2 K-2 \ell+2,\bar{p}+2 K - 2 \ell + 3)}$ for $\ell = \{ 1, \cdots, \lfloor (K+1)/2 \rfloor \}$,
one can find the alien derivative to $\eta(\kappa)$ from Eq.(\ref{eq:Del_E_Del_eta}) as
\be
&& \bul{\Delta}_0[\eta^{-1}(\kappa)]\nl
&=& \left.  \frac{\log \left( 1 + B_{(\bar{p},\bar{p}+1)} \right)  + \log \left( 1 + B_{(\bar{p}+K,\bar{p}+K+1)} \right) \delta_{K \, {\rm mod} \, 2, 0} + 2 \sum_{\ell=1}^{\lfloor (K-1)/2 \rfloor} \log \left( 1 + B_{(\bar{p}+2\ell,\bar{p}+2\ell+1)} \right)}{i \sum_{n \in 2 {\mathbb N}_0 -1} n v_{n} \sin \frac{\pi K n}{N} \cdot \eta^{n+1}}\right|_{\eta = \eta(\kappa)} \nl
&=& \left.  \frac{ B_{(\bar{p},\bar{p}+1)} + B_{(\bar{p}+K,\bar{p}+K+1)} \delta_{K \, {\rm mod} \, 2, 0} + 2 \sum_{\ell=1}^{\lfloor (K-1)/2 \rfloor}  B_{(\bar{p}+2\ell,\bar{p}+2\ell+1)}}{i \sum_{n \in 2 {\mathbb N}_0 -1} n v_{n} \sin \frac{\pi K n}{N} \cdot \eta^{n+1}}\right|_{\eta = \eta(\kappa)} + O(B^2).  \label{eq:Del_eta_P_evenN}
\ee
Notice that, although the non-perturbative sector associated to $B_{(\bar{p},\bar{p}+1)}$, which corresponds to $\sigma_{(\lfloor K/2 \rfloor +1)}$ in Eq.(\ref{eq:sig_evenN_Kgen}), does not arise alone in the energy solution, it appears in the alien derivative in Eq.(\ref{eq:Del_eta_P_evenN}).
In addition, the alien derivative has information related to only $B_{(\bullet,\bullet)}$ but does not contain the other cycles, $C_{(\bullet,\bullet)}$.
Moreover, the alien derivative is pure imaginary and does not provide information of $(-1)^k$ in Eq.(\ref{eq:sig_evenN_Kgen}).
These facts imply that it is impossible to fully figure out information of the non-perturbative sectors from the perturbative sector in the transseries  solution of the energy.

Then, let us try to extract some information due to the other cycles in Eq.(\ref{eq:sig_evenN_Kgen}) from the perturbative sector in Eq.(\ref{eq:D0_p}) by introducing non-zero $\theta$ into Eq.(\ref{eq:aliden_evenN_Kgen}).
Actually, Stokes phenomena occur at not only $\theta = 0$ but also non-zero $\theta$, and the similar analysis is applicable to the other cycles.
For any $\theta = \arg(\eta)$ inducing Stokes phenomena, the alien derivative to any cycles $C_{(n_1,n_2)} \notin {\bf C}_{{\rm NP},\theta}$, 
where ${\bf C}_{{\rm NP},\theta}$ is a set of non-perturbative cycles going around degenerated Stokes lines induced at a $\theta$, is written as
\be
&& \bul{\Delta}_{\eta,\theta}  [C_{(n_1,n_2)}] = C_{(n_1,n_2)} \cdot \log \left[ \prod_{B_{j} \in {\bf C}_{{\rm NP},\theta}} (1 + B_{j} )^{\langle C_{(n_1,n_2)}, B_{j} \rangle} \right], \qquad C_{(n_1,n_2)} \notin {\bf C}_{{\rm NP},\theta}, \label{eq:Del_C_evenN_Kgen} \\
&& \theta = \frac{\pi n}{N} \ \ \mbox{with} \ \ n \in \{-N,-N+1,\cdots, N-1\}. \label{eq:Del_C_theta_evenN}
\ee
By this DDP formula, Eq.(\ref{eq:alieneta_C_pp_Kgen}) is modified as\footnote{
  The DDP formula generally changes the cycle-representation of the exact QC depending on each Stokes phenomenon, but in this procedure the phase is introduced only to the alien derivative as fixing  the exact QC, ${\frak D}^0$.
  It is because our purpose is extracting non-perturbative information from the solution, i.e., $\eta(\kappa)$, which has been already obtained by ${\frak D}^0$.
}
\be
&& \bul{\Delta}_{\eta,\theta }  [{\frak D}_{\rm P}^0] = P \cdot \log \left[ \prod_{B_{j} \in {\bf C}_{{\rm NP},\theta}} (1 + B_{j} )^{\langle P, B_{j} \rangle} \right].  \label{eq:Del_DP_evenN_Kgen}
\ee
Finally, the generalization of the alien derivative of $\eta^{-1}$ to an arbitrary $\theta$ is written as
\be
\bul{\Delta}_\theta[\eta^{-1}(\kappa)] &=& \left. \frac{ \log \left[ \prod_{B_{j} \in {\bf C}_{{\rm NP},\theta}} (1 + B_{j} )^{\langle P, B_{j} \rangle} \right]}{2 i \sum_{n \in 2 {\mathbb N}_0 -1} n v_{n} \sin \frac{\pi K n}{N} \cdot \eta^{n+1}}\right|_{\eta = \eta(\kappa)} \nl
&=& \left. \frac{\sum_{B_{j} \in {\bf C}_{{\rm NP},\theta}} \langle P, B_{j} \rangle \log (1+ B_{j})}{2 i \sum_{n \in 2 {\mathbb N}_0 -1} n v_{n} \sin \frac{\pi K n}{N} \cdot \eta^{n+1}}\right|_{\eta = \eta(\kappa)} + O(B_{j}^2). \label{eq:Del_eta_theta_evenN}
\ee
Notice that, by taking $\theta=0$, any elements $B_{j}$ in ${\bf C}_{{\rm NP},\theta}$ at a certain non-zero $\theta$ becomes some cycles expressed by $C_{(n_1,n_2)}$ in Eq.(\ref{eq:D0_evenN_Kgen}).
Thus, Eq.(\ref{eq:Del_eta_theta_evenN}) can extract non-perturbative information due to $C_{(n_1,n_2)}$ from the perturbative sector (\ref{eq:D0_p}) by introducing a non-zero $\theta$.
However, it is extremely though to make relations of the alien derivatives for each $\theta$ to the full transseries solution of the energy (or $\eta$) at once.

\subsection{Odd $N$}
In this part, we briefly describe the odd $N$ cases.
Since no Stokes phenomenon occurs at $\theta = 0$, so that one has to take a non-zero $\theta$ to induce a Stokes phenomenon.
The phases $\theta$ inducing Stokes phenomena are given by
\be
&& \theta = \frac{\pi n}{N} \ \ \mbox{with} \ \ n \in \{-N+\tfrac{1}{2},-N+\tfrac{3}{2},\cdots, N-\tfrac{1}{2} \}. \label{eq:Del_C_theta_oddN}
\ee
We do not argue the specific alien calculus for the odd $N$ cases in more detail because the procedure is the same to the even $N$ cases. It can be easily derived by using Eqs.(\ref{eq:Del_C_evenN_Kgen})(\ref{eq:Del_DP_evenN_Kgen})(\ref{eq:Del_eta_theta_evenN}) and taking $\theta$ in Eq.(\ref{eq:Del_C_theta_oddN}).

\bibliographystyle{utphys}
\bibliography{n_pt_sym.bib}

\end{document}